\documentclass[fleqn,usenatbib]{mnras}

\usepackage{newtxtext,newtxmath}
\usepackage{bm}
\usepackage[T1]{fontenc}

\DeclareRobustCommand{\VAN}[3]{#2}
\let\VANthebibliography\thebibliography
\def\thebibliography{\DeclareRobustCommand{\VAN}[3]{##3}\VANthebibliography}

\usepackage{graphicx}	
\usepackage{amsmath}	

\title[Black hole growth and nuclear star formation]{The importance of nuclear star clusters for massive black hole growth and nuclear star formation in simulated low-mass galaxies}

\author[C. Partmann et al.]{Christian Partmann$^{1}$\thanks{E-mail: partmann@mpa-garching.mpg.de},
Thorsten Naab$^{1},$
Natalia Lah\'en$^{1},$
Antti Rantala$^{1},$
Michaela Hirschmann$^{2},$
\newauthor
Jessica M. Hislop$^{3},$
Jonathan Petersson$^{2},$
and Peter H. Johansson$^{3}$
\\
$^{1}$Max Planck Institute for Astrophysik, Karl-Schwarzschild-Str. 1, 85748, Garching\\
$^{2}$Institute for Physics, Laboratory for Galaxy Evolution and Spectral modelling, Ecole Polytechnique Federale de Lausanne,\\
Observatoire de Sauverny, Chemin Pegasi 51, 1290 Versoix, Switzerland\\
$^{3}$Department of Physics, University of Helsinki, Gustaf Hällströmin katu 2, FI-00014, Helsinki, Finland\\
}

\date{Accepted XXX. Received YYY; in original form ZZZ}

\pubyear{2023}

\begin{document}
\label{firstpage}
\pagerange{\pageref{firstpage}--\pageref{lastpage}}
\maketitle

\begin{abstract}
Observed low-mass galaxies with nuclear star clusters (NSCs) can host accreting massive black holes (MBH). We present simulations of dwarf galaxies ($M_{\mathrm{baryon}} \sim 0.6 - 2.4 \times 10^8 \rm \, M_\odot$) at solar mass resolution ($0.5\rm \, M_\odot < m_{\mathrm{gas}} < 4 \rm \, M_\odot$) with a multi-phase interstellar medium (ISM) and investigate the impact of NSCs on MBH growth and nuclear star formation (SF). The {\sc griffin} simulation model includes non-equilibrium low temperature cooling, chemistry and the effect of H{\sc ii} regions and supernovae (SN) from massive stars. Individual stars are sampled down to 0.08 $\rm M_\odot$ and their non-softened gravitational interactions with MBHs are computed with the regularised {\sc KETJU} integrator. MBHs with masses in the range of $10^2 - 10^5 \, \rm M_\odot$ are represented by accreting sink particles without feedback. We find that the presence of NSCs boost nuclear SF (i.e. NSC growth) and MBH accretion by funneling gas to the central few parsecs. Low-mass MBHs grow more rapidly on $\sim 600$ Myr timescales, exceeding their Eddington rates at peak accretion. MBH accretion and nuclear SF is episodic (i.e. leads to multiple stellar generations), coeval and regulated by SN explosions. On 40 – 60 Myr timescales the first SN of each episode terminates MBH accretion and nuclear SF. Without NSCs, low-mass MBHs do not grow and MBH accretion and reduced nuclear SF become irregular and uncorrelated. This study gives the first insights into the possible co-evolution of MBHs and NSCs in low-mass galaxies and highlights the importance of considering dense NSCs in galactic studies of MBH growth.  
\end{abstract}



\section{Introduction}

Recent observations have established that many low-mass or dwarf  galaxies host massive black holes (MBHs with $100 \, {\rm M_\odot} \lesssim M_{\rm MBH} \lesssim 10^6 \, \rm M_\odot$, or intermediate mass black holes, IMBHs, with $100 \, {\rm M_\odot} \lesssim M_{\rm MBH} \lesssim 10^5 \, \rm M_\odot$) in their centers \citep{2007ApJ...670...92G,Greene2020ARA&A..58..257G,Reines_2022, askar2024intermediatemassblackholesstar}, albeit with lower masses than their supermassive black hole (SMBH, $M_{\rm SMBH} \gtrsim 10^6 \, \rm M_\odot$) counterparts \citep[see e.g.][]{2013ARA&A..51..511K} in massive galaxies. Also following the trend of the BH mass - velocity dispersion relation \citep[$M_{\rm MBH}-\sigma_{*}$ relation, see e.g.][]{2006ApJ...641L..21G}, dwarf galaxies are plausible host galaxies for central black holes (BHs) above the stellar mass regime ($\geq 10^2 \, \rm M_\odot$). So far a small but steadily increasing number of MBH candidates have been observationally detected \citep[see][for a review]{Greene2020ARA&A..58..257G, askar2024intermediatemassblackholesstar}. 

Such low-mass BHs are difficult to detect for several reasons \citep[e.g.][]{Silk_2017}. Due to their low masses the dynamical impact on the surrounding environment is small. This makes a dynamical detection, which requires a resolved gravitational sphere of influence, very challenging. In addition, the low BH masses result in low accretion rates and hence lower luminosities from the accretion disc, when compared to the active galactic nuclei (AGN) of massive galaxies. Therefore, X-ray surveys can only detect the most massive and actively accreting BHs in dwarf galaxies \citep[e.g.][]{2007ApJ...656...84G,Sharma_2020}. In addition, it is unclear what fraction of time the BHs in dwarf galaxies are active and even a full census of actively accreting BHs might miss the majority of BHs in dwarf galaxies \citep{Pacucci_2021}. Furthermore, dwarf galaxies have very shallow gravitational potential wells, such that their central BHs can be easily displaced from the galactic center and might hence be harder to find \citep[e.g.][]{2020ApJ...898L..30M}. 

Many recent studies have overcome observational complications and the sample size of active BHs in dwarf galaxies at high and low redshifts is steadily increasing \citep[e.g.][]{Ibata_2009, Reines_2012, Mezcua_2018, Mezcua_2019, 2019MNRAS.489L..12K, Greene2020ARA&A..58..257G, Mezcua_2020, Reines_2020, Birchall_2020, Zaw_2020, Latimer_2021, Davis_2022, Reines_2022, Schutte_2022, Mezcua_2023, Zheng_2023, bykov2023srgerosita, 2023A&A...677A.145U,sacchi2024xray}. Most of the detected BH candidates are in the range of $10^5-10^6 \, \rm M_\odot$ which is close to the SMBH regime. The majority of lower mass BHs, if they exist, would still be missed by current observations. A number of MBH candidates at lower masses have also been detected in star clusters \citep[e.g.][]{K_z_ltan_2017, L_tzgendorf_2015, 2013A&A...552A..49L, 2013A&A...555A..26L, 2002AJ....124.3270G}, including a dynamically detected MBH candidate with a mass of at least {$\sim 8200 \, \rm M_\odot$} in the center of  $\omega$ Centauri \citep{H_berle_2024}.

While the population of BHs in dwarf galaxies is still elusive, it has been established that the majority of dwarf galaxies host massive star clusters at their center \citep[e.g.][]{Neumayer_2020, 2024A&A...682A..36H}. These very bright centrally located massive clusters are termed nuclear star clusters (NSC). In galaxies with stellar masses above $\geq 10^8 \, \rm M_\odot$, $30-80 \%$ of the galaxies host a NSC at their center \citep{Hoyer_2021}. The typical stellar surface densities are between $10^3$ and $10^4 \, \rm M_\odot/pc^2$ inside the NSC effective radius \citep{Pechetti_2020, Hoyer_2023} and their expected mass scales with the mass of the host galaxies \citep{Neumayer_2020}. The origin of NSCs is still unclear, in particular it is an ongoing debate whether they form {\it{in-situ}} via star formation in the galactic center or if they are the remnants of globular clusters (or massive star clusters) that formed outside the galactic center and have migrated inwards due to dynamical friction. In general, a combination of both formation mechanisms is possible. For low-mass NSCs (i.e. in galaxies with stellar masses below $M_{*} < 10^9 \, \rm M_\odot$), the sinking and merging of globular clusters is currently the preferred formation channel for the majority of NSCs \citep{Neumayer_2020, Fahrion_2022}. Even though the NSCs are typically very old, there is clear evidence for multiple stellar populations and ongoing star formation \citep{Fahrion_2021, Fahrion_2022b, Fahrion_2024}.

There is a clear connection between NSCs and BHs \citep[see e.g.][for a review]{Neumayer_2020}. Many BHs in the intermediate mass to the supermassive regime are embedded in a NSC and the NSC mass is tightly correlated with the mass of the BH (\citet{Neumayer_2020}. \citet{hoyer2024massiveblackholesnuclear} also find such a relation for galaxies with masses $\gtrsim 10^{8.5} \, \rm M_\odot$. Observationally it is not clear whether all NSCs host BHs or if all accreting BHs in  dwarf galaxies are embedded in a NSC (e.g. it is unclear if the X-ray detected BH candidates in the dwarf galaxies presented in \citealt{bykov2023srgerosita} are embedded in NSCs). In contrast to massive galaxies, the NSCs of dwarf galaxies can be several orders of magnitude more massive than their central BH. For theoretical studies of galactic centers and the physical processes of BH growth in a dwarf galaxy, a NSC is a crucial component. It can dominate the entire central gravitational dynamics while the gravitational sphere of influence of the central BH might not even extend beyond the NSC.

The physical connection between BHs and NSCs might also be related to the solution of several open theoretical problems regarding the formation and growth of MBHs in general. The origin of the observed population of SMBHs is still debated, but many theoretical studies find that the growth of low-mass BH seeds (e.g. BH seeds from the remnants of PopIII stars, the first generation of stars in the high-redshift Universe) is too inefficient in order to grow them to the IMBH or SMBH mass scale \citep[see e.g.][for a review on MBH formation]{Inayoshi_2020}. The additional potential well from a dense NSC might promote the growth of low-mass seeds and provide a pathway for the ``light BH seeding'' scenario. In addition, many theoretical studies find that low-mass BHs might not be able to sink to the center of their host galaxy during the hierarchical assembly of galaxies \citep[e.g.][]{Islam_2003, 2003ApJ...593..661V, Volonteri_2005, Tremmel_2017, Pfister_2019, Ma_2021, 2021MNRAS.505.5129B, 2022MNRAS.513..670N, Beckmann_2023, 2024MNRAS.532.4681P}. If these BHs were embedded in a NSC, the enhanced dynamical friction would result in more rapid sinking and BH coalescence. Finally, the NSC itself might be an environment where MBHs can form. For high cluster densities, stellar collisions and the tidal disruption of stars can lead to the formation and growth of MBHs \citep[e.g.][]{2004Natur.428..724P, Rizzuto_2022,2023MNRAS.526..429A,rantala2024frostclusters}. Gas accretion might accelerate the process even further \citep[e.g.][]{2023PhRvD.108h3012K, 2024Sci...384.1488F}.

The role of BHs in dwarf galaxies has also been discussed in several numerical studies. Using large-scale cosmological simulations, many studies have found that the early growth of BHs in galaxies with bulge masses below $\sim 10^9 \, \rm M_\odot$ is suppressed by supernova (SN) feedback \citep{2015MNRAS.452.1502D, 2017MNRAS.468.3935H, Angl_s_Alc_zar_2017, Trebitsch_2018, Koudmani_2021}. As already pointed out by \cite{Sivasankaran_2022}, resolving the multi-phase, turbulent nature of the ISM is critical for understanding the interplay between BH accretion, star formation and the stellar feedback mechanisms. They present a simulation with a gas resolution of up to $\sim 860 \, \rm M_\odot$ including an Eddington limited Bondi-Hoyle accretion prescription and found that a better resolved ISM leads to significantly more bursty accretion histories. Hence, resolving the turbulent multi-phase structure of the ISM is crucial for simulating the growth of MBHs, however this is currently not feasible in large-scale cosmological simulations.

In this work, we explore the physical processes at the centers of low-mass galaxies around BHs in the intermediate mass regime. We demonstrate that the presence of a NSC has a strong impact on nuclear star formation and BH growth and that a NSC component, well supported by observations, will have to be considered in further galactic studies of MBH growth. Our simulations with the {\sc griffin}\footnote{\url{https://wwwmpa.mpa-garching.mpg.de/~naab/griffin-project}} simulation model have very high dynamical fidelity at solar mass gas particle resolution, a multi-phase ISM with HII region formation and well resolved supernova blast waves, individually sampled stellar masses down to $0.08 \, \rm M_\odot$, and accurate (unsoftened) gravitational interactions between stars and BHs. All of these features are crucial for the simulations and results presented in this work. 

The paper is structured as follows: First, we present the simulation model in section \ref{sec:model}, including the ISM model (section \ref{sec:ISM}), the BH accretion prescription (section \ref{sec:BHmodel}) and the improved gravity solver {\sc Ketju} (section \ref{sec:Ketjutec}). We introduce the suite of simulations in section \ref{sec:sims} and discuss a selected simulation in detail (section \ref{sec:flagship}) before we present a parameter study for different BH and NSC masses in section \ref{sec:parameterstudy}. Finally, we analyse the impact of the NSC and MBH on the larger scale environment in section \ref{sec:largescaleimpact}. Our results are discussed in section \ref{sec:discussion} before we conclude in section \ref{sec:concl}.

\section{Simulations}
\label{sec:model}
We use the {\sc griffin} simulation model that is designed to represent the three major phases (cold, warm, and hot) of the complex low metallicity multi-phase interstellar medium as accurately as currently possible in a galactic-scale simulation. The model includes a sufficiently high gas resolution to follow the evolution of individual supernova (SN) explosions (a gas resolution of $\lesssim \, 4 \, \rm M_\odot$), the realization of individual stars and detailed models for the ISM chemistry, cooling, heating, star formation, and stellar feedback. Initially presented in \cite{Hu_2014, Hu_2016, Hu_2017}, the model has been continuously improved and was used in several studies of isolated \citep[e.g.][]{Steinwandel_2020, 2022MNRAS.509.5938H, Lah_n_2023} and interacting galaxies (e.g. \citealt{2020ApJ...904...71L, 2024MNRAS.530..645L} and \citealt{2024MNRAS.534..215F} for an analysis of star forming cloud properties in a dwarf galaxy starburst). For this project, we implemented a sink particle accretion prescription for BHs and added {\sc Ketju}, an improved gravity scheme for close encounters between stars and BHs \citep{Rantala_2017,Mannerkoski_2023}. We will briefly present the code and the additions made for this paper in the next sections.

\subsection{ISM model}
\label{sec:ISM}
The {\sc griffin} simulation model is built into the {\sc sphgal} simulation code, an extension of the tree-smoothed particle hydrodynamics (SPH) gravity code {\sc gadget-3} \citep{Springel_2005} with an improved hydrodynamics scheme. This modern SPH solver is based on the reconstruction of the pressure and energy field \citep{Hu_2014, ediss19451} using a Wendland C4 \citep{2012MNRAS.425.1068D} kernel with $N_{\rm ngb} = 100$ neighbors inside the SPH smoothing length. Detailed validation and convergence tests including e.g. the description of artificial viscosity, the treatment of shocks, and the evolution of HII regions and SN explosions are presented in \cite{Hu_2014}, \citet{Hu_2016}, \citet{Hu_2017}, \citet{2019MNRAS.483.3363H}, \citet{Steinwandel_2020}, and \citet{Lah_n_2023}.

The star formation algorithm is described in \cite{Lah_n_2023}. Gas elements with a Jeans mass lower than half of an SPH kernel mass, i.e. $M_{\rm J}(T, \rho) < 0.5 \times 100 \times m_{\rm SPH}$ are labelled star-forming and are immediately converted into "reservoir" particles that are decoupled from the hydrodynamics solver. This reservoir of star-forming particles is then used to realize individual stars. After one dynamical timescale $t_{\rm dyn} = (4 \pi G \rho)^{-1/2}$, accounting for the timescale of the gravitational collapse on unresolved scales, the reservoir particles draw stellar masses from the \cite{2001MNRAS.322..231K} initial mass function (IMF) between $0.08$ and $500 \, \rm M_\odot$. If there is enough reservoir mass within a search radius of $1 \, \rm pc$, a star with the drawn initial mass is spawned in the simulation and the corresponding mass is removed from the reservoir particles to ensure mass conservation. The sampled stars get assigned lifetimes based on their initial mass from \citet{2013A&A...558A.103G} at a metallicity of $0.1\,\mathrm{Z}_\odot$ and massive stars immediately start releasing ionising radiation as outlined below. If there is not enough reservoir mass within the search radius, a new stellar mass is drawn from the IMF until a mass compatible with the local reservoir has been found. Individual low-mass stars can be spawned from the same reservoir particle while massive stars can originate from many reservoir particles. To approximate the unresolved gas dynamics in star-forming clouds, every star receives a small Gaussian distributed displacement in position and momentum space. Detailed tests and discussions of this star formation prescription are presented in \cite{Lah_n_2023}.

Non-equilibrium cooling, heating, and chemistry of gas between temperatures of $10 \, \rm K$ and $3 \times 10^4 \, \rm K$ is modelled with a chemical network, that evolves the six chemical species $\rm H_2 , H^{+}, H, CO, C^{+}, O$ and free electrons as introduced in detail in \cite{Hu_2016} \citet{1997ApJ...482..796N}, \citet{2007ApJS..169..239G} and \citet{2012MNRAS.421..116G}. We assume a fixed dust to gas mass ratio of $0.1\%$ and use metal dependent equilibrium cooling tables from \cite{2009MNRAS.393...99W} above $3 \times 10^4 \, \rm K$. As described in \cite{Hu_2016, Hu_2017}, photoelectric heating rates are computed from a spatially and temporally varying interstellar far-ultraviolet radiation field ($6-13.6 \, \rm eV$), using the FUV and PI rates from the \textsc{BaSeL} spectral library at $Z\sim 0.1 \, \rm Z_\odot$ \citep{1997A&AS..125..229L, 1998A&AS..130...65L, 2002A&A...381..524W} combined with the Geneva stellar models \citep{2013A&A...558A.103G}, extrapolated to high- and lower IMF masses outside of the Geneva stellar mass range from $0.8$ to $120 \,\rm M_\odot$. Every gas particle receives FUV radiation from individual stars within $50 \, \rm pc$, attenuated by the optical depth that is computed along 12 sight-lines with the {\sc TreeCol} algorithm \citep{Clark_2011}. In addition to the spatially varying FUV radiation, we model the effect of photoionizing radiation from stars more massive than $8 \, \rm M_\odot$ with a Strömgen sphere approach as described in \cite{2012MNRAS.421.3488H, Hu_2017}. Based on the rate of ionizing photons, gas particles are marked as ionized and heated to $10^4 \, \rm K$, which corresponds to the typical temperature in H{\sc ii} regions. As demonstrated by several studies, early stellar feedback (before the first SN sets in after a few million years) is crucial for regulating star formation and the properties of star clusters in ISM and galaxy scale simulations \citep[e.g.][]{2021MNRAS.504.1039R,2021MNRAS.506.3882S,2022MNRAS.509.5938H,Andersson_2023, 2024A&A...681A..28A}. We also include feedback from asymptotic giant branch stars with masses between $0.5$ and $8 \, \rm M_\odot$ as a single burst at the end of their lifetime \citep{2010MNRAS.403.1413K}.

Stars between $8-50 \rm \, M_\odot$ explode as type II supernovae at the end of their lifetimes. For each supernova, a fixed thermal energy of $E_{\rm SN} = 10^{51} \rm ergs$ is distributed isotropically into the ISM using a {\sc healpix} map \citep{Gorski_2005} to find the closest $8 \pm 2$ particles in each of the 12 {\sc healpix} bins \citep{Lah_n_2023}. In contrast to lower-resolution simulations, the injection of thermal energy is well suited for resolving the Sedov phase of each SN explosion at the target resolution of $< 4 \rm \, M_\odot$ without suffering from "overcooling" issues \citep{Hu_2016,Steinwandel_2020}. Supernovae enrich the ISM isotropically by distributing the generated metals into the $8 \pm 2$ closest particles in each {\sc healpix} bin using the metal yields from \cite{2004ApJ...608..405C} and we track the abundance of 11 elements. For the few star particles with initial stellar masses of $> 50 \rm \, M_\odot$, we assume a direct collapse into a stellar mass BH without a supernova explosion.

\begin{table*}
\centering
\begin{tabular}{c c || c c c c c }
 IC & component & $M \, \rm [M_\odot]$ & size [kpc] & $ m_{\rm res} \, \rm [M_\odot]$ & $\epsilon_{\rm soft} [$\rm pc$] $ & $N_{\rm part}$ \\ 
 \hline
 \hline
{\tt massive} dwarf galaxy &dark matter & $1.1 \times 10^{11}$ & $27.9\,(r_{1/2})$ & 680.0 & 20.0 & $ 1.6 \times 10^8$ \\
& disc gas & $1.6 \times 10^{8}$ & 2.3\,($r_{\rm s})$ & 4.0 & 0.1 & $ 4.0 \times 10^7$ \\
  & disc stars & $ 8.0 \times 10^{7}$ & 1.2 ($r_{\rm s}$) & 4.0 & 0.1 & $ 2.0 \times 10^7$ \\
 &NSC & $1.0  \times 10^{6}$ & $3\times 10^{-3}$ ($r_{1/2}$) & 10.0 & 0.5  & $ 1.0 \times 10^5$ \\
 \hline
 \hline
{\tt fiducial} dwarf galaxy &dark matter & $2.7 \times 10^{10}$ & 17.6 ($r_{1/2}$) & 340.0 & 20.0 & $ 8.0 \times 10^7$ \\
& disc gas & $4.0 \times 10^{7}$ & 0.7 ($r_{\rm s}$) & 4.0 & 0.1  & $ 1.0 \times 10^7$ \\
  & disc stars & $ 2.0 \times 10^{7}$ & 0.7 ($r_{\rm s}$)  & 4.0 & 0.1  & $ 5.0 \times 10^6$ \\
&NSC & $5.0  \times 10^{5}$ & $3\times 10^{-3}$ ($r_{1/2}$) & 10.0 & 0.3  & $ 5.0 \times 10^4$ \\

 \end{tabular}
\caption{Properties of the two isolated dwarf galaxy initial conditions (ICs). Both ICs have dark matter halos with a spin parameter of $\lambda = 0.02$, the disc scale height of the pre-existing stellar component is $0.35 \, \rm kpc$ and the initial metallicity is $Z = 0.1 Z_\odot$. The exponential scale radii and half-mass radii are $r_s$ and $r_{1/2}$, respectively. The gravitational softening length is $\epsilon_\mathrm{soft}$ and $N_\mathrm{part}$ is the particle number in each component. If not stated otherwise, simulations have a gas resolution and NSC mass as shown in this table. All simulations (including the MBH mass, physical model and resolution variations) are listed in Table \ref{table:simulations}.}
\label{table:ics}
\end{table*}

\begin{figure*}
	\includegraphics[width=1\columnwidth]{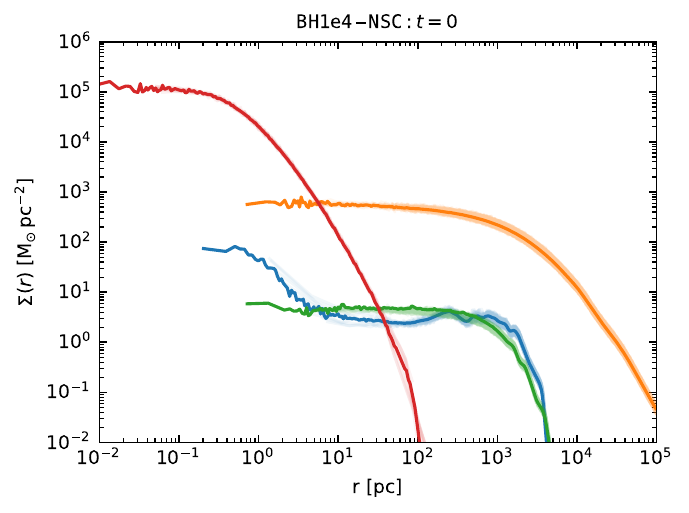}
	\includegraphics[width=1\columnwidth]{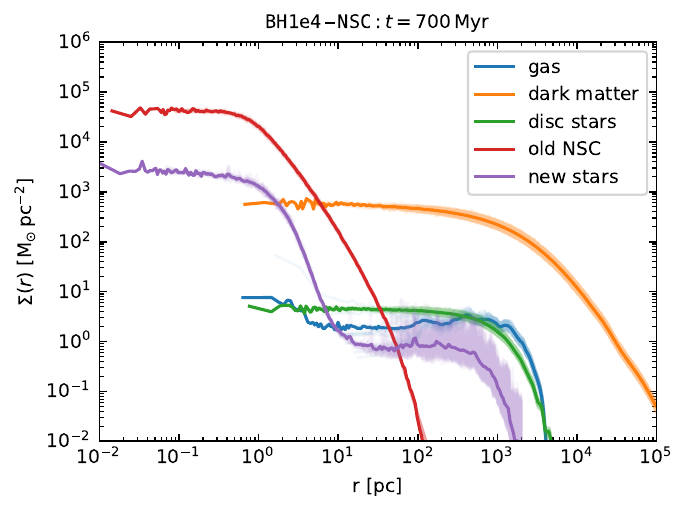}
    \caption{Surface density profiles of the {\tt fiducial} dwarf galaxy {\tt BH1e4-NSC} at the start (left) and at the end of the simulation ($t = 700 \, \rm Myr$, right). The initial condition consists of a dark matter halo (orange), a gas component (blue), a stellar disc (green) and a NSC (red). The final surface density profiles are obtained by stacking 10 snapshots within $10 \, \rm  Myrs$ to reduce sampling effects. The colored background indicates the typical scatter between the stacked snapshots. The scatter is especially large for the stars that formed during the simulation ("new stars", purple) due to their strongly clustered spatial distribution. During the simulation, a young population of new stars forms in the nucleus, adding mass to the pre-existing nuclear star cluster (red). The pre-exising NSC develops a density core on the scale of a few gravitational softening lengths $\epsilon_{\rm soft} \sim 0.3 \rm \, pc$.}
    \label{fig:sufracedensity}
\end{figure*}

\subsection{Black hole accretion model}
\label{sec:BHmodel}

In this study we explore under which conditions central MBHs in dwarf galaxies can grow via accretion of gas from the turbulent ISM. The resolution in our simulation is sufficient to resolve the Bondi-Hoyle-Lyttleton radius of an MBH. For example, for a BH with $M_{\rm BH} = 10^4 \, \rm M_\odot$, the Bondi radius of gas in the warm ISM phase with $T=10^4 \, \rm K$ is $r_{\rm Bondi} \sim 0.9 \, \rm pc$ and increases to $r_{\rm Bondi} > 29 \, \rm pc$ for gas in the cold ISM phase ($T<300 \, \rm K$). It is, however, not possible to resolve a BH accretion disc. Thus, as an approximation, we use a sink particle prescription, remove gas from the simulation at the resolution limit, and assign it to an unresolved accretion disc reservoir. Following \cite{Bate_1995}, our sink particle accretion prescription depends on three criteria which can be expressed in terms of the relative distance $\bm{r} = \bm{r}_{\rm gas} - \bm{r}_{\rm BH}$ and relative velocity $\bm{v} = \bm{v}_{\rm gas} - \bm{v}_{\rm BH}$ between the BH and the gas particles, the BH mass $(M_{\rm BH})$, and the specific internal energy of the gas $(u_{\rm gas})$. First we impose a maximum accretion radius $r_{\rm acc}$ that is chosen to be close to the resolution limit, but is still well resolved within the simulation. We do not allow for accretion outside this region to avoid removal of gas from spatial scales that are still well resolved and where gas might still become star-forming or affected by stellar feedback instead of falling into the BH. For the fiducial simulations, we choose $r_{\rm acc} = 0.5 \, \rm pc$, which corresponds to a typical SPH smoothing length of star-forming gas (i.e. the scale below which gravitational collapse is not well resolved anymore). For every gas particle inside $r_{\rm acc}$, we check whether it is gravitationally bound to the BH 
\begin{equation}
\label{eqn:sinkbound}
    \frac{1}{2}  \lvert   \bm{v} \rvert^2 + u_{\rm gas} < \frac{G  M_{\rm BH}}{\lvert  \bm{r} \rvert} \, ,
\end{equation}
and allow accretion if the angular momentum of the gas particle with respect to the sink particle cannot anymore sustain a circular orbit at the accretion radius, i.e. $l = | \boldsymbol{r} \times \boldsymbol{v} | < \sqrt{G \, M_{\rm BH} \, r_{\rm acc}}$. SPH particles that fulfill the accretion criteria are removed from the simulation and added to an un-resolved accretion disc reservoir. We are not considering BH feedback in this work and therefore only the mass of the BH sink particle is dynamically important. It represents the combined mass of the BH, $M_{\rm BH}$, and the mass in the unresolved sub-grid accretion disc, $M_{\rm disc}$. In this study we make no additional assumptions about the interaction between the accretion disc and the BH and use the dynamical mass of the sink particle as the BH mass throughout the paper. To smooth out the discreetness due to the accretion of SPH particle with given discrete mass, we compute the sink particle accretion rate from a running mean over $2 \, \rm Myrs$. Furthermore, we log several properties of the accreted gas particles (time of accretion, mass, particle ID, temperature, angular momentum), for further analysis in post-processing, thus taking advantage of the Lagrangian nature of the hydrodynamical scheme.

\begin{table*}
\begin{tabular}{c | c c c c c c }
 label & $M_{\rm BH}\, \rm [M_\odot]$ & $M_{\rm NSC}\, \rm [M_\odot]$ & $r_{1/2}^{\rm NSC} \, \rm [pc]$ & $\Delta m_{\rm gas} \, \rm [M_\odot]$   & & \\ 
 \hline
 \hline
 {\tt BH1e3-NSC-massive}   & $10^3$ & $ 1 \times 10^6 $   & 3  & 4.0 & - & \\ 
 {\tt BH1e3-NSC-massive-hr}   & $10^3$ & $ 1 \times 10^6 $   & 3  & 1.0 & only one SF cycle & \\ 
 \hline
 \hline
 {\tt BH1e2-NSC}   & $10^2$   & $ 5\times10^5 $   & 3  & 4.0 & - & \\ 
 {\tt BH1e3-NSC}   & $10^3$   & $ 5\times10^5 $   & 3  & 4.0 & - & \\ 
 {\tt BH1e4-NSC}   & $10^4$   & $ 5\times10^5 $   & 3  & 4.0 & - & \\ 
 {\tt BH1e5-NSC}   & $10^5$   & $ 5\times10^5 $   & 3  & 4.0 & - & \\ 
 \hline
 {\tt BH1e2}   & $10^2$   & -   & -  & 4.0 & - & \\ 
 {\tt BH1e3}   & $10^3$   & -   & -  & 4.0 & - & \\ 
 {\tt BH1e4}   & $10^4$   & -   & -  & 4.0 & - & \\ 
 {\tt BH1e5}   & $10^5$   & -   & -  & 4.0 & - & \\ 
 \hline
 \hline
 {\tt BH1e4-NSC-2.0}   & $10^4$   & $ 5\times10^5 $   & 3  & 2.0 & - & \\ 
 {\tt BH1e4-NSC-1.0}   & $10^4$   & $ 5\times10^5 $   & 3  & 1.0 & - & \\ 
 {\tt BH1e4-NSC-1.0-fix}   & $10^4$   & $ 5\times10^5 $   & 3  & 1.0 & fixed SF Jeansmass & \\ 
 {\tt BH1e4-NSC-0.5}   & $10^4$   & $ 5\times10^5 $   & 3  & 0.5 & - & \\ 
 \hline
 \hline
 {\tt BH1e4-NSC-racc}   & $10^2$   & $ 5\times10^5 $   & 3  & 4.0 &$r_{\rm acc} = 1.0, 0.3, 0.1 \, \rm pc $& \\ 
 \hline
 \hline
 
\end{tabular}
\caption{Overview of all simulations presented in this paper. The massive galaxy {\tt BH1e3-NSC-massive} hosts a $10^6 \, \rm M_\odot$ NSC with a $10^3 \, \rm M_\odot$ MBH at the center, consistent with observational scaling relations. A high resolution version ({\tt -hr}) with one solar mass gas resolution was simulated for one star formation cycle. The {\tt fiducial dwarf galaxy} is simulated with both nuclear star clusters ({\tt -NSC}) and without (--), and with central MBH masses in the range of $100 - 10^5 \, \rm M_\odot$ ({\tt BH1e2
- BH1e6}). To test convergence, we also run simulations at different mass resolutions ({\tt -2.0, -1.0, -1.0-fix, -0.5}) and varying sink accretion radii ({\tt -racc}).}
\label{table:simulations}
\end{table*}

\subsection{Accurate integration with KETJU}
\label{sec:Ketjutec}

To accurately follow the dynamical interaction of MBHs with their stellar environment, we have implemented the regularized integration scheme {\sc Ketju} into the {\sc griffin} code-base. {\sc Ketju} is an extension of the tree gravity solver {\sc Gadget-3} and {\sc Gadget-4} and has been introduced in \citet{Rantala_2017,2018ApJ...864..113R, 2021ApJ...912L..20M} and \citet{Mannerkoski_2023}. {\sc Ketju} allows to capture the sinking of BHs via dynamical friction and can resolve the orbits of stars around BHs as well as the scattering of stars with BHs and BH binaries. Although not relevant for our study of isolated galaxies with one MBH, {\sc Ketju} also includes post-Newtonian corrections up to order $3.5$ for the forces between BHs, allowing for gravitational wave driven coalescence \citep[see e.g.][]{2024MNRAS.532.4681P}. 

In regions close to BHs (the "{\sc Ketju} region"), {\sc Ketju} uses the accurate {\sc MSTAR} integrator based on algorithmic regularization \citep[see][and references therein]{2020MNRAS.492.4131R} instead of the standard {\sc Gadget-3} leapfrog integrator to follow the orbits of stellar particles around BHs without gravitational softening. In this study, the forces between star particles and BHs are unsoftened (star-BH), while the forces between stars inside the {\sc Ketju} region remain softened (star-star). To allow for a smooth transition of particles in and out of the {\sc Ketju} region surrounding the BH, the gravitational softening $\epsilon_{\rm soft}$ of the star particles must be sufficiently smaller than the size of the Ketju region $r_{\rm Ketju}$ (at least by a factor of 2.8). The details of the regularization method are described in \cite{2020MNRAS.492.4131R}, see also \cite{1965SJNA....2..384G,Bulirsch1966, 1999MNRAS.310..745M,1999AJ....118.2532P,2006MNRAS.372..219M,2008AJ....135.2398M}. In short, the method is based on a time transformation of the equations of motion and a suitable leapfrog integrator that algorithmically circumvent the Newtonian singularity at small particle separations. A minimum spanning tree coordinate system is used to reduce numerical round-off errors and the Gragg–Bulirsch–Stoer (GBS) extrapolation method \citep{1965SJNA....2..384G,Bulirsch1966} ensures extremely high user-desired integration accuracy.

While previous studies have used {\sc Ketju} to model the collisional interaction of SMBHs with stellar population particles in massive galaxies \citep[e.g.][]{2018ApJ...864..113R, 2024MNRAS.528.5080L, 2024MNRAS.530.4058L} as well as between dark matter and stellar particles in low-mass galaxies \citep{2024MNRAS.532.4681P}, in this work we use {\sc ketju} to accurately follow the encounters between individual stars and MBHs. In particular, individual stars can be tidally disrupted by the MBH. To account for this potential MBH growth channel, we implement a check for possible tidal disruption events \citep[TDEs, see e.g.][]{1988Natur.333..523R}, based on the tidal radius $r_{\rm tidal}$
\begin{equation}
\label{eqn:tidal}
    r_{\rm pericenter} < r_{\rm tidal} = 1.3 \, r_{\rm star} \left(\frac{M_{\rm BH} + M_{\rm star}}{2 \,M_{\rm star}}\right)^{1/3} \, .
\end{equation}
We compute the stellar radii assuming a main sequence scaling relation $r_{\rm star} \sim r_\odot \,(M_{\rm star}/M_{\odot})^{0.8}$. This condition is checked at the end of each GBS time-step using the Keplerian elements to compute the pericenter distance $r_{\rm pericenter}$. For a solar mass star, this corresponds to a tidal radius of $r_{\rm tidal} \sim 5 \times 10^{-7} \, \rm pc$. 

In a forthcoming work (Lah{\'{e}}n et al. in prep), this implementation will be expanded by adding {\sc ketju} regularised integration regions around all forming massive stars in an entire galaxy to improve the stellar dynamics of individual stars in star clusters. 

\subsection{Initial conditions}
\label{sec:sims}

We simulate two isolated galaxy models with the global properties specified in Table \ref{table:ics}. We will use the computationally expensive {\tt massive} dwarf galaxy as an example to discuss the physical mechanisms in detail and the lower mass {\tt fiducial} dwarf galaxy for parameter exploration.

The {\tt massive} dwarf galaxy has a total mass of $1.1 \times 10^{11} \, \rm M_\odot$ with a total baryonic mass of $2.4 \times 10^8 \, \rm M_\odot$. The dark matter component follows a \cite{1990ApJ...356..359H} profile with a half-mass radius of $r_{\rm 1/2} = 27.9 \, \rm kpc$ at a particle mass resolution of $680 \, \rm M_\odot$. The baryonic component consists of a gas disc ($M_{\rm gas}=1.6  \times 10^8 \, \rm M_\odot$ in a disc with scale length of $r_s = 2.3 \, \rm kpc$) and an old stellar population of stars ($M_{\rm disc}=8 \times 10^7 \, \rm M_\odot$ in a disc with a scale length of $r_s = 1.2 \, \rm kpc$), both at a particle mass resolution of $4 \, \rm M_\odot$. The {\tt fiducial} galaxy has $25 \, \%$ of the mass of the {\tt massive} galaxy, i.e. a dark matter mass of $M_{\rm dm} = 2.7 \times 10^{10} \, \rm M_\odot$ (with $r_{\rm 1/2} = 17.6 \, \rm kpc$ and a particle resolution of $340 \, \rm M_\odot$), a disc gas mass of $M_{\rm gas} = 4 \times 10^{7} \, \rm M_\odot$ and a stellar mass of $M_{\rm disc} = 2 \times 10^{7} \, \rm M_\odot$ (both with disc scale lengths are $r_s = 0.7 \, \rm kpc$). These initial conditions are inspired by the observed galaxy Holmberg II \citep{Weisz_2009} that hosts an MBH candidate \citep{2007MNRAS.376.1317M}. For both initial conditions, all components are set-up in equilibrium with the method presented in \cite{2005MNRAS.361..776S}. To avoid the initial artificial starburst before the system is relaxed, we employ "peak turbulent driving" for $20 \, \rm Myrs$, i.e. we inject thermal energy in regions that would otherwise be star-forming as described in \cite{Hu_2017}. The initial metallicity is set to $Z = 0.1 \, \rm Z_\odot$ globally.

\begin{figure*}
	\includegraphics[width=2\columnwidth]{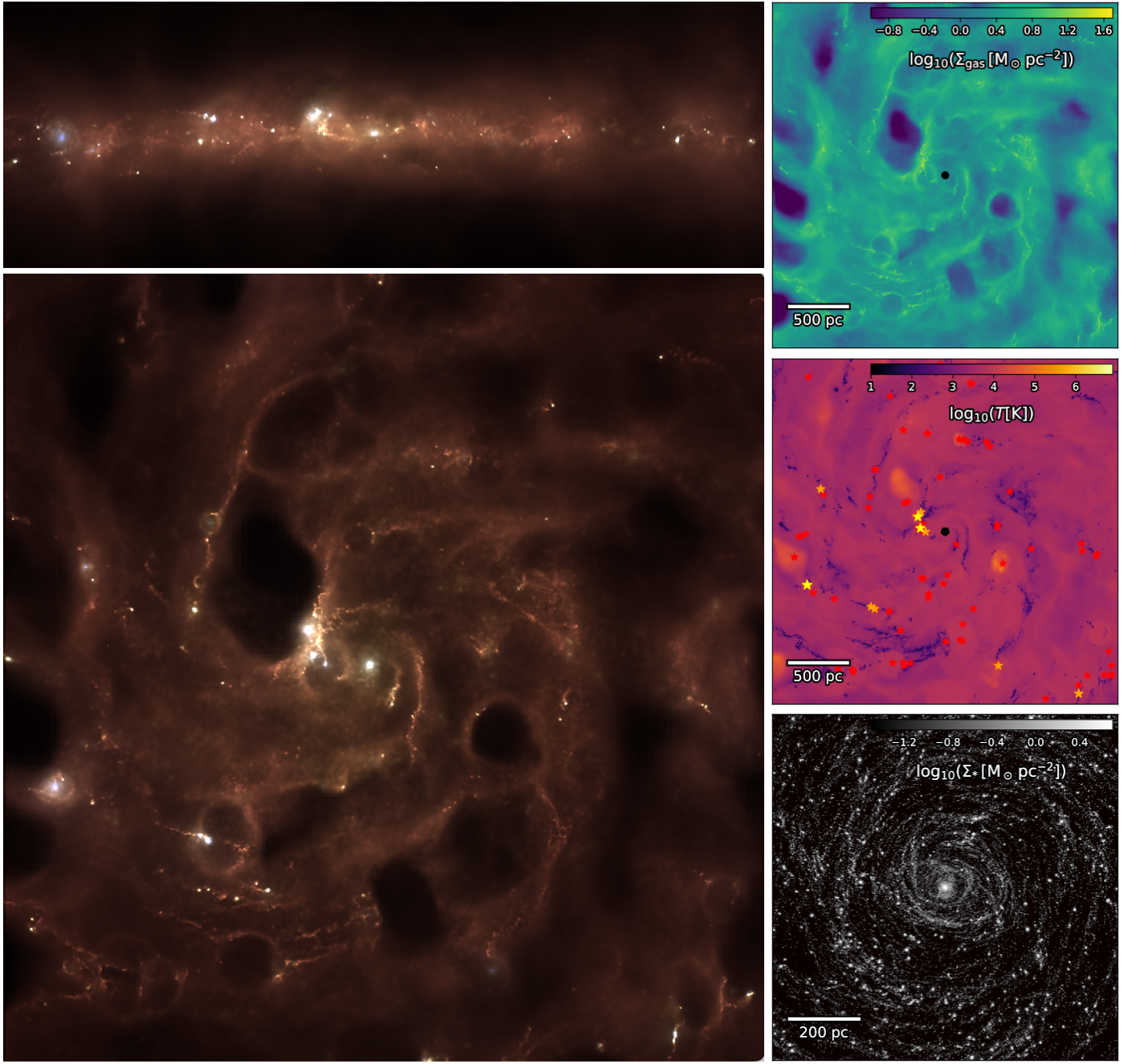}
    \caption{The figure shows the state of the massive dwarf galaxy {\tt BH1e3-NSC-massive} after $\sim 250 \, \rm Myr$ of evolution. The left panels show a colour composite image of dust re-emission in \mbox{24 $\mu$m} (blue), \mbox{160 $\mu$m} (green) and \mbox{250 $\mu$m} (red) bands that was generated with the radiative transfer code \textsc{skirt} (face-on view on the central $3 \, \rm kpc$ in the bottom panel, edge-on in the top panel). The NSC is visible as a bright spot in the center of the galaxy. The gas surface density, temperature distribution and stellar surface density as they are inferred from the simulation directly are shown in the right panel (from top to bottom). The MBH that is placed at the galactic center is represented by a black dot, stars more massive than $8 \, \rm M_\odot$, $15 \, \rm M_\odot$ and $25 \, \rm M_\odot$ are shown as red, orange and yellow star symbols, respectively. The figure is best viewed on a computer screen.}
    \label{fig:overviewplot}
\end{figure*}

\begin{figure*}
	\includegraphics[width=2\columnwidth]{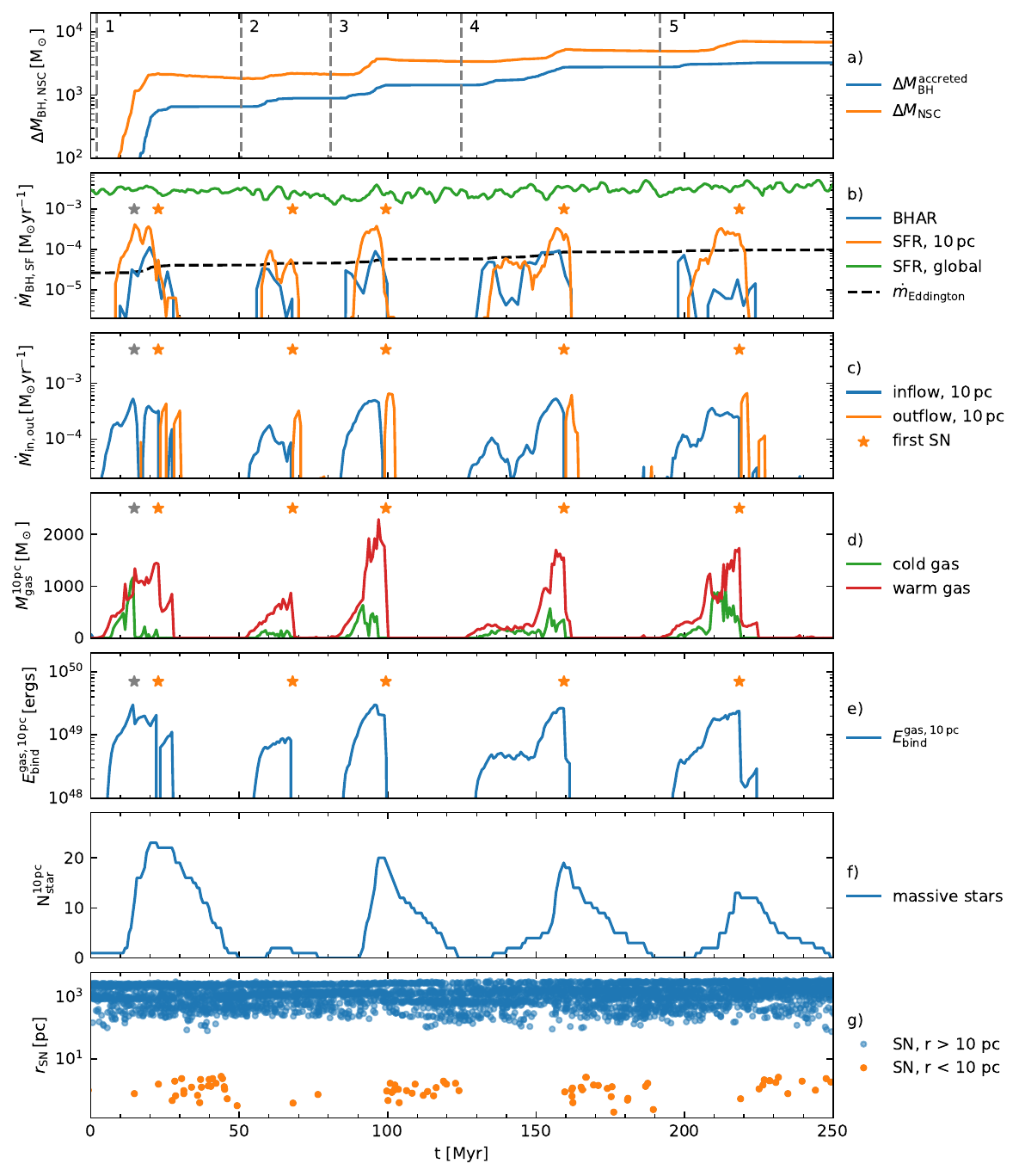}
    \caption{Star formation and MBH accretion cycles in the galactic center (inner 10 pc) of simulation {\tt BH1e3-NSC-massive} with a $10^6 \, \rm M_\odot$ NSC and a $10^3 \, \rm M_\odot$ MBH. The top panel (a) shows the MBH (blue) and the NSC (orange) mass growth, separated into 5 episodes. Within 250 Myr, the MBH grows by $\sim 3200 \, \rm M_\odot$, while the NSC grows by $\sim 7000 \, \rm M_\odot$. Panel (b) shows the MBH accretion rate (BHAR) and the nuclear SFR. The MBH accretion rate can exceed the Eddington limit for short periods of time. The episodic growth cycles are typically terminated by the first SN of the cycle (orange stars). The inflow (blue) and outflow (orange) rates at $r=10\, \rm pc$ are shown in panel (c). Brief outflow phases start after the first SN. Panel (d) shows the nuclear content of cold ($\rm T < 300 \rm \, K$, green) and warm ($300 \rm \, K < \rm T < 2 \times 10^4 \rm \, K$, red) gas. The binding energy (panel e) of the central gas reservoir with respect to the NSC and MBH never exceeds $\sim 4 \times 10^{49} \, \rm ergs$ and thus one single SN can unbind the gas. The number of stars more massive than $8 \, \rm M_\odot$ is shown in panel (f). Gas can only flow back to the center (panel c) as soon as the last massive star has exploded. The nuclear SNe (orange dots, bottom panel g) are spatially separated and clustered in time compared to events in the main galaxy (blue dots). In the first cycle, there is one long-lived massive star with $\sim 8 \, \rm M_\odot$ leftover from the previous cycle that explodes (gray star marker) before the first SN associated with stars born in this cycle fully shuts down BH accretion.}
    \label{fig:overview_main_simulation}
\end{figure*}

To account for a pre-existing NSC, we initialize collision-less particles at a fiducial particle resolution of $10 \, \rm M_\odot$, following a \cite{1993MNRAS.265..250D} profile with $\gamma = 1.75$ and a scale radius of $2.2 \, \rm pc$. This profile corresponds to a half-mass radius of $r_{1/2} \sim 3 \, \rm pc$. Consistent with observed scaling relations presented in \cite{Neumayer_2020}, we choose a NSC mass of $M_{\rm NSC} = 10^6 \,\rm M_\odot$ ($N = 10^5$ particles) for the {\tt massive} and $M_{\rm NSC} = 5 \times 10^5 \,\rm M_\odot$ ($N = 5 \times 10^4$) for the {\tt fiducial} dwarf galaxy, respectively. The gravitational softening lengths of the NSC particles are set to $\epsilon_{\rm soft} = 0.3 \, \rm pc$ and $\epsilon_{\rm soft} = 0.5 \, \rm pc$ for the {\tt fiducial} and the {\tt massive} dwarf galaxy, respectively. The softening leads to a flattening of the initially very cuspy NSC density profile on the softening scale which is comparable to the accretion radius of $r_{\rm acc} = 0.5 \rm \, pc$, below which we do not follow the gas dynamics anymore. The NSCs are initialised in equilibrium with central MBHs in the range of $100 - 10^6 \, \rm M_\odot$ using the method described in \cite{Rantala_2017}. We run the ICs with the full ISM model for $500 \, \rm Myr$ ($300 \, \rm Myr$ for the {\tt massive} galaxy) to relax the system, before adding the MBH/NSC and turning on BH accretion. Stars that form during the initial relaxation process are added to the population of "old" disc stars. All stars that form during the simulation ("new stars") and "old" disc stars have a gravitational softening of $\epsilon_{\rm soft} = 0.1 \, \rm pc$ and are integrated with the {\sc Ketju} integrator in a region of $r_{\rm Ketju} = 0.3 \, \rm pc$ around the MBH, such that all MBH-star forces are resolved without gravitational softening. The dense NSC component that cannot currently be realized in the limit of individual stars is evolved with the standard {\sc Gadget-3} integration scheme.

These ICs are designed to represent a realistic nucleated dwarf galaxy with a "live" NSC and a MBH at its center. Such a NSC is expected to impact the gas, stellar and MBH dynamics as well as star formation and gas accretion onto the MBH in its sphere of influence. Excluding the MBHs, the sphere of influence of the NSC is $r_{\rm NSC}^{\rm infl} \sim 70 \, \rm pc$ (for $M_{\rm NSC} = 5 \times 10^5 \, \rm M_\odot$) in the {\tt fiducial} galaxy. This is slightly smaller than the sphere of influence of $r_{\rm NSC}^{\rm infl} \sim 80 \, \rm pc$ of the NSC in the {\tt massive} dwarf galaxy ($M_{\rm NSC}=10^6 \, \rm M_\odot$). We define the sphere of influence as the radius that encloses a total mass (excluding the NSC itself) of $2 \times M_{\rm NSC}$.

We list all simulations in Table \ref{table:simulations}. For the simulation of the {\tt massive} galaxy, the NSC and MBH properties are chosen to be in the expected mass range based on the NSC-MBH mass relation, i.e. $M_{\rm BH} = 10^3 \, \rm M_\odot$ and $M_{\rm NSC} = 10^6 \, \rm M_\odot$. Then we use the {\tt fiducial galaxy} to test MBH masses between $M_{\rm BH}=100$ and $10^6 \, \rm M_\odot$, representing MBHs from the under-massive to the over-massive regime compared to the expectation for dwarf galaxies with and without a NSC \citep{Neumayer_2020}. The simulations of the {\tt fiducial} dwarf galaxy are simply labelled {\tt BH1e2} - {\tt BH1e6}, indicating the MBH mass, while simulations with a NSC have the extension {\tt NSC}. As an example, we show the surface density profiles of all components of the simulation {\tt BH1e4-NSC} at the beginning and end ($t = 700 \, \rm Myr$) of the simulation in Fig. \ref{fig:sufracedensity}.

\section{Results}

In this section, we first analyse the growth of a central BH inside the NSC of the {\tt massive} dwarf galaxy and show how individual massive stars and their SN explosions (SNe) regulate MBH growth and nuclear star formation cycles. Then, we present a parameter study using the {\tt fiducial} dwarf galaxy to determine under which conditions MBHs and NSCs can grow and how they interact with their environment.

The model {\tt BH1e3-NSC-massive} has a stellar mass of $\sim 10^8 \, \rm M_\odot$ and represents a star-forming dwarf galaxy in the local Universe. It hosts a NSC with a mass of $10^6 \, \rm M_\odot$ and a MBH with an initial mass of $10^3 \, \rm M_\odot$ in the galactic center. We follow the evolution of the system over $\sim 250 \, \rm Myr$. In Figure \ref{fig:overviewplot}, we show the state of the galaxy at the end of the simulation. The colour composite images (left panels) of dust re-emission have been produced with the dusty radiative transfer code \textsc{skirt} \citep{2020A&C....3100381C}. The radiative transfer post-processing was performed following the method outlined in \citet{Lah_n_2022, Lah_n_2023} with small changes in assigning the stellar spectra. The gaseous properties (metallicity, temperature) recorded in the simulation snapshot were used as input for the dust grid built by \textsc{skirt}. The dust-to-metals ratio was set to $\sim 0.8$ to correspond to the fixed mass fraction of 0.1 per cent of the gas mass used in the simulation. The effective temperature and surface gravity that set the spectral energy distributions at a given metallicity in the \citet{2003IAUS..210P.A20C} stellar atmosphere models provided in \textsc{skirt} have been computed from the initial mass and current stellar age using the Geneva stellar models at a metallicity of $Z=0.1\,\mathrm{Z}_\odot$. The old stellar disk particles were supplemented with simple stellar population spectral energy distributions from \citet{2003MNRAS.344.1000B} at a metallicity of $Z=0.1\,\mathrm{Z}_\odot$ and randomly distributed ages according to a linearly decaying star formation history starting \mbox{$\sim13$ Gyr} ago and ending at the present with a star formation rate (SFR) of a few $10^{-4}\,\mathrm{M}_\odot\,\mathrm{yr}^{-1}$. Similarly, we assigned spectral energy distributions according to \citet{2003MNRAS.344.1000B} with a fixed age of 10 Gyr and a metallicity of $Z=0.1\,\mathrm{Z}_\odot$ to the NSC particles. Radiation from the accreting BH is not considered here. Broadband filters readily available in \textsc{skirt} used in Fig. \ref{fig:overviewplot} are the \textit{Spitzer Space Telescope} \mbox{24 $\mu$m} and \mbox{$160$ $\mu$m} equivalent bands and the \textit{Herschel Space Observatory} 250 $\mu$m equivalent band. The pixel resolution has been selected as $\sim 0.5\, \mathrm{pc}$ which corresponds to $\sim0.03$\arcsec{} at a distance of 3 Mpc (typical \textit{Hubble Space Telescope} or \textit{James Webb Space Telescope} resolution) or $\sim2.5$\arcsec{} at a distance of 40 kpc (\mbox{24 $\mu$m} \textit{Spitzer} pixel scale). The images have been degraded with a Gaussian point spread function with a full width at half maximum of 2 pixels. The right panels of Fig. \ref{fig:overviewplot} show the face on gas surface density (top), gas temperature (middle), and the stellar surface density (bottom).

\subsection{Black hole growth and nuclear star formation cycles}
\label{sec:flagship}

In the top panel (a) of Fig. \ref{fig:overview_main_simulation}, we show the mass-growth of the MBH, $\Delta M_{\rm BH}^{\rm accreted} = M_{\rm BH}(t) - M_{\rm BH}(t_0)$, and the growth of the NSC, $\Delta M_{\rm NSC}$. We define $\Delta M_{\rm NSC}$ as the new stellar mass that forms inside the central $10 \, \rm pc$ after the simulation has started. Throughout this paper, we choose a spherical region of $10 \, \rm pc$ around the MBH to analyse the physical processes at the galactic center (inside the NSC/MBH sphere of influence). The MBH and the NSC grow simultaneously during short episodes (separated into five numbered cycles by dashed lines). The absolute NSC growth is larger than the MBH growth as most of the gas in the galactic center is turned into stars before it can be accreted by the MBH. Nevertheless, the MBH grows from $1000 \, \rm M_\odot$ to $\sim 4200 \, \rm M_\odot$ and has more than quadrupled its mass by the end of the simulation. This corresponds to an e-folding timescale of $\sim 180 \, \rm Myr$, much faster than in comparable simulations without a NSC (see Section \ref{sec:parameterstudy}). In the same time, the NSC has grown by $\sim 7 \times 10^3 \, \rm M_\odot$ which corresponds to a relative growth of only $\sim 0.7$ percent. The gas accreted onto the MBH is roughly at a constant fraction of $\sim 40$ percent of the gas turned into stars in the nuclear region.

\begin{figure*}
	\includegraphics[width=2\columnwidth]{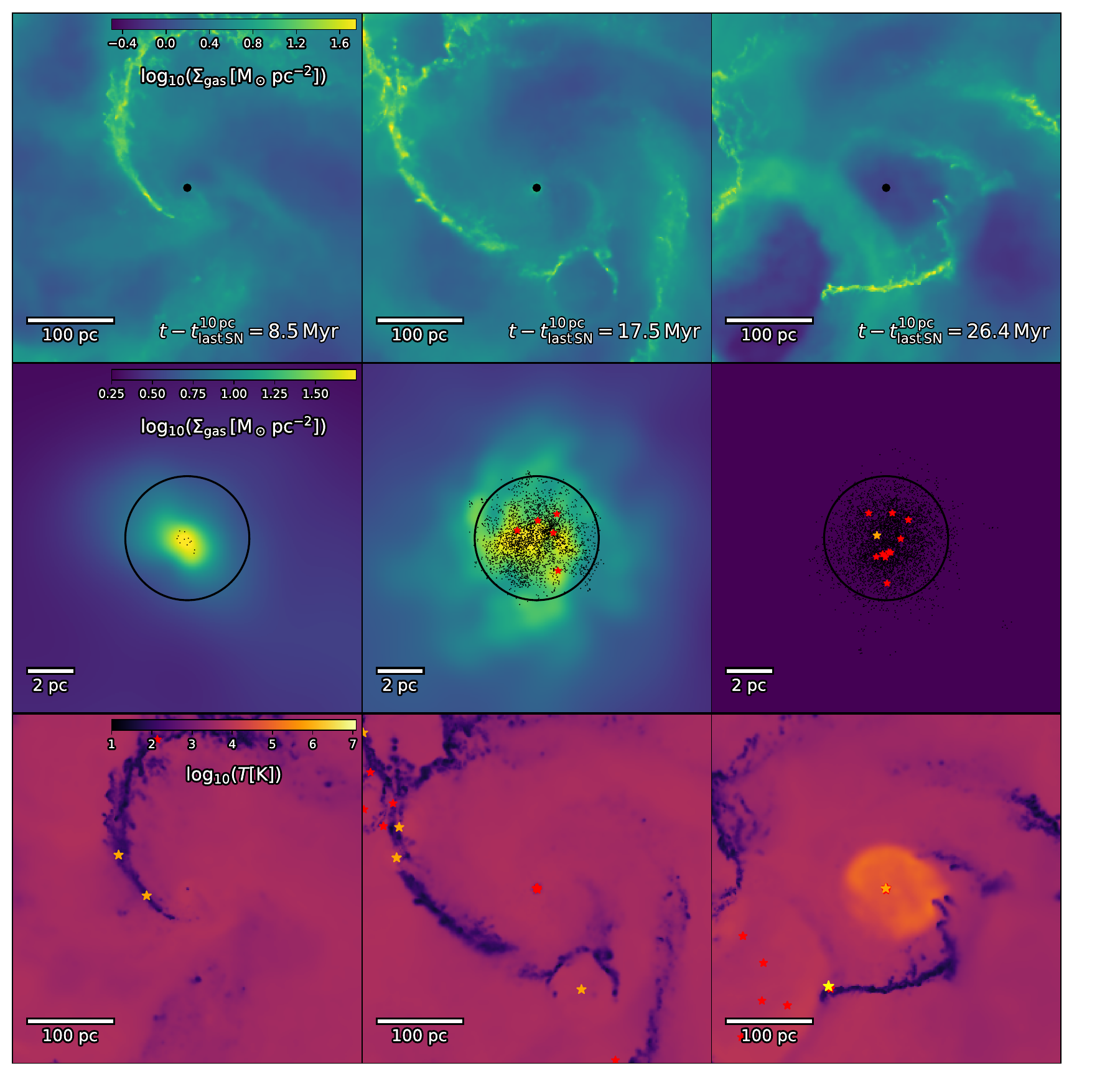}
    \caption{Illustration of one accretion cycle in the {\tt massive} dwarf galaxy {\tt BH1e3-NSC-massive-hr} (i.e. a re-simulation of {\tt BH1e3-NSC-massive} at a global resolution of $1 \, \rm M_\odot$). The top panel shows the gas surface density in the $\sim 200 \, \rm pc$ centered on the BH, the middle panel a zoom-in on the gas and stellar distribution in a $\sim 6 \, \rm pc$ region around the BH. The half-mass radius of the NSC is shown as a black circle, stars that form during this cycle are displayed as black dots ($M_{*} < 8 \, \rm M_\odot$). Massive stars are highlighted as red, orange or yellow star markers for initial stellar masses of $8-15 \, \rm M_\odot$, $15-25 \, \rm M_\odot$ and $>25 \, \rm M_\odot$, respectively. The bottom panel shows the temperature distribution in the larger scale environment $(\sim 200 \, \rm pc)$. The left column shows how gas starts to re-accumulate inside the NSC region $\sim 8.5 \, \rm Myrs$ after the most recent SN explosion in the central $10 \, \rm pc$ that has terminated the previous nuclear accretion cycle. After $\sim 17.5 \, \rm Myrs$, the nuclear SF has led to the formation of several massive stars. The cycle is terminated after $\sim 26 \, \rm Myr$ when the first SN ejects the gas from the galactic center (visible as a hot, low density bubble in the bottom right panel).}
    \label{fig:accretioncycleexample}
\end{figure*}

\begin{figure}
	\includegraphics[width=1.0\columnwidth]{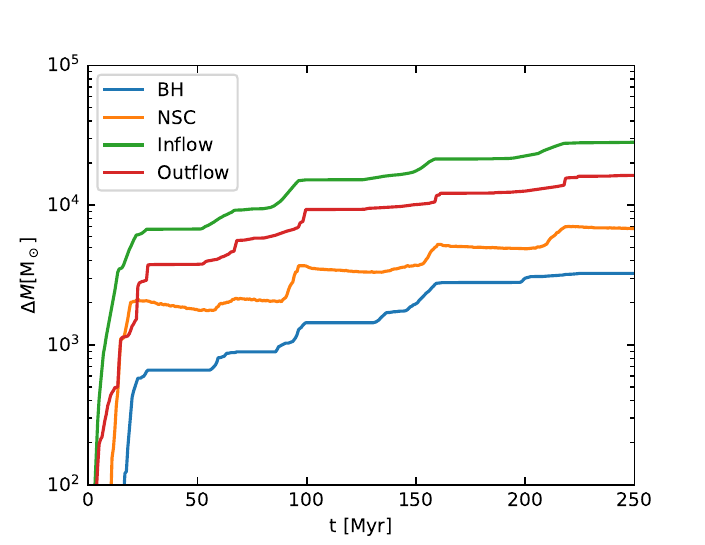}
    \caption{The mass budget in the central $10 \, \rm pc$ around the BH as a function of time for simulation {\tt BH1e3-NSC-massive}. We show the BH mass growth (blue line), NSC mass growth (orange line), and the integrated gas inflow (green line) and outflow (red line). A total of $\sim 3 \times 10^4 \, \rm M_\odot$ flows into the center of the galaxy, $\sim 11$ percent is accreted by the BH, $\sim 25$ percent is converted into stars. The remaining mass is lost in stellar feedback driven outflows.}
    \label{fig:conversioneff_example}
\end{figure}

The second panel (b) shows the time-evolution of the MBH accretion rate and SFR inside the central $10 \, \rm pc$. The concurrent episodes of star formation (orange) and MBH accretion (blue) cycles are clearly visible. While the global SFR remains constant at $\sim 4 \times 10^{-3} \, \rm M_\odot\, yr^{-1}$, the nuclear SFR and MBH accretion rate are regulated by massive stars. As soon as the first supernova of the respective star formation episode explodes (orange star symbols) star formation and gas accretion cease. The cycles of MBH accretion and star formation typically last for $10 - 30 \, \rm Myr$, followed by $20 - 40 \, \rm Myr$ of quiescence. During the accretion phases, the MBH accretion rate can become as high as $10^{-4} \, \rm M_\odot\, yr^{-1}$ and sometimes exceed the Eddington rate (black dashed line). The nuclear SFR is typically larger than the BH accretion rate and can become as high as $\sim  5 \times 10^{-4} \, \rm M_\odot\, yr^{-1}$.

The inflow (blue) and outflow (orange) rates of gas measured at a radius of $10\, \rm pc$ are shown in panel (c) of Fig. \ref{fig:overview_main_simulation}. The inflow typically starts before the SFR and MBH accretion rate increase (see panel b) and continues for 10 - 30 Myr. The inflow is terminated by the first nuclear supernova (orange star) which drives a short phase of outflow for a few Myr ending the respective cycle.  

Up to $\sim 2500 \, \rm M_\odot$ of warm and $\sim  1000 \, \rm M_\odot$ of cold gas can accumulate in the central $10 \, \rm pc$ during the accretion cycles (red and green lines in panel d of Fig. \ref{fig:overview_main_simulation}). The cold gas (${\rm T} < 300 \, \rm K$) is also the reservoir for star formation while the warm ($300 \rm \, K < \rm T < 2 \times 10^4 \,\rm K$) gas dominates the nuclear mass budget. There is little gas at higher temperatures. The MBH accretes from both phases (cold and warm phase, see Section \ref{sec:parameterstudy} for a more detailed discussion), as long as the gas is bound to the MBH and has sufficiently low angular momentum. For the simulation time shown here $\sim 1800\, \rm M_\odot$ of cold gas and $\sim 1600 \, \rm M_\odot$ of warm gas are accreted onto the MBH. In the fifth panel (e) we show the total binding energy of all gas inside $10 \, \rm pc$ as a function of time. As the maximum binding energy never exceeds $\sim 4 \times 10^{49} \,\rm ergs$ a single supernova event with $E_{\rm SN} = 10^{51}\,\rm ergs$ can easily unbind the nuclear gas component and thus terminate star formation and MBH accretion.

Supernovae explode in the simulation whenever a massive star with $8\, \rm M_\odot < M_{\rm star} < 50 \, \rm M_\odot$ comes to the end of its lifetime. The number of massive stars in the central $10 \, \rm pc$ is shown in panel (f). Each star formation cycle leads to the formation of up to $\sim$ 25 stars with initial masses greater than $8 \, \rm M_\odot$ (blue). Typically, after the last massive star has exploded, a new accretion cycle starts. The stellar lifetimes range between a few Myr and $\sim 32 \, \rm Myr$, such that every accretion and star formation cycle leads to a number of spatially clustered SNe in the galactic center (bottom panel g, orange dots). Only after the last nuclear SN has exploded, inflow to the center starts again and the next cycle can begin. 

The only exception here is the first cycle (shown as cycle one in panel a), where a long-lived massive star with $\sim 8 \, \rm M_\odot$ is left-over from the previous accretion cycle (gray star marker). Due to its low mass, the output of radiation from this star is not sufficient to prevent the accumulation of gas for the next accretion cycle (see Sec. \ref{sec:ISM} for the modelling of stellar radiation). It explodes $\sim 10 \, \rm Myr$ before the first SN explosion associated with stars that were born in cycle one (orange star marker) happens. Because this SN explosion takes place in a relatively dense and cold environment that has not yet been dispersed by the early stellar feedback and where the cooling timescale is short, this SN leads to a temporary drop in the inflow rate and heats up the gas, but it does not terminate the cycle. Cycle one only ends after the first SN associated with stars that form during this has exploded in the warm ISM, that has already been heated up by the stellar feedback \citep[for a discussion of the effect of early stellar feedback on the impact of SN explosions, see e.g.][]{2022MNRAS.509.5938H}. We also note that stellar radiation alone can in some cases drive gas outflows from the central $10 \, \rm pc$. For example in cycle three, the outflow rate already increases slightly before the first SN explosion as a result of strong radiation feedback from a star with $\sim 40 \, \rm M_\odot$.

In the bottom panel (g) of Fig. \ref{fig:overview_main_simulation}, we show the radial distribution of SNe inside (orange) and outside (blue) $10 \, \rm pc$. There is a clear separation between the two populations of SNe and the central SNe are clustered in time.

To better understand the effect of nuclear star formation and stellar feedback, we re-simulate {\tt BH1e3-NSC-massive} at four times higher global gas resolution ($1 \, \rm M_\odot$, {\tt BH1e3-NSC-massive-hr} ) and show one accretion cycle in Fig. \ref{fig:accretioncycleexample}. The top and bottom panels show the larger scale gas surface density and temperature distribution, respectively, at three different times after the last nuclear SN has exploded. A zoom-in on the NSC region is shown in the middle row. Only $\sim 8 \, \rm Myr$ after the most recent SN has terminated the previous accretion cycle, the NSC captures gas from the turbulent ISM (left). The gas cools and the first low-mass stars ($<8 \, \rm M_\odot$, faint black dots) already start to form. At $17.5 \, \rm Myr$, several massive stars have formed inside the NSC (red and orange symbols). Like in all simulations with NSCs presented in this paper, the nuclear star formation is almost entirely confined to a region within the half mass radius of the NSC (black circle, the old star cluster stars are not shown here). After $26.4 \, \rm Myr$, the first massive star has exploded as a SN and has cleared out the gas from the NSC, creating a hot, low density bubble in the galactic center (right). The hot bubble terminates the accretion cycle and partially destroys the stream of cold gas that is moving towards the NSC, preventing further inflows to the NSC.

\begin{figure*}
	\includegraphics[width=1.95\columnwidth]{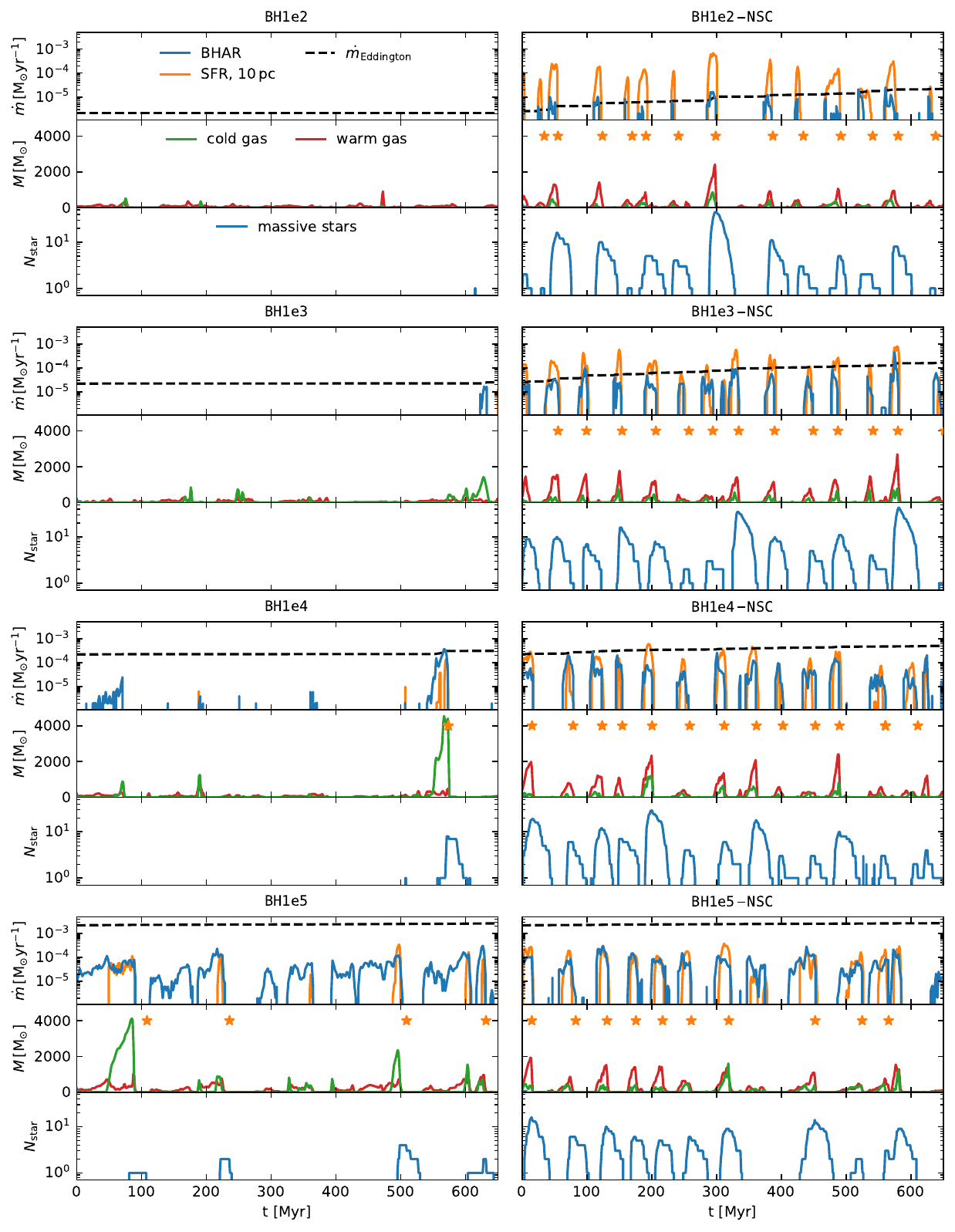}
    \caption{The nuclear properties for all {\tt fiducial} simulations as a function of time with different BH masses (increasing from top to bottom, $M_{\rm BH} = 10^2, 10^3, 10^4, 10^5 \, \rm M_\odot$) without (left) and with a NSC (right). For each simulation, we show the BH accretion rate (BHAR) and the SFR inside the central $10 \, \rm pc$ (top), the warm and cold gas mass (middle), and the number of massive stars in the central region (bottom). The first SNe following a star formation cycle are marked with an orange star. Low-mass BHs ($\leq 10^4 \, \rm M_\odot$) without a NSC cannot capture gas. More massive BHs occasionally capture gas, but the BH accretion is not always connected to star formation. With a NSC, even low-mass BHs accrete episodically with simultaneous star formation. The BH growth rate depends on the initial BH mass, but the central SFR is similar in all simulations with NSCs.}
    \label{fig:overview_NSCnoNSC_simulation}
\end{figure*}

\begin{figure*}
	\includegraphics[width=2\columnwidth]{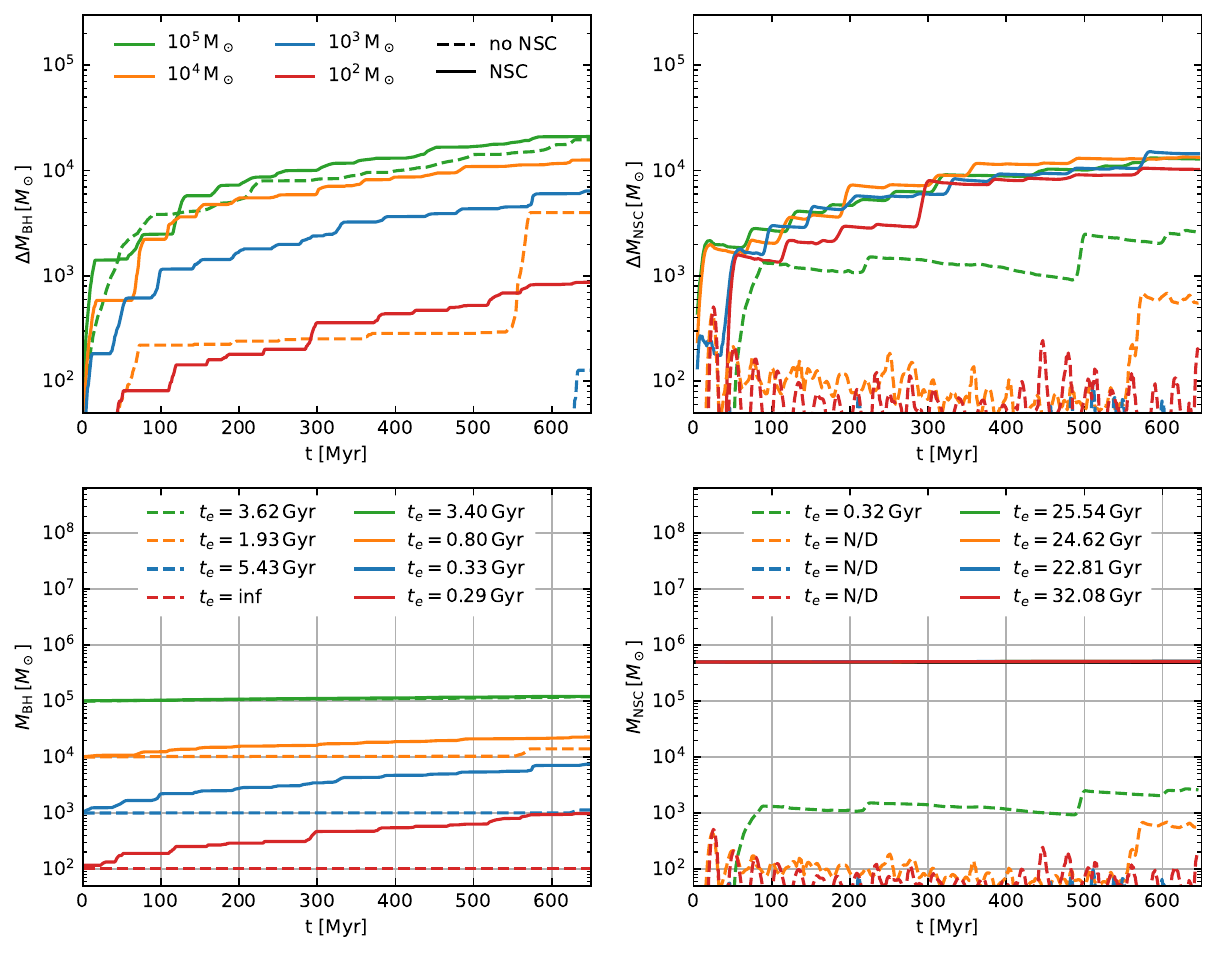}
    \caption{Growth histories of the BHs and NSCs for all {\tt fiducial} simulations starting from different initial MBH masses with (solid lines) and without (dashed line) a pre-existing NSC. The top left panel shows the accreted BH mass $\Delta M_{\rm BH} =  M_{\rm BH}(t) - M_{\rm BH}(t_0)$ as a function of time. The pre-existing NSC boosts growth of low-mass BHs ($M_{\rm BH} < 10^4 \, \rm M_\odot$) by several orders of magnitude, but does not change significantly the mass growth history for massive BHs ($M_{\rm BH} \geq 10^5 \, \rm M_\odot$). For low-mass BHs, the total mass growth $M_{\rm BH}(t)$ (bottom left panel) has short e-folding timescales $t_e$. For high mass BHs, the growth timescales become very long. The growth of stellar mass $\Delta M_{\rm NSC}(t)$ inside $10 \, \rm pc$ (solid lines, top right panel) is independent of the BH mass if embedded in a NSC. Even without the pre-existing NSC, a small cluster forms around the BH. The absolute growth timescale of the pre-exsisting NSC is very long (bottom right panel). The e-folding growth timescales for the BH and NSC are reported in the bottom left and right panel, respectively.}
    \label{fig:massgrowth_NSCnoNSC}
\end{figure*}

\begin{figure}
	\includegraphics[width=1\columnwidth]{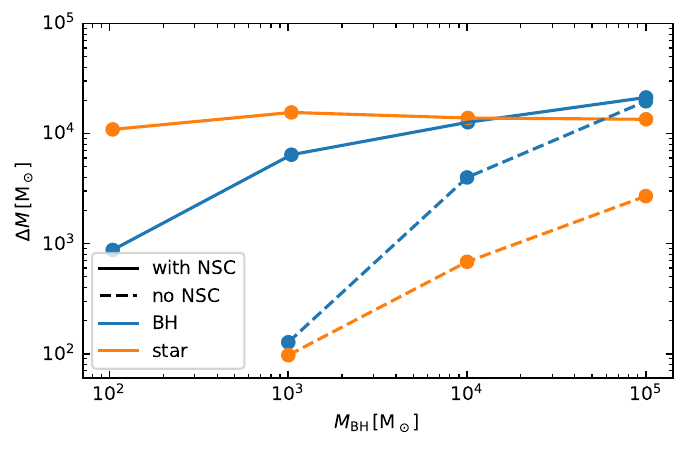}
    \caption{Comparison of the accreted BH mass $\Delta M_{\rm BH}$ and the new stellar mass $\Delta M_{\rm NSC}$ inside $10 \, \rm pc $ at the end of the simulation. Without the NSC (dashed line), MBH growth is usually dominant compared to nuclear SF. With a NSC (solid line), central SF dominates over BH growth, even though the BH growth is significantly larger than without a NSC. For large BH masses ($\geq 10^5 \, \rm M_\odot$), the NSC does not promote BH growth anymore.}
    \label{fig:massgrowth}
\end{figure}

To show that only a small fraction of the available gas in the center can be accreted by the BH, we show the mass budget in the central $10 \, \rm pc$ in Fig. \ref{fig:conversioneff_example}. In particular, we plot the integrated inflow and outflow, $\Delta M_{\rm in/out} = \int \dot{m}_{\rm in/out} \rm \, dt$, in the central $10 \, \rm pc$, the young NSC mass $\Delta M_{\rm NSC}$, and the BH growth $\Delta M_{\rm BH}$ as a function of time. Only a small fraction of the gas that flows into the central region is accreted by the BH ($\Delta M_{\rm BH} / \Delta M_{\rm in} \sim 11 \, \%$). About $\sim 25 \, \%$ of the in-flowing mass budget is converted into stars while the largest part of the mass budget is lost to outflows. This indicates that most of the gas is lost to SF or expelled by stellar FB before it can reach the accretion radius of the BH.

\begin{figure*}
	\includegraphics[width=2\columnwidth]{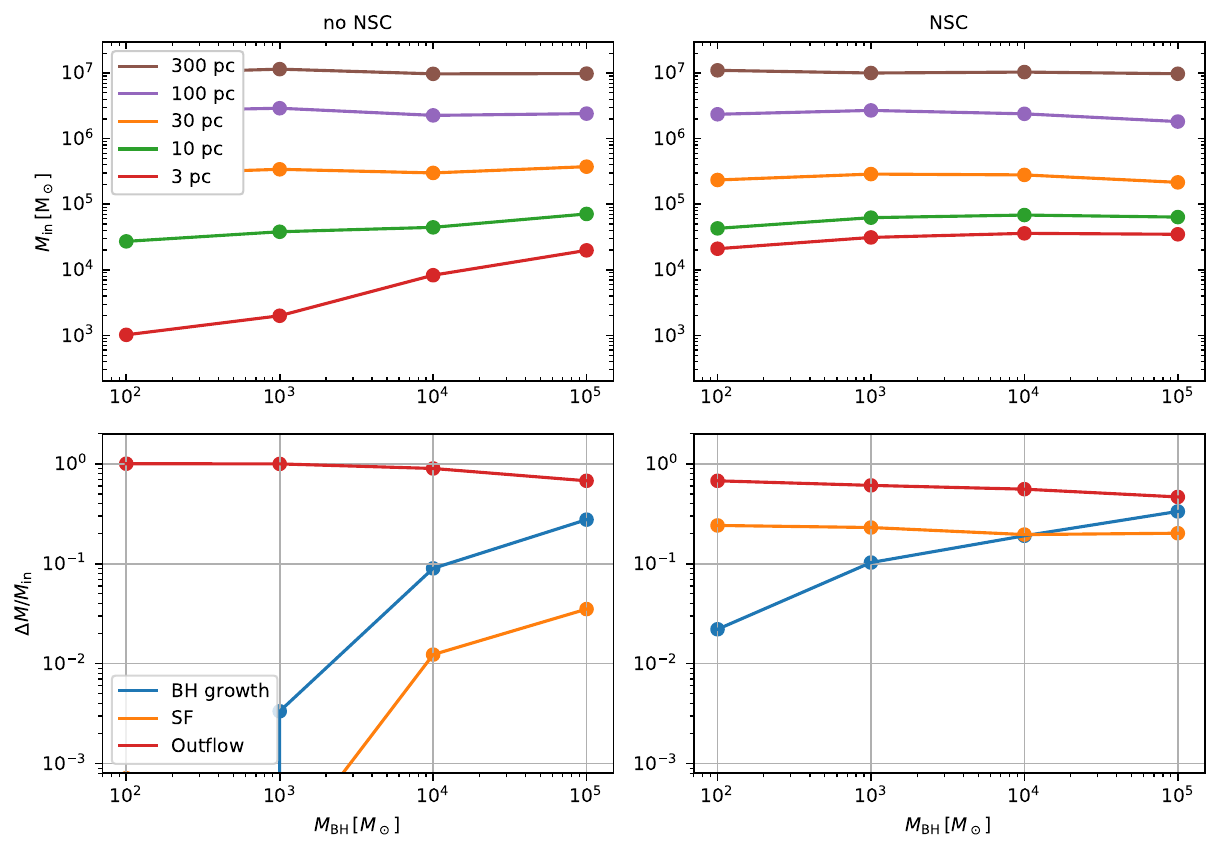}
    \caption{Analysis of the mass budgets available for BH accretion and SF. In the top panel, we show the inflow of gas mass measured at radii of $r=300, 100, 30, 10$ and $3 \, \rm pc$ integrated over time. With a pre-existing NSC, the accumulated inflows are similar for all initial MBH masses for radii down to $3 \, \rm pc$. Without a NSC, the inflows decrease significantly for low MBH masses, indicating that a lack of gas in the galactic center is the reason for inefficient MBH growth in this regime. In the bottom panel, we show which fraction of this in-flowing gas supply ($r = 10 \, \rm pc$) is converted into stars (orange line), is accreted by the BH (blue line) or is lost in outflows (red).}
    \label{fig:conversioneff}
\end{figure*}

\subsection{Black hole accretion cycles with and without nuclear star clusters}
\label{sec:parameterstudy}

In this section, we consider the {\tt fiducial} dwarf galaxy with and without a $5 \times 10^5 \, \rm M_\odot $ NSC and with initial BH masses between $10^2$ and $10^5 \, \rm M_\odot$. All simulations discussed in this section are listed in Table \ref{table:simulations}. BH masses with $\lesssim 10^4 \, \rm M_\odot$ are within the expected range for the stellar and NSC mass of the host galaxy \citep[see e.g.][]{Neumayer_2020}, more massive BHs would be considered over-massive \citep{Mezcua_2023}. 

In Fig. \ref{fig:overview_NSCnoNSC_simulation}, we show the accretion and star formation histories for eight simulations (similar to Fig. \ref{fig:overview_main_simulation}). Simulations without and with ({\tt -NSC}) a pre-existing NSC are shown in the left and right columns respectively. The initial BH masses increase by factors of ten from $100 \, \rm M_\odot$ ({\tt BH1e2}) to $10^5 \, \rm M_\odot$ ({\tt BH1e5}) from top to bottom. 

Each simulation is shown on an individual panel split into three subpanels, with the top showing the BH accretion rate and SFR in the central $10 \, \rm pc$. The red curve in the middle shows the amount of warm gas in the center ($300 \, {\rm K} < T < 2 \times 10^4 \, \rm K$), while the green curve represents gas in the cold ISM phase ($T < 300 \, \rm K$). The number of massive stars ($M_{\rm star} > 8 \, \rm M_\odot$) as a function of time is given in the bottom panel. As in Fig. \ref{fig:overview_main_simulation}, we mark the time of the first SN of each accretion cycle with an orange star. 

Without a NSC, the low-mass BHs ({\tt BH1e2}, {\tt BH1e3}) do not grow within the 650 Myr of evolution shown here. The small amount of gas in the center also does not form any stars. For {\tt BH1e4}, the BH grows during three accretion events when more than $\sim 1000 \, \rm M_\odot$ of gas accumulates in the center. Only during one cycle (at $\sim 550 \, \rm Myr$), the BH reaches an accretion rate above $\dot{m}_{\rm BH} \geq 10^{-4} \, \rm M_\odot \,yr^{-1}$, comparable to the Eddington accretion rate. This cycle also leads to star formation and is terminated by a supernova explosion, while the other cycles with smaller gas masses in the center do not lead to the formation of massive stars and are not obviously terminated by a SN explosion close to the MBH. As the gas is weakly bound in the absence of a NSC, feedback from more distant stars is likely sufficient to unbind the small gas reservoirs here. For the most massive BH without a NSC, {\tt BH1e5}, the star formation rate and the BH accretion rates are decoupled. Gas is captured frequently and the BH accretes gas most of the time, only interrupted by short and irregular periods of quiescence. In our Jeans instability based star formation model, stars can only form from sufficiently cold and dense gas. There are several accretion events when the gas is warm and not dense enough to form stars but can still be accreted by the BH.
This results in relatively continuous, less bursty BH growth episodes until the supply of gas is depleted. However, there are a few cycles during which larger amounts of cold gas accumulate in the center and the gas is able to form stars. These cycles are typically terminated already before the first central SN explosion, indicating that radiation feedback from massive stars is enough to disperse the cold gas.

The simulations with NSCs (right column in Fig. \ref{fig:overview_NSCnoNSC_simulation}) show a much more regular behaviour. Independent of the initial BH mass, all accretion cycles are terminated at latest by the first nuclear SN. This is very similar to the {\tt massive} galaxy discussed in Section \ref{sec:flagship}. The nuclear SFR is similar for all NSC simulations and regulated only by the presence of the NSC potential and star formation. The absolute BH accretion rate, however, increases with BH mass as expected from their larger sphere of influence inside the NSC. Still, all but the most massive BH ({\tt BH1e5-NSC}) can have peak accretion rates close to the Eddington rate. 
Another striking difference to the simulations without NSCs is that in the presence of a NSC, every gas accretion cycle leads to simultaneous star formation and BH growth. As discussed for the {\tt massive} dwarf galaxy, this star formation produces massive stars and subsequent SNe. This makes the cycles very regular with a period of $\sim 40 - 50 \, \rm Myrs$, where the minimum is set by the lifetime of massive stars of $\sim 8 \,  \rm M_\odot$ ($\sim 33$ Myr) plus the time it takes for the NSC to capture new gas from the ISM again, i.e. the dynamical time at $\sim 10 \ \rm pc$, which is $t_{\rm dyn}\sim 5 \ \rm Myr$.

\subsubsection{Black hole and nuclear star cluster growth}
\label{sec:NSCnoNSC}
We compare the BH and NSC growth histories of the eight simulations of the {\tt fiducial} dwarf galaxies in Fig. \ref{fig:massgrowth_NSCnoNSC}. 
The top left panel shows the BH mass growth $\Delta M_{\rm BH} =  M_{\rm BH}(t) - M_{\rm BH}(t_0)$ in the respective simulations as a function of time. The NSC boosts the BH growth significantly for low-mass BHs which otherwise accrete no ($M_{\rm BH}(t_0) = 100 \, \rm M_\odot$, red) or only very little gas ($M_{\rm BH}(t_0) = 1000 \, \rm M_\odot$, blue). For more massive BHs the effect becomes smaller and almost negligible if the BH mass becomes comparable to the NSC mass ($M_{\rm BH} = 10^5 \, \rm M_\odot$, green).

The absolute mass $M_{\rm BH}(t)$ is shown in the bottom left panel of Fig. \ref{fig:massgrowth_NSCnoNSC}. Especially for the low-mass BHs, the boosted accretion due to the NSC leads to short growth timescales. We use the e-folding time as a metric for the growth timescale, defined by $t_e = t \, \ln{(M(t)/M(t_0))}^{-1}$, and report it for each initial BH mass in the plot. For example, the BH with an initial mass of $10^3 \, \rm M_\odot$ grows with a timescale of $>5 \, \rm Gyr$. If the same BH is embedded into a NSC, the growth timescale drops by more than one order of magnitude to $0.33 \, \rm Gyr$. For more massive BHs, the differences between growth timescales in simulations with and without NSC become smaller. 

In the presence of a pre-existing NSC, the formation of new nuclear stellar mass $\Delta M_{\rm NSC}$ (i.e. new stars inside $10 \, \rm pc$, solid lines, top right panel of Fig. \ref{fig:massgrowth_NSCnoNSC}) is relatively independent of the initial BH mass and varies only by a factor two. Simulations with BHs that are initially not embedded in a NSC but are massive enough to capture gas can grow a small NSC (orange and green dashed lines). The mass of this cluster scales with the BH mass and, as discussed in the previous section, not every BH accretion cycle leads to the formation of stars, such that most of the stellar growth can be attributed to a small number of events during the evolution of $\sim 650 \, \rm Myr$. 

As the absolute growth $M_{\rm NSC}(t) = M_{\rm NSC}(t_0) + \Delta M_{\rm NSC}$ in the bottom right panel shows, the star formation inside the pre-existing NSC does not lead to significant absolute growth (only a few percent). In contrast, the small star clusters that form around the BH without pre-existing NSC grow on a short timescale. This timescale (computed using $t_0 = 400 \, \rm Myr$ to obtain the e-folding timescale since the mass is zero initially) is significantly shorter than the growth timescale of the BH.

This suggests that if a low-mass BH is initially under-massive compared to its host star cluster, it will have a shorter growth timescale compared to the NSC growth and "catch up" in mass. As the BH grows, the advantage of the NSC potential well becomes smaller and the BH growth timescale increases. This is a possible explanation for the relatively tight correlation between NSC and BH masses. On the other hand, if there is no star cluster around the BH initially, a star cluster forms self-consistently, suggesting that in this case the NSC mass "catches up" to the BH mass. However, we note that BH FB is not included here and might lead to a suppression of star formation around the BH.  

We show an overview of the accreted BH mass and new central stellar mass after $650 \, \rm Myr$ as a function of initial BH mass for all {\tt fiducial} simulations in Fig. \ref{fig:massgrowth}. For simulations with NSCs (solid lines) the BH mass growth scales with initial BH mass (blue solid line) while the stellar mass growth (solid orange line) is independent of the initial BH mass and set by the presence of the initial NSC. For simulations without initial NSCs, stellar (orange dashed line) and BH mass growth (blue dashed line) scale with the initial BH mass. The BH mass growth dominates for high initial BH masses. For large initial BH masses, the BH growth is similar with and without NSC.

To understand why a NSC is boosting BH growth, we show the integrated inflow rates $\Delta M_{\rm in} = \int \dot{m}(t) \rm \, dt$ for the different initial BH masses at the end of the simulations in the top panels of Fig. \ref{fig:conversioneff}. The inflow is measured at five different radii and we use it as an estimate for the available gas budget for central star formation and BH growth (see Fig. \ref{fig:conversioneff_example}). In the bottom panel, we show which fraction of the gas flowing to the central $10 \, \rm pc$ is converted into stars (orange line), is accreted by the BH (blue line), or is lost to outflows (red).

Down to the central 10 pc, the gas inflow is independent of the presence of a NSC and similar for all initial conditions. Without a NSC the inflow into the central 3 pc increases for initial BH masses greater than $10^3 \, \rm M_\odot$ (top left panel). This gas can form stars (orange line, bottom left panel) and up to 30 per cent can be accreted onto the central BH. For BH masses lower than $10^4 \, \rm M_\odot$ almost all gas is leaving the central region again. With a NSC, about 50 percent of the gas funneled to the central 10 pc flows all the way to the central 3 pc (green and red lines, top right panel). A BH independent fraction of $\sim$ 25 per cent of the 10 pc inflow is converted into stars (orange line bottom right panel). The gas fraction accreted onto the central BH increases from $\sim$ 2 percent to $\sim$ 30 percent for the most massive BH of $10^5 \, \rm M_\odot$ (blue line).
This indicates that the main effect of the NSC in the simulations is to boost the nuclear gas inflow rates. The effect on BH growth is particularly strong in models with MBH masses which are much lower than the NSC mass. Due to the increased inflow rates the nuclear gas can become dense, cool and form stars which regulate the accretion cycles as discussed above. The nuclear gas can also be accreted onto the BHs.  

\begin{figure*}
	\includegraphics[width=2\columnwidth]{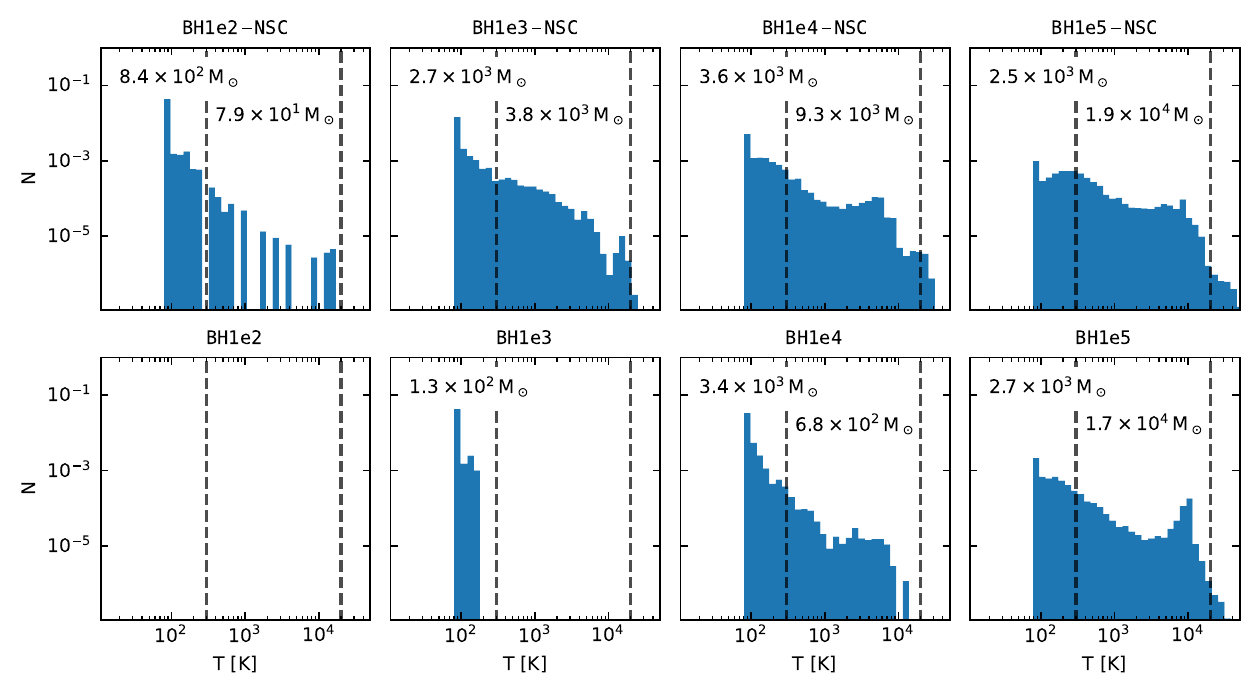}
    \caption{The temperature distribution of accreted gas particles for simulations with (top) and without a NSC (bottom). BHs more massive than $10^4 \, \rm M_\odot$ accrete from the cold ($T < 300 \, \rm K$) and warm ($300 \, {\rm K} < T < 2 \times 10^4 \, \rm K$) ISM phase while accretion from the hot phase ($T > 2 \times 10^4 \, \rm K$) is always subdominant. These three ISM phases are visually separated by a dashed black line. Low-mass BHs preferentially accrete from the cold ISM phase, that has smaller internal energy and can be captured by the BH more easily. The total masses accreted from the cold and warm ISM phase are given in the top left and right corner of each panel, respectively.}
    \label{fig:accreted_temp}
\end{figure*}

\subsubsection{Properties of the accreted gas}

In Fig. \ref{fig:accreted_temp}, we show the temperature distribution of gas particles which are accreted onto the central BH . With our sink based accretion model, BHs can in general accrete from all phases of the ISM as long as the gas is gravitationally bound (see Eq. \ref{eqn:sinkbound}) and meets the additional radius and angular momentum criteria. Overall, we find that gas with temperatures in the range $\sim 80 - 4 \times 10^4 \ \rm K$ can be accreted in the simulations presented here. For low-mass BHs with $M_{\rm MBH} \leq 10^3 \, \rm M_\odot$ mostly cold gas ($T \leq 300 \, \rm K$) is accreted. With increasing BH mass increasingly more gas at higher temperatures becomes bound and can also be accreted. For BH masses $\geq 10^4 \, \rm M_\odot$, the distributions become bi-modal (reflecting the multiphase structure of the ISM) and the BH growth becomes dominated by accretion from the warm ($300 \, \rm K < T \leq 2 \times 10^4 \, \rm K$) ISM gas phase.

\begin{figure*}
	\includegraphics[width=1\columnwidth]{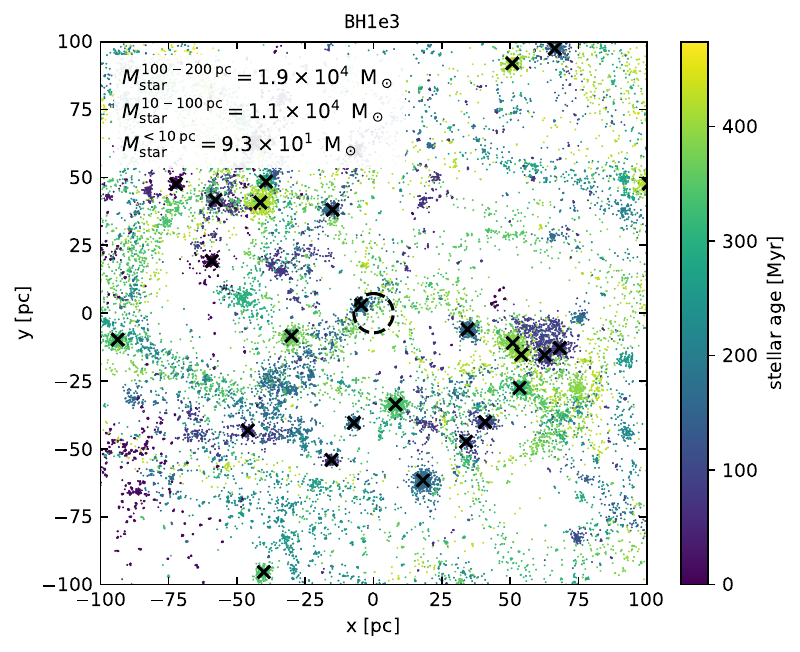}
	\includegraphics[width=1\columnwidth]{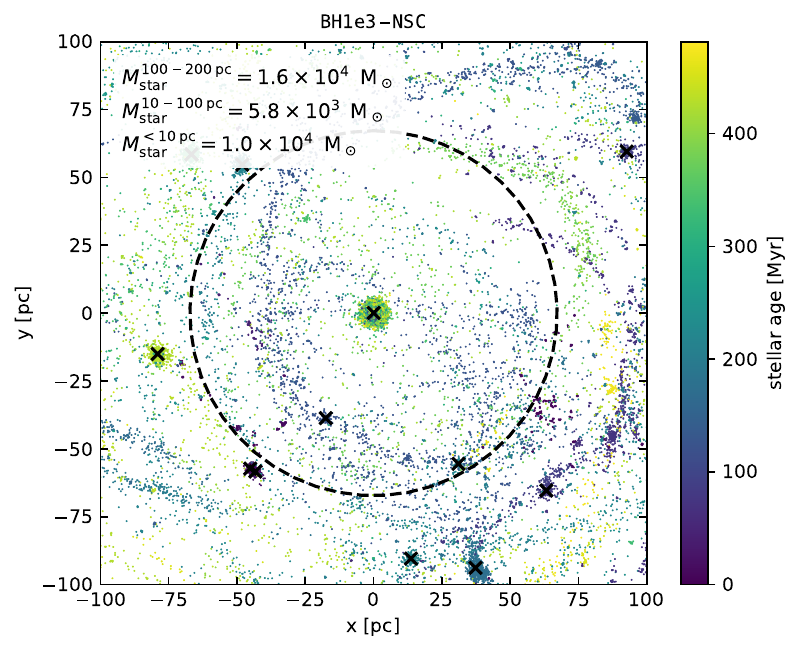}
	\includegraphics[width=1\columnwidth]{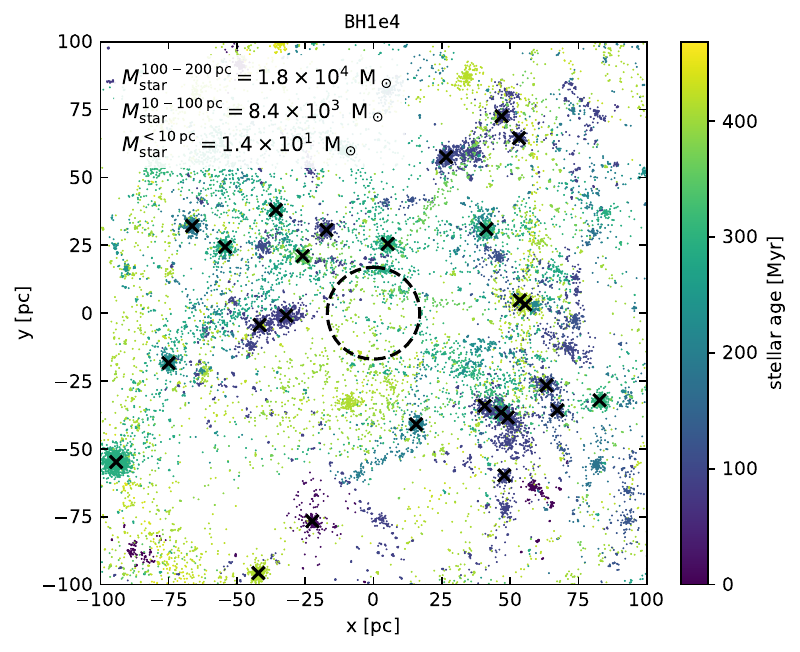}
	\includegraphics[width=1\columnwidth]{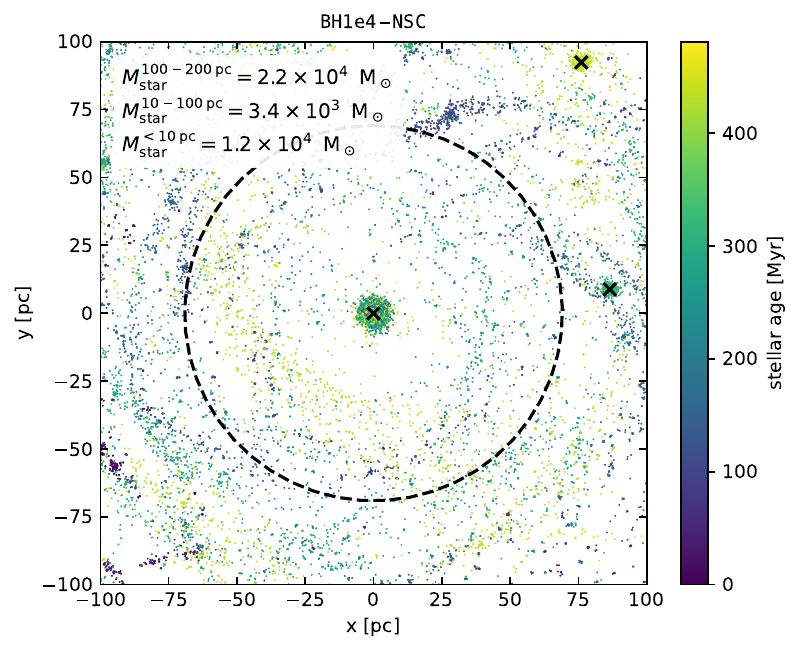}
	\includegraphics[width=1\columnwidth]{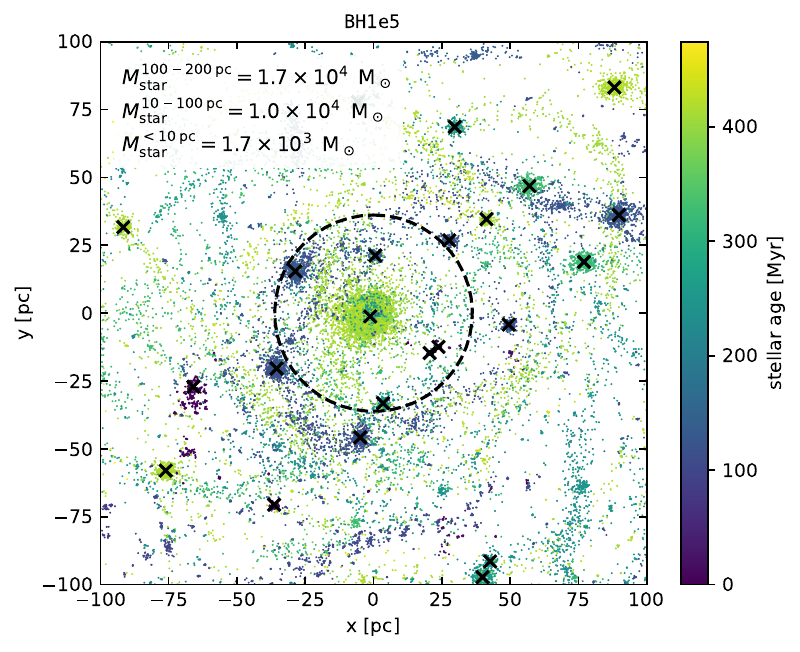}
	\includegraphics[width=1\columnwidth]{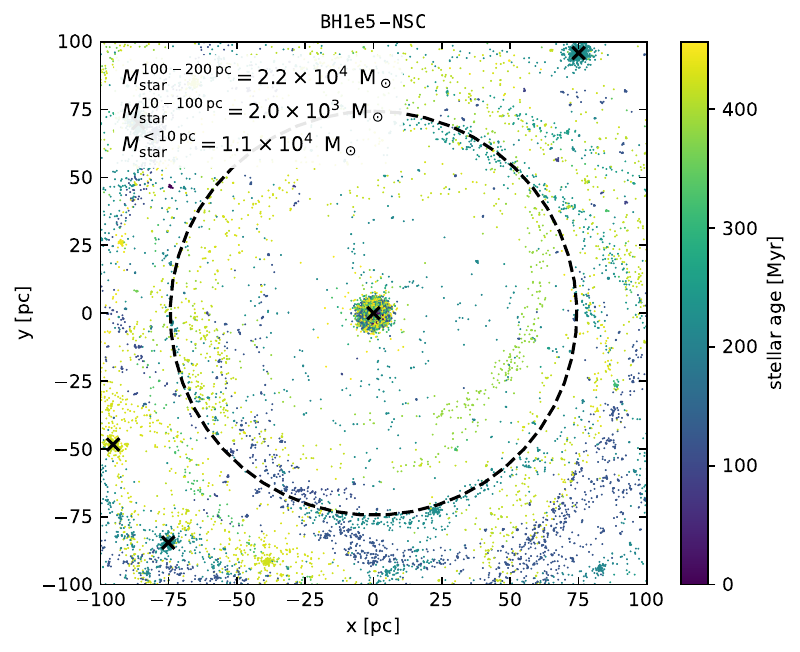}
    \caption{The distribution of stars in the central 100 pc around the BH at $t = 500 \, \rm M_\odot$ for six {\tt fiducial} simulations with (right) and without (left) a NSC and initial BH masses of $10^3, 10^4, 10^5 \, \rm M_\odot$ from top to bottom. The sphere of influence of the central object (NSC+BH) is represented as a black dashed circle. The age of each star is color-coded. Black crosses mark the position of identified star clusters. The total formed stellar mass in radial bins of $r < 10 \ \rm pc$, $10 - 100 \ \rm pc$ and $100 - 200 \ \rm pc$ are given in each panel. The presence of NSCs has a strong impact on the survival of star clusters (most get disrupted) and nuclear star formation (strongly enhanced).}
    \label{fig:starclusters_in_galactic_center}
\end{figure*}

\subsection{Star cluster properties and nuclear star cluster growth}
\label{sec:largescaleimpact}
In this section, we present an analysis of how the NSC and/or BH interacts with its larger scale environment. In particular, we will show that the potential well of the central object changes the star formation rate and the stellar clustering in its sphere of influence and that NSC growth through the accretion of sinking star clusters ("ex-situ growth") is inefficient in our simulations.
\begin{figure}
    \includegraphics[width=1.0\columnwidth]{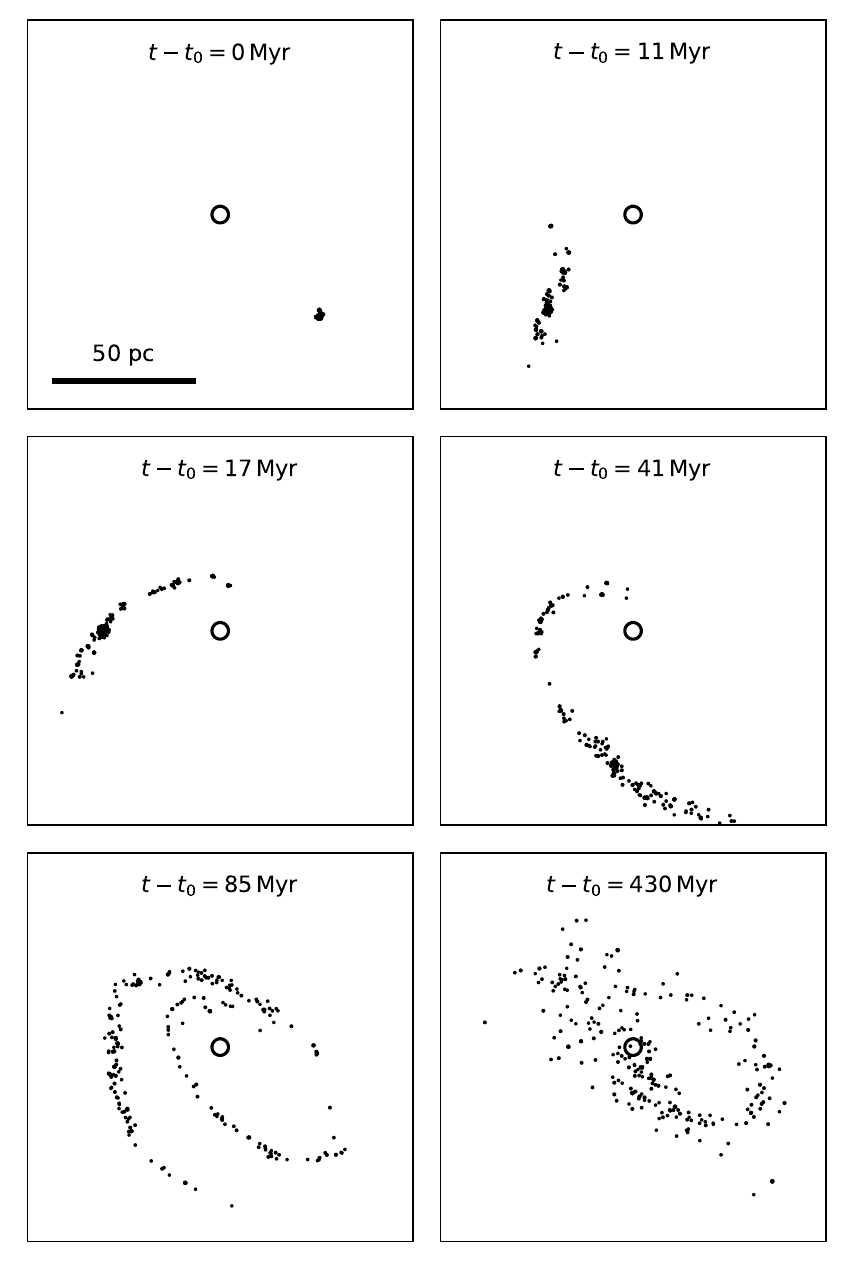}
    \caption{An example of the disruption of a small star cluster with initially $\sim 300$ stars. The cluster is bound initially and has a mass of $250 \, \rm M_\odot$ with a half-mass radius of $r=1.1 \, \rm pc$. A leading and trailing tail already develops after $11 \, \rm Myr$. After $17 \, \rm Myr$ (and $41 \, \rm Myr$), a small star cluster is still visible, although many stars have already been stripped from the cluster. At $85 \, \rm Myr$, the cluster is fully disrupted. At the final time, even the stellar stream that is left behind by the disrupted cluster is fully dissolved and the stars have mixed randomly with the stars in the central $\sim 50 \, \rm  pc$.}
    \label{fig:starcluster_disruption}
\end{figure}

\subsubsection{Star formation and star clusters around the black holes}
\label{sec:clusterdisruption}

In Fig. \ref{fig:starclusters_in_galactic_center} we show the distribution of newly formed stars in the central $100 \, \rm pc$ for the {\tt fiducial} dwarf galaxy without (left) and with (right) a NSC and initial BH masses of $10^3$ (top), $10^4$ (middle) and $10^5\, \rm M_\odot$ (bottom) after $t = 500 \, \rm Myr$ of evolution. The gravitational sphere of influence, defined by $M(r_{\rm infl}) = 2 \times M_{\rm central}$ inside which the central object is expected to dominate the dynamics, is shown as a black dashed circle. Here, $M_{\rm central}$ represents the sum of BH and NSC masses and $M(r)$ denotes the cumulative total mass in the galaxies without a NSC and a BH. We color-code the stars by age to highlight which stars formed simultaneously. This allows us to visually keep track of stars that initially formed together from the same gas cloud. The position of star clusters as identified with a friends-of-fiends algorithm (linking length 1 pc and at least 100 stars per cluster) are marked with a black cross. We also report the total stellar mass formed inside $10 \, \rm pc$, between $10$ and $100 \, \rm pc$ as well as $100$ and $200 \, \rm pc$ in each panel.

At large distances $ > 100 \, \rm pc$, outside the sphere of influence of the NSC, the differences between the stellar mass in simulations with and without NSC is very small. The presence of a NSC (right panel) leads to a suppression of total star formation by a factor of 2 - 5 in the 10 - 100 pc region (e.g. $M_{\rm star}^{10-100 \, \rm pc} \sim 2.0 \times 10^3 \, \rm M_\odot$ for {\tt BH1e5-NSC} versus $M_{\rm star}^{10-100 \, \rm pc} \sim 1.0 \times 10^4 \, \rm M_\odot$ for {\tt BH1e5}). The situation is reversed in the central 10 pc. Here all simulations with NSCs show the formation of new nuclear cluster stars with at least $M_{\rm star}^{<10\, \rm pc} \sim 10^4 \, \rm M_\odot$. This is comparable to the total stellar mass formed in the simulations without NSC inside 100 pc and indicates that the shear from the NSC suppresses star formation until the inflowing gas assembles at the center and becomes too dense and cold for shear to suppress it any further.
Also the spatial clustering of stars is strongly affected by the NSC. In the absence of a NSC (left panels), most stars are part of low-mass star clusters. The uniform colors inside each cluster indicate that their stars formed at the same time from the same dense gas cloud. The number of star clusters inside the NSC sphere of influence (right panels) is very small and no cluster is closer than $\sim 40 \, \rm pc$ to the BH. The few clusters residing inside the influence radius are younger than $\sim 100 \, \rm Myr$ (as indicated by the dark color). The simulations suggest that older star clusters are disrupted in the tidal field of the NSC. For example, simulation {\tt BH1e3-NSC} (top right panel) clearly shows stellar streams from a disrupted cluster that share the same age and extend over $\sim 150 \, \rm pc$. Such streams can also be seen in the other two NSC simulations.

In Fig. \ref{fig:starcluster_disruption}, we follow the time evolution of one of the disrupted star clusters with initially $\sim 300$ bound stars, approximately $\sim 50 \, \rm pc$ away from the NSC. The cluster formed from a gas cloud on a clock-wise orbit in the $x-y-$plane, even though the galaxy rotates counterclockwise. After less than half an orbit around the NSC, the star cluster develops a leading and a trailing tidal tail, but most stars are still inside a small star cluster after $17 \, \rm Myr$. After $41 \, \rm Myr$ and an encounter with the NSC at distance $r \sim 10\, \rm pc$, the majority of stars are not part of the cluster anymore. At $t=85 \, \rm Myr$, there is only an extended stellar stream left, and the cluster has lost its structure entirely at the end of the simulation. 

\begin{figure}
	\includegraphics[width=1.0\columnwidth]{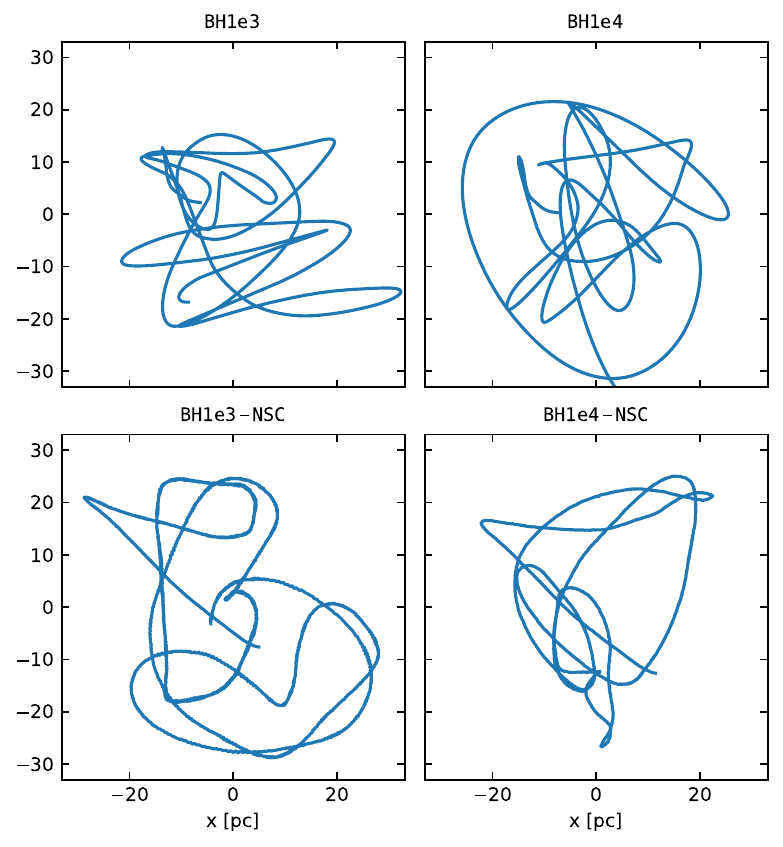}
    \caption{Typical trajectories ($200 \, \rm Myr$) of the BHs in four example simulations without (top panel) and with (bottom panel) NSCs. The wandering radius ($\sim30 \, \rm pc$) is similar in all cases. The BH orbits for the simulations with NSCs (bottom panels) appear slightly thicker due to oscillations of the BH around the NSC center. }
    \label{fig:encounter}
\end{figure}

The BHs and NSCs move around freely in the centers of the dwarf galaxies and respond to the gravitational interaction with the surrounding star clusters and other galactic matter. In simulations with NSCs the BH always stays in the center of the NSC. In Fig. \ref{fig:encounter} we show typical trajectories of BHs for models {\tt BH1e3} and {\tt BH1e4} in the top panel and the trajectory of BHs at the center of NSCs in models {\tt BH1e3-NSC} and {\tt BH1e4-NSC} in the bottom panels. The wandering radii are typically $\sim 30 \, \rm pc$. We consider this to be an extremely accurate estimate as we sample individual stellar masses and dynamical interactions between BHs and stars are computed without softening at high precision with {\sc Ketju}. Due to the shallow potential well of the dwarf galaxy, there is no strong restoring force such that even the relatively massive NSC with $5 \times 10^5 \, \rm M_\odot$ oscillates in the center without sinking back if it gets pushed away from the galactic center.

\begin{figure*}
	\includegraphics[width=1\columnwidth]{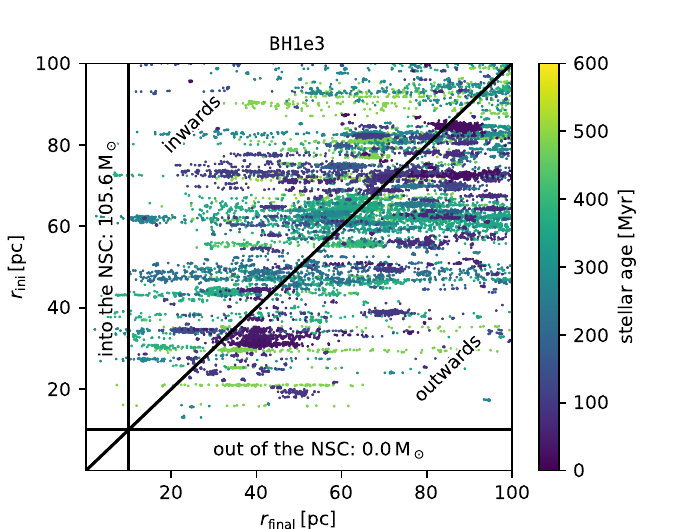}
	\includegraphics[width=1\columnwidth]{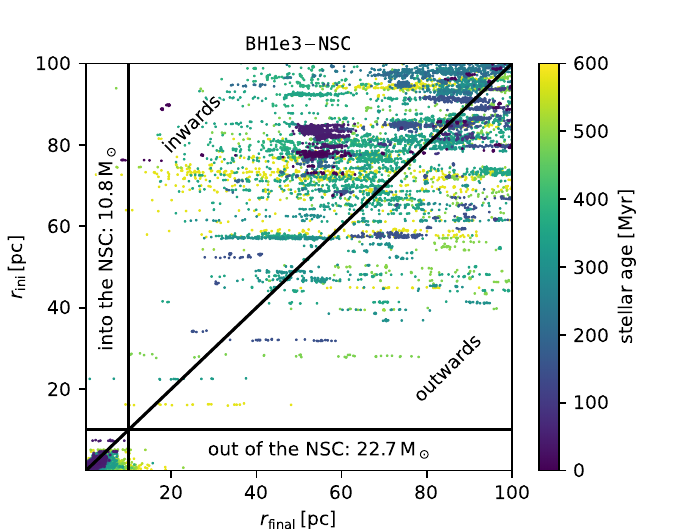}
	\includegraphics[width=1\columnwidth]{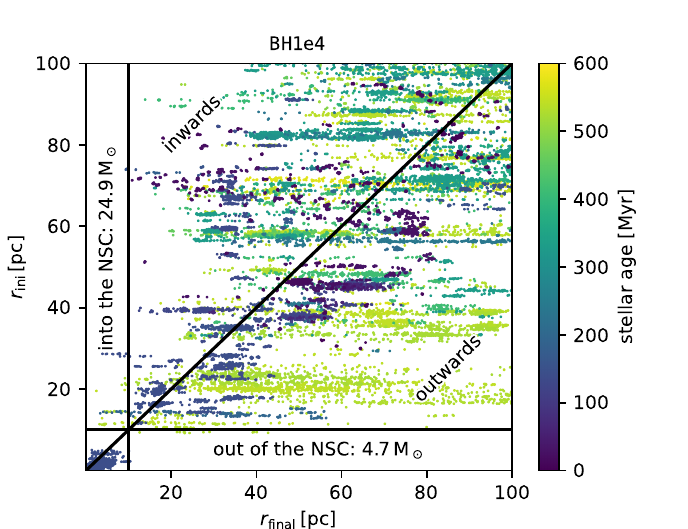}
	\includegraphics[width=1\columnwidth]{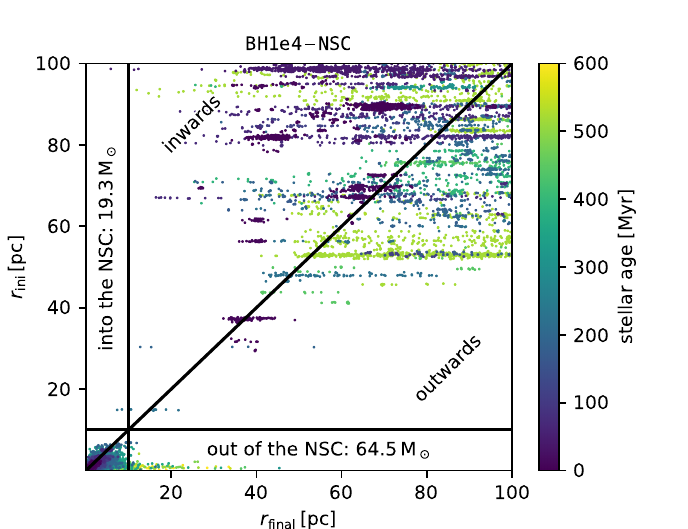}
	\includegraphics[width=1\columnwidth]{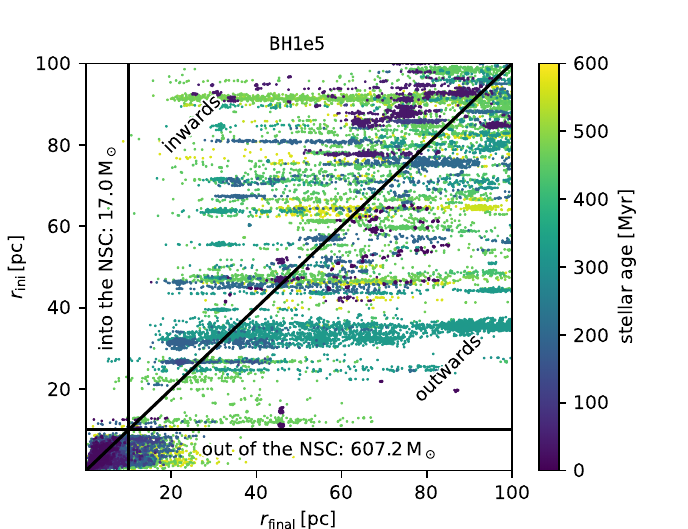}
	\includegraphics[width=1\columnwidth]{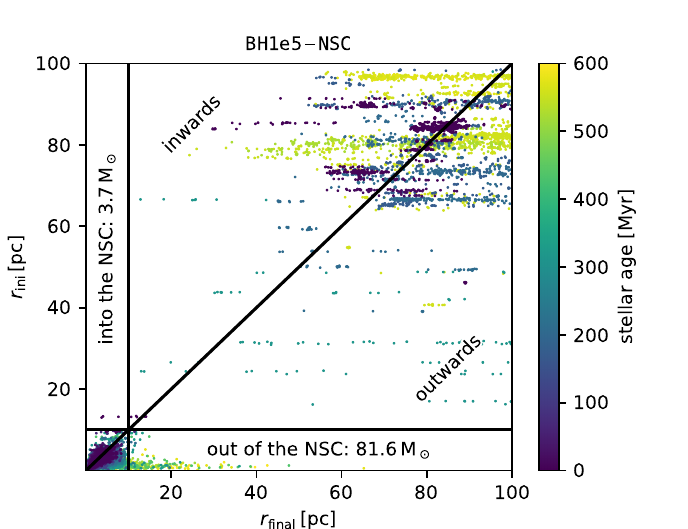}
    \caption{The radial distance of stars from the center at birth, $r_{\rm ini}$, vs. their radial distance at the end of the simulation, $r_{\rm final}$. The panels are arranged as in Fig. \ref{fig:starclusters_in_galactic_center} and each star is again colour coded by age. Stars above the unity line have moved inwards, stars below have moved outwards. Horizontal patterns at the same color indicate cluster disruption. Most stars in the nuclear region ($\gtrsim 10^4 \, \rm M_\odot$ within a radius of 10 pc, lower left corners in each panel) have formed there. The masses of incoming and leaving stars are given at the left and bottom part of the plot, respectively. There is no evidence for ex-situ NSC growth. }
    \label{fig:NSC_migration}
\end{figure*}

\subsubsection{Ex-situ vs.in-situ growth of the nuclear star cluster}

It is still debated if NSCs predominantly form via "in-situ" star formation in the galactic center or if they originate from sinking and merging massive star clusters in the "ex-situ" formation scenario \citep{1975ApJ...196..407T, 2004ApJ...605L..13M, 2019A&A...628A..92F}. Although the typical star clusters that form in our isolated dwarf galaxy have lower masses than typical massive clusters, the sinking and accretion of star clusters might be an additional growth channel for the NSC in our simulations. Here we explore which fraction of new stars in the NSC has formed in-situ inside the central region in contrast to star clusters that might have sunk to the center ("ex-situ").

In Fig. \ref{fig:NSC_migration}, we compare the distances of stars to the central BH at their time of birth $r_{\rm ini} = |\boldsymbol{r}_{\rm star, ini} - \boldsymbol{r}_{\rm BH, ini}|$ to the distances at the end of the simulation $r_{\rm final} = |\boldsymbol{r}_{\rm star, final} - \boldsymbol{r}_{\rm BH, final}|$. The panels are arranged as in Fig. \ref{fig:starclusters_in_galactic_center}. Stars with $r_{\rm final}$ < $r_{\rm ini}$ have moved inwards (top left), while stars $r_{\rm final}$ > $r_{\rm ini}$ moved outwards (bottom right).

\begin{figure}
	\includegraphics[width=0.9\columnwidth]{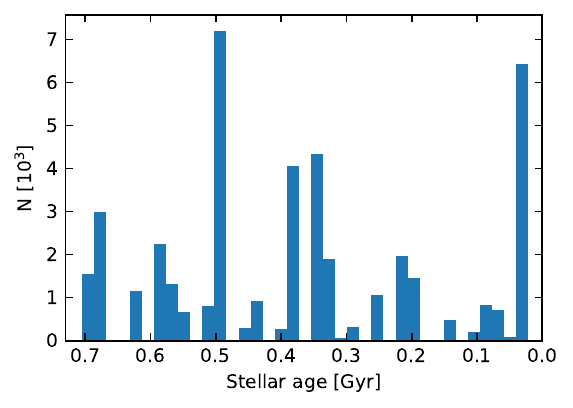}
    \caption{The age distribution of new nuclear stars contributing to the NSC in simulation {\tt BH1e4-NSC}. The episodic star formation as discussed in section \ref{sec:flagship} and \ref{sec:parameterstudy} leads to a broad distribution of stellar ages. The new NSC stars form in multiple generations.}
    \label{fig:NSC_age}
\end{figure}
Especially in the simulations with a massive central object (i.e. the simulation {\tt BH1e5} and the simulations with NSC in the bottom right panel), it is clear that stars with the same age typically form at the same radius, but then become spread out over a wider range of radii at final time. This gives rise to horizontal patterns with constant stellar age and constant $r_{\rm ini}$, but different $r_{\rm final}$. This is a result of the cluster disruption process described in the previous section \ref{sec:clusterdisruption} and in Fig. \ref{fig:starcluster_disruption}. 
The majority of stars that end up in the NSC (here defined by $r_{\rm final} < 10 \, \rm pc$) have already formed in the central region with $r_{\rm ini} < 10 \, \rm pc$. Only a small fraction of stars has formed at larger radii ($r_{\rm ini} > 10 \, \rm pc$) and are found in the center at final time ($r_{\rm final} < 10 \, \rm pc$) such that the stellar mass gain of $\Delta M_{\rm NSC}^{\rm ex-situ} \leq 100 \, \rm M_\odot$ via sinking stars is always negligible compared to the stellar mass from in-situ star formation ($\Delta M_{\rm NSC}^{\rm in-situ} > 10^4 \, \rm M_\odot$ for all simulations with NSCs). Mass loss from the NSC ($r_{\rm ini} < 10 \, \rm pc$ and $r_{\rm final} > 10 \, \rm pc$) is typically also negligible (except for simulation {\tt BH1e5} where some stars formed on eccentric orbits that extend beyond $10 \, \rm pc$).

As expected from the periodic star formation cycles discussed in sections \ref{sec:flagship} and \ref{sec:parameterstudy}, the in-situ NSC stars have very broad age distribution. As a typical example, we show the histogram of stellar ages of simulation {\tt BH1e4-NSC} in Fig. \ref{fig:NSC_age}. Such age distributions are consistent with observations of multiple stellar generations in NSCs \citep[see e.g.][]{Neumayer_2020, Fahrion_2021, Fahrion_2022b}.

In conclusion, in our simulations the majority of stars adding to the pre-existing NSC form in-situ while sinking star cluster do not contribute. Most stars that form inside the central $10 \, \rm pc$ remain bound to the NSC and/or BH and do not propagate in and out, which makes $r = 10 \, \rm pc$ a well-motivated boundary for the nuclear region. The inefficient ex-situ growth is expected here. Because the sinking timescale of star clusters depends on the cluster mass, low-mass clusters or individual stars after the cluster is disrupted have extremely long sinking timescales, even if they are already relatively close to the NSC. The disruption of star clusters in the galactic centers makes "ex-situ" growth even less efficient. Hence, our simulations do not contradict studies favouring the "ex-situ" formation scenario of NSCs in dwarf galaxies as we do not form massive star clusters \citep[star clusters formed in the simulations typically have masses below $10^4 \, \rm M_\odot$, see e.g.][]{2022MNRAS.509.5938H}. More extreme environments (starbursts, galaxy mergers) might give rise to star clusters that are massive enough to sink to the galaxy center on relatively short timescales.

\section{Discussion}
\label{sec:discussion}

With this high resolution study of accreting BHs in dwarf galaxies with and without NSCs, we test under which conditions and to which extent BHs can grow in the centers of low-mass galaxies. The simulations suggest that the growth of BHs in a dwarf galaxy without a NSC is very inefficient, even without considering BH feedback. This is in agreement with high-resolution studies of isolated gas clouds as presented in \cite{Shi_2022}. They found that surface densities of more than $\gtrsim 1000 \, \rm M_\odot / pc^2$ and cloud masses above $\gtrsim 10^6 \, \rm M_\odot$ are required to sustain efficient accretion onto MBHs in the presence of stellar feedback. These are conditions that are not typically found in the dwarf galaxies that we are simulating here. As for example shown by \cite{2024ApJ...969L..31S} and \cite{2024arXiv240908326M}, gas rich environments where efficient accretion onto low-mass MBHs is possible might be found in low-mass galaxies at high-redshift.

In our simulations, the SN explosions of individual massive stars lead to bursty BH accretion histories. This is consistent with the findings of \cite{Sivasankaran_2022}. They presented simulations of similar galaxies, although at about two orders of magnitude lower gas mass resolution (up to $\sim 860 \, \rm M_\odot$). They found very bursty BH accretion histories as a result of stellar feedback and the clumpy ISM structure, emphasizing the importance of resolving the multi-phase ISM. 

To reduce the model complexity in this study, we have not considered the effect of BH feedback. It is still an ongoing debate to what extent the feedback from MBHs in dwarf galaxies with their typically low accretion rates and luminosities shapes their environment and impacts the accretion rate \citep{2019MNRAS.484.2047K, Barai_2019, Sharma_2020, Koudmani_2021, Koudmani_2022, arjonagalvez2024roleagnfeedbackevolution}. However, aside from theoretical arguments, there are several observational examples for BH feedback processes in dwarf galaxies \citep[e.g.][]{Schutte_2022}.

While it is clear that feedback from BHs should be included in simulations including gas accretion \citep[see e.g.][]{2024arXiv240512164S}, it is not clear how to correctly model its effect on the environment in simulations of dwarf galaxy evolution. As the strong effect of individual SNe indicates, the larger expected injection of energy and momentum from AGN discs might completely terminate nuclear star formation and further gas accretion onto BHs. However, the accurate amount of energy and momentum released by the accretion disc and how it couples to the ISM (e.g. as radiation, bipolar wind, or collimated jet) is highly uncertain and depends on the assumed physical processes on the accretion disc scale.

\begin{figure}
	\includegraphics[width=1\columnwidth]{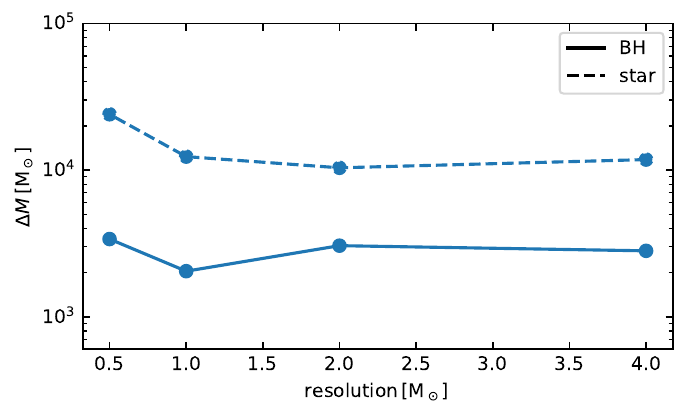}
    \caption{The impact of varying gas mass resolution $0.5, 1, 2, 4 \, \rm M_\odot$ based on simulations {\tt BH1e4-NSC-0.5}, {\tt BH1e4-NSC-1.0}, {\tt BH1e4-NSC-2.0}, and {\tt BH1e4-NSC} on the BH and nuclear stellar mass growth after $200 \rm \, Myr$. The BH mass growth is well converged, while the stellar mass increases slightly (a factor of 3) at the highest resolution. The Jeans mass limit for star formation $M_{\rm J} < 0.5 \times 100 \times m_{\rm SPH}$ varies with resolution here.}
    \label{fig:massgrowth_res}
\end{figure}
\begin{figure}
	\includegraphics[width=1\columnwidth]{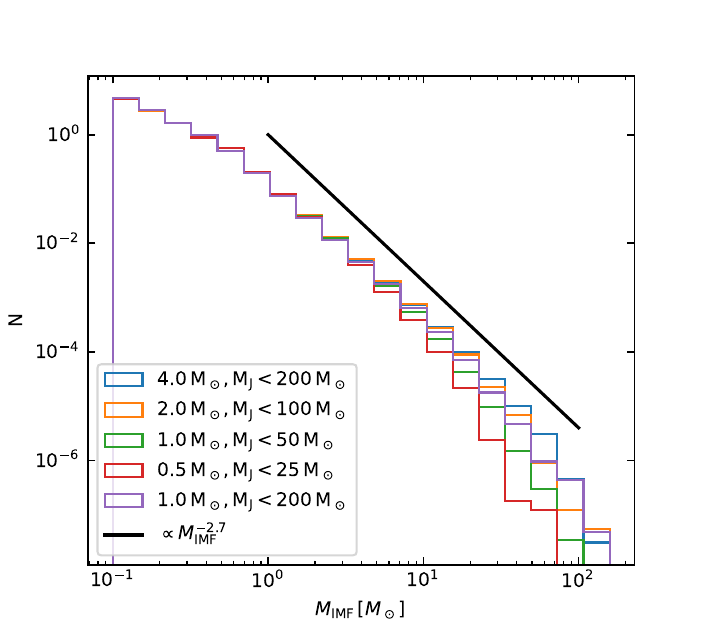}
    \caption{Histogram of realised stellar masses in the central $100 \, \rm pc$ at gas phase resolutions of $0.5, 1, 2$ and $4 \, \rm M_\odot$ with the adjusted Jeans mass limits for star formation ($M_{\rm J} < 0.5 \times 100 \times m_{\rm SPH}$ , same simulations as in Fig. \ref{fig:massgrowth_res}). The number of sampled massive stars decreases for higher resolutions. The original high-mass slope of the Kroupa initial mass function is shown as a black line. The purple line shows a simulation with $1 \, \rm M_\odot$ resolution and the fiducial Jeans mass limit of $200 \, \rm M_\odot$ ({\tt  BH1e4-NSC-1.0-fix}). In this case the realised stellar mass distribution is similar to the fiducial model.}
    \label{fig:massgrowth_res2}
\end{figure}
\begin{figure}
	\includegraphics[width=1\columnwidth]{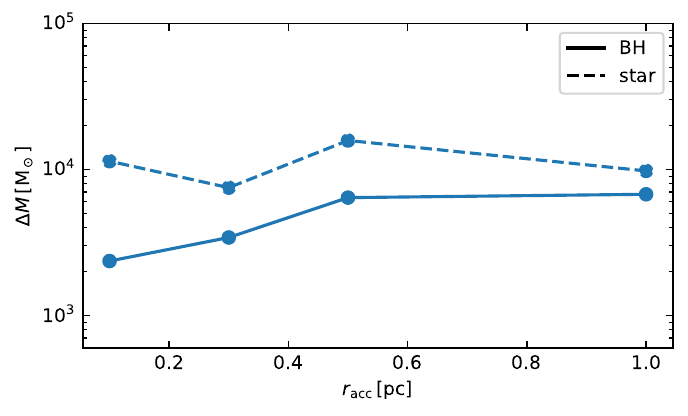}
    \caption{The impact of varying BH accretion radii $r_{\rm acc} = 0.1, 0.3, 0.5, 1.0 \, \rm pc$ on the BH and nuclear stellar mass growth after $\rm 350 \, \rm Myr$ based on simulations {\tt BH1e4-NSC-racc} and {\tt \tt BH1e4-NSC}. The BH and stellar mass growth is well converged (better than a factor of 2) for accretion radii larger than the gravitational softening length of $0.1 \, \rm pc$. The fiducial value is $r_{\rm acc} = 0.5$.}
    \label{fig:massgrowth_racc}
\end{figure}

Another challenge for simulations of accreting BHs in galaxies is that it is typically not possible to follow the gas dynamics down to accretion disc scales. Instead, gas must be removed from scales that are still resolved in the simulation and the evolution of the unresolved accretion disc has to be followed with simplified sub-grid models. In our simulations, we set an accretion radius of $0.5 \, \rm pc$, below which we do not follow the gas dynamics anymore. We assume that all gas accreted by the sink particle on this scale contributes to the growth of the BH. However, in even higher resolution simulations gas might still form stars on its way down to the accretion disc and the BH. For example, recent attempts to bridge the gap between galaxy and BH accretion disc scales suggest that star formation becomes inefficient close to the BH (on sub-parsec scales for a $\sim 10^7 \, \rm M_\odot$ BH), but it is not obvious how this result translates to the dwarf galaxy regime \citep{2024OJAp....7E..18H}. Even higher resolution would be required to follow the dynamics to smaller scales. We have tested the impact of higher gas mass resolution ($m= 2;\,1;\,0.5 \, \rm M_\odot$) on the growth of the BH and the NSC (models {\tt BH1e4-NSC-2.0} , {\tt BH1e4-NSC-1.0}, {\tt BH1e4-NSC-0.5}). As we show in Fig. \ref{fig:massgrowth_res}, the BH growth remains mostly unaffected and does not show a clear trend with increasing resolution. The nuclear star formation mildly increases with resolution by about a factor two. Interestingly, the shape of the sampled IMF becomes steeper at the high mass end (less massive stars) at higher gas mass resolution if the star formation Jeans threshold is increased accordingly ($M_{\rm J} < 0.5 \times 100 \times m_{\rm SPH}$, i.e. the gas is allowed to collapse to higher densities at higher resolution, see Fig. \ref{fig:massgrowth_res2}). If the Jeans threshold is kept fixed ($M_{\rm J} < 200 \, \rm M_\odot$, purple curve), the sampled IMF becomes independent of resolution again (purple vs. blue curve). Whether this effect is physical or dominated by numerical effects or sub-grid modelling remains to be investigated. We have also tested the effect of smaller and larger accretion radii ($r_{\rm acc}=1, 0.3, 0.1 \, \rm pc$) than the fiducial value of $r_{\rm acc}=0.5$ at the fiducial mass resolution of $4 \, \rm M_\odot$ on BH mass growth and nuclear star formation (Fig. \ref{fig:massgrowth_racc}). For numerically reasonable values that are sufficiently larger than the gravitational softening length of 0.1 pc the masses are well converged.

With the {\sc Ketju} integrator described in Section \ref{sec:Ketjutec}, we can follow the accurate orbit of newly formed stars around the BH and check whether stars cross the tidal radius (as defined in equation \ref{eqn:tidal}). TDEs are a viable detection channel for the otherwise hidden population on non-accreting BHs. We typically find not more than 10 TDEs within 800 Myr. This translates into a TDE rate of $\sim 10^{-8}\, \rm yr^{-1}$. However, this number can only be used as a first estimate (and most likely a lower bound) from high resolution galaxy simulations like the ones presented here. Despite the accurate modelling of BH - star interactions with {\sc Ketju}, the formation of stars inside the BH accretion radius is artificially suppressed and the forces between stars are still softened. Our simulations are a step towards a realistic modelling of TDEs in a galaxy simulation, but we can only report lower bounds on the expected TDE rate at this point. In the forthcoming work by Lah{\'{e}}n et al. in prep, the {\sc ketju} integrator is used to compute the unsoftened interactions between between all stars in the vicinity of massive stars and their compact remnants for more accurate modelling of the internal evolution of the star clusters (e.g. relaxation, mass-segregation, dynamical binary star formation and core-collapse) driven by two-body and multi-body interactions. This method will also allow for more accurate modelling of the star cluster that forms around a central MBH and a better estimate of the expected TDE rates.

\section{Conclusions}
\label{sec:concl}
We have presented high-resolution simulations of accreting central BHs in dwarf galaxy models with and without NSCs ($\sim 10^{5.5} - 10^6 \, \rm M_\odot$) to explore the co-evolution of nuclear star formation and BH accretion in a realistic galactic environment. The BHs have varying masses of $100 - 10^5 \, \rm M_\odot$. Such BHs are also termed massive black holes (MBH) or intermediate mass black holes (IMBH). The simulations highlight the strong impact of NSCs. They boost nuclear star formation (i.e. NSC growth) and gas accretion onto low-mass BHs ($10^2 - 10^3 \, \rm M_\odot$) in particular. The high-resolution ISM model includes the effect of HII region formation and SN explosions from massive stars and non equilibrium thermochemistry. The star formation model realizes individual stars down to $0.08 \, \rm M_\odot$ and the accurate regularized integration scheme {\sc ketju} computes gravitational interactions with BHs and individual stars. Our main conclusions can be summarized as follows:
\begin{description}
  \item [$\bullet$]In the presence of a NSC, warm and cold gas is captured from the turbulent ISM and funnelled to the central $\sim$ 3 pc. This leads to enhanced BH growth and nuclear star formation (i.e. NSC growth). The BH and NSC growth is coeval and episodic. Each growth episode is terminated by the first nuclear SN followed by a quiescent phase of $\sim 30 - 40 \, \rm Myrs$ determined by the longest massive star lifetime (i.e. the last SN of the star formation event) and the local dynamical time for gas re-accretion. This leads to star formation cycles with peak rates separated by $\sim 40-60 \, \rm Myrs$.  

  \item [$\bullet$] Low-mass BHs ($10^2 - 10^3 \, \rm M_\odot$) embedded in a NSC can grow rapidly with e-folding timescales of 200 - 400 Myr and peak accretion rates exceeding the Eddington rate. This effect disappears for the highest mass BHs included in this study. Note however, that we do not consider feedback from the accreting BHs. The peak BH accretion rates are about an order of magnitude lower than the nuclear star formation rate (which is determined by the NSC and is largely independent of the BH mass). For the highest mass BH the rates become comparable. 

  \item[$\bullet$] In the absence of a NSC low-mass BHs ($10^2 - 10^3 \, \rm M_\odot$) cannot grow and there is no enhanced nuclear star formation. Only the most massive BHs ($10^5 \rm M_\odot$) can accrete larger amount of gas and trigger nuclear star formation which could be considered as the onset of nuclear cluster formation. However, star formation and BH accretion episodes are not regular and not coeval. The BH can accrete warm and cold gas in the absence of nuclear star formation.  
  
  \item[$\bullet$] The presence of a NSC suppresses extended star formation at a separation of 10 - 100 pc from the center. Gas is instead funnelled to the center and forms stars there. Most star clusters in this region are disrupted by the tidal field of the NSC.  
  
  \item[$\bullet$] With the Ketju integrator, we resolve the unsoftened interaction between the new stars and the BHs whose orbits we can now track more accurately than before. The typical wandering radius of the central BHs is $\sim 30\, \rm pc$. 
  
 \item[$\bullet$] Repeated star formation events inside the NSC give rise to a broad age distribution of newly formed nuclear cluster stars (i.e. multiple generations). We do not see evidence for NSC mass growth via sinking star clusters in the simulations. This is not in tension with "ex-situ" scenarios of NSC growth, because the mass of the star clusters that form in the simulation are low such that the related sinking timescales are long.
  
\end{description}

\section*{Acknowledgements}

We thank Marta Volonteri, Nils Hoyer, Katja Fahrion, Sergei Bykov, Aniket Bhagwat and Volker Springel for valuable discussions and scientific input. T.N. acknowledges support from the Deutsche Forschungsgemeinschaft (DFG, German Research Foundation) under Germany’s Excellence Strategy - EXC-2094 - 390783311 from the DFG Cluster of Excellence "ORIGINS". Computations were performed on the HPC system Raven, Viper and Orion at the Max Planck Computing and Data Facility. C.P., N.L and T.N. acknowledge the computing time granted by the LRZ (Leibniz-Rechenzentrum) on SuperMUC-NG under project number pn49qi. P.H.J. acknowledges the support by the European Research Council via ERC Consolidator Grant KETJU (no. 818930) and the support of the Academy of Finland grant 339127.

\section*{Data Availability}

 The data will be made available based on reasonable request to the corresponding author.

\bibliographystyle{mnras}
\bibliography{literature} 

\begin{thebibliography}{}
\makeatletter
\relax
\def\mn@urlcharsother{\let\do\@makeother \do\$\do\&\do\#\do\^\do\_\do\%\do\~}
\def\mn@doi{\begingroup\mn@urlcharsother \@ifnextchar [ {\mn@doi@} {\mn@doi@[]}}
\def\mn@doi@[#1]#2{\def\@tempa{#1}\ifx\@tempa\@empty \href {http://dx.doi.org/#2} {doi:#2}\else \href {http://dx.doi.org/#2} {#1}\fi \endgroup}
\def\mn@eprint#1#2{\mn@eprint@#1:#2::\@nil}
\def\mn@eprint@arXiv#1{\href {http://arxiv.org/abs/#1} {{\tt arXiv:#1}}}
\def\mn@eprint@dblp#1{\href {http://dblp.uni-trier.de/rec/bibtex/#1.xml} {dblp:#1}}
\def\mn@eprint@#1:#2:#3:#4\@nil{\def\@tempa {#1}\def\@tempb {#2}\def\@tempc {#3}\ifx \@tempc \@empty \let \@tempc \@tempb \let \@tempb \@tempa \fi \ifx \@tempb \@empty \def\@tempb {arXiv}\fi \@ifundefined {mn@eprint@\@tempb}{\@tempb:\@tempc}{\expandafter \expandafter \csname mn@eprint@\@tempb\endcsname \expandafter{\@tempc}}}

\bibitem[\protect\citeauthoryear{Andersson, Agertz, Renaud  \& Teyssier}{Andersson et~al.}{2023}]{Andersson_2023}
Andersson E.~P.,  Agertz O.,  Renaud F.,   Teyssier R.,  2023, \mn@doi [\mnras] {10.1093/mnras/stad692}, 521, 2196–2214

\bibitem[\protect\citeauthoryear{{Andersson}, {Mac Low}, {Agertz}, {Renaud}  \& {Li}}{{Andersson} et~al.}{2024}]{2024A&A...681A..28A}
{Andersson} E.~P.,  {Mac Low} M.-M.,  {Agertz} O.,  {Renaud} F.,   {Li} H.,  2024, \mn@doi [\aap] {10.1051/0004-6361/202347792}, \href {https://ui.adsabs.harvard.edu/abs/2024A&A...681A..28A} {681, A28}

\bibitem[\protect\citeauthoryear{Anglés-Alcázar, Faucher-Giguère, Quataert, Hopkins, Feldmann, Torrey, Wetzel  \& Kereš}{Anglés-Alcázar et~al.}{2017}]{Angl_s_Alc_zar_2017}
Anglés-Alcázar D.,  Faucher-Giguère C.-A.,  Quataert E.,  Hopkins P.~F.,  Feldmann R.,  Torrey P.,  Wetzel A.,   Kereš D.,  2017, \mn@doi [\mnras] {10.1093/mnrasl/slx161}, 472, L109–L114

\bibitem[\protect\citeauthoryear{{Arca Sedda}, {Kamlah}, {Spurzem}, {Rizzuto}, {Naab}, {Giersz}  \& {Berczik}}{{Arca Sedda} et~al.}{2023}]{2023MNRAS.526..429A}
{Arca Sedda} M.,  {Kamlah} A. W.~H.,  {Spurzem} R.,  {Rizzuto} F.~P.,  {Naab} T.,  {Giersz} M.,   {Berczik} P.,  2023, \mn@doi [\mnras] {10.1093/mnras/stad2292}, \href {https://ui.adsabs.harvard.edu/abs/2023MNRAS.526..429A} {526, 429}

\bibitem[\protect\citeauthoryear{Arjona-Galvez, Cintio  \& Grand}{Arjona-Galvez et~al.}{2024}]{arjonagalvez2024roleagnfeedbackevolution}
Arjona-Galvez E.,  Cintio A.~D.,   Grand R. J.~J.,  2024 (\mn@eprint {arXiv} {2402.00929})

\bibitem[\protect\citeauthoryear{Askar, Baldassare  \& Mezcua}{Askar et~al.}{2024}]{askar2024intermediatemassblackholesstar}
Askar A.,  Baldassare V.~F.,   Mezcua M.,  2024 (\mn@eprint {arXiv} {2311.12118})

\bibitem[\protect\citeauthoryear{Barai \& de Gouveia Dal Pino}{Barai \& de Gouveia Dal Pino}{2019}]{Barai_2019}
Barai P.,  de Gouveia Dal Pino E.~M.,  2019, \mn@doi [\mnras] {10.1093/mnras/stz1616}, 487, 5549–5563

\bibitem[\protect\citeauthoryear{Bate, Bonnell  \& Price}{Bate et~al.}{1995}]{Bate_1995}
Bate M.~R.,  Bonnell I.~A.,   Price N.~M.,  1995, \mn@doi [\mnras] {10.1093/mnras/277.2.362}, 277, 362–376

\bibitem[\protect\citeauthoryear{Beckmann et~al.,}{Beckmann et~al.}{2023}]{Beckmann_2023}
Beckmann R.~S.,  et~al., 2023, \mn@doi [\mnras] {10.1093/mnras/stad1544}, 523, 5610–5623

\bibitem[\protect\citeauthoryear{{Bellovary} et~al.,}{{Bellovary} et~al.}{2021}]{2021MNRAS.505.5129B}
{Bellovary} J.~M.,  et~al., 2021, \mn@doi [\mnras] {10.1093/mnras/stab1665}, \href {https://ui.adsabs.harvard.edu/abs/2021MNRAS.505.5129B} {505, 5129}

\bibitem[\protect\citeauthoryear{Birchall, Watson  \& Aird}{Birchall et~al.}{2020}]{Birchall_2020}
Birchall K.~L.,  Watson M.~G.,   Aird J.,  2020, \mn@doi [\mnras] {10.1093/mnras/staa040}, 492, 2268

\bibitem[\protect\citeauthoryear{{Bruzual} \& {Charlot}}{{Bruzual} \& {Charlot}}{2003}]{2003MNRAS.344.1000B}
{Bruzual} G.,  {Charlot} S.,  2003, \mn@doi [\mnras] {10.1046/j.1365-8711.2003.06897.x}, \href {https://ui.adsabs.harvard.edu/abs/2003MNRAS.344.1000B} {344, 1000}

\bibitem[\protect\citeauthoryear{Bulirsch \& Stoer}{Bulirsch \& Stoer}{1966}]{Bulirsch1966}
Bulirsch R.,  Stoer J.,  1966, Numerische Mathematik, 8, 1

\bibitem[\protect\citeauthoryear{Bykov, Gilfanov  \& Sunyaev}{Bykov et~al.}{2023}]{bykov2023srgerosita}
Bykov S.~D.,  Gilfanov M.~R.,   Sunyaev R.~A.,  2023, SRG/eROSITA catalogue of X-ray active SDSS dwarf galaxies (\mn@eprint {arXiv} {2310.00303})

\bibitem[\protect\citeauthoryear{{Camps} \& {Baes}}{{Camps} \& {Baes}}{2020}]{2020A&C....3100381C}
{Camps} P.,  {Baes} M.,  2020, \mn@doi [Astronomy and Computing] {10.1016/j.ascom.2020.100381}, \href {https://ui.adsabs.harvard.edu/abs/2020A&C....3100381C} {31, 100381}

\bibitem[\protect\citeauthoryear{{Castelli} \& {Kurucz}}{{Castelli} \& {Kurucz}}{2003}]{2003IAUS..210P.A20C}
{Castelli} F.,  {Kurucz} R.~L.,  2003, in {Piskunov} N.,  {Weiss} W.~W.,   {Gray} D.~F.,  eds,  IAU Symposium Vol. 210, Modelling of Stellar Atmospheres. p.~A20 (\mn@eprint {arXiv} {astro-ph/0405087}), \mn@doi{10.48550/arXiv.astro-ph/0405087}

\bibitem[\protect\citeauthoryear{{Chieffi} \& {Limongi}}{{Chieffi} \& {Limongi}}{2004}]{2004ApJ...608..405C}
{Chieffi} A.,  {Limongi} M.,  2004, \mn@doi [\apj] {10.1086/392523}, \href {https://ui.adsabs.harvard.edu/abs/2004ApJ...608..405C} {608, 405}

\bibitem[\protect\citeauthoryear{Clark, Glover  \& Klessen}{Clark et~al.}{2011}]{Clark_2011}
Clark P.~C.,  Glover S. C.~O.,   Klessen R.~S.,  2011, \mn@doi [\mnras] {10.1111/j.1365-2966.2011.20087.x}, 420, 745–756

\bibitem[\protect\citeauthoryear{Davis et~al.,}{Davis et~al.}{2022}]{Davis_2022}
Davis F.,  et~al., 2022, \mn@doi [\mnras] {10.1093/mnras/stac068}, 511, 4109

\bibitem[\protect\citeauthoryear{{Dehnen}}{{Dehnen}}{1993}]{1993MNRAS.265..250D}
{Dehnen} W.,  1993, \mn@doi [\mnras] {10.1093/mnras/265.1.250}, \href {https://ui.adsabs.harvard.edu/abs/1993MNRAS.265..250D} {265, 250}

\bibitem[\protect\citeauthoryear{{Dehnen} \& {Aly}}{{Dehnen} \& {Aly}}{2012}]{2012MNRAS.425.1068D}
{Dehnen} W.,  {Aly} H.,  2012, \mn@doi [\mnras] {10.1111/j.1365-2966.2012.21439.x}, \href {https://ui.adsabs.harvard.edu/abs/2012MNRAS.425.1068D} {425, 1068}

\bibitem[\protect\citeauthoryear{{Dubois}, {Volonteri}, {Silk}, {Devriendt}, {Slyz}  \& {Teyssier}}{{Dubois} et~al.}{2015}]{2015MNRAS.452.1502D}
{Dubois} Y.,  {Volonteri} M.,  {Silk} J.,  {Devriendt} J.,  {Slyz} A.,   {Teyssier} R.,  2015, \mn@doi [\mnras] {10.1093/mnras/stv1416}, \href {https://ui.adsabs.harvard.edu/abs/2015MNRAS.452.1502D} {452, 1502}

\bibitem[\protect\citeauthoryear{{Fahrion} et~al.,}{{Fahrion} et~al.}{2019}]{2019A&A...628A..92F}
{Fahrion} K.,  et~al., 2019, \mn@doi [\aap] {10.1051/0004-6361/201935832}, \href {https://ui.adsabs.harvard.edu/abs/2019A&A...628A..92F} {628, A92}

\bibitem[\protect\citeauthoryear{Fahrion et~al.,}{Fahrion et~al.}{2021}]{Fahrion_2021}
Fahrion K.,  et~al., 2021, \mn@doi [\aap] {10.1051/0004-6361/202140644}, 650, A137

\bibitem[\protect\citeauthoryear{Fahrion, Leaman, Lyubenova  \& van~de Ven}{Fahrion et~al.}{2022a}]{Fahrion_2022}
Fahrion K.,  Leaman R.,  Lyubenova M.,   van~de Ven G.,  2022a, \mn@doi [\aap] {10.1051/0004-6361/202039778}, 658, A172

\bibitem[\protect\citeauthoryear{Fahrion et~al.,}{Fahrion et~al.}{2022b}]{Fahrion_2022b}
Fahrion K.,  et~al., 2022b, \mn@doi [\aap] {10.1051/0004-6361/202244932}, 667, A101

\bibitem[\protect\citeauthoryear{Fahrion et~al.,}{Fahrion et~al.}{2024}]{Fahrion_2024}
Fahrion K.,  et~al., 2024, \mn@doi [\aap] {10.1051/0004-6361/202449629}, 687, A83

\bibitem[\protect\citeauthoryear{{Fotopoulou} et~al.,}{{Fotopoulou} et~al.}{2024}]{2024MNRAS.534..215F}
{Fotopoulou} C.~M.,  et~al., 2024, \mn@doi [\mnras] {10.1093/mnras/stae2072}, \href {https://ui.adsabs.harvard.edu/abs/2024MNRAS.534..215F} {534, 215}

\bibitem[\protect\citeauthoryear{{Fujii}, {Wang}, {Tanikawa}, {Hirai}  \& {Saitoh}}{{Fujii} et~al.}{2024}]{2024Sci...384.1488F}
{Fujii} M.~S.,  {Wang} L.,  {Tanikawa} A.,  {Hirai} Y.,   {Saitoh} T.~R.,  2024, \mn@doi [Science] {10.1126/science.adi4211}, \href {https://ui.adsabs.harvard.edu/abs/2024Sci...384.1488F} {384, 1488}

\bibitem[\protect\citeauthoryear{{Georgy} et~al.,}{{Georgy} et~al.}{2013}]{2013A&A...558A.103G}
{Georgy} C.,  et~al., 2013, \mn@doi [\aap] {10.1051/0004-6361/201322178}, \href {https://ui.adsabs.harvard.edu/abs/2013A&A...558A.103G} {558, A103}

\bibitem[\protect\citeauthoryear{{Gerssen}, {van der Marel}, {Gebhardt}, {Guhathakurta}, {Peterson}  \& {Pryor}}{{Gerssen} et~al.}{2002}]{2002AJ....124.3270G}
{Gerssen} J.,  {van der Marel} R.~P.,  {Gebhardt} K.,  {Guhathakurta} P.,  {Peterson} R.~C.,   {Pryor} C.,  2002, \mn@doi [\aj] {10.1086/344584}, \href {https://ui.adsabs.harvard.edu/abs/2002AJ....124.3270G} {124, 3270}

\bibitem[\protect\citeauthoryear{{Glover} \& {Clark}}{{Glover} \& {Clark}}{2012}]{2012MNRAS.421..116G}
{Glover} S. C.~O.,  {Clark} P.~C.,  2012, \mn@doi [\mnras] {10.1111/j.1365-2966.2011.20260.x}, \href {https://ui.adsabs.harvard.edu/abs/2012MNRAS.421..116G} {421, 116}

\bibitem[\protect\citeauthoryear{{Glover} \& {Mac Low}}{{Glover} \& {Mac Low}}{2007}]{2007ApJS..169..239G}
{Glover} S. C.~O.,  {Mac Low} M.-M.,  2007, \mn@doi [\apjs] {10.1086/512238}, \href {https://ui.adsabs.harvard.edu/abs/2007ApJS..169..239G} {169, 239}

\bibitem[\protect\citeauthoryear{Gorski, Hivon, Banday, Wandelt, Hansen, Reinecke  \& Bartelmann}{Gorski et~al.}{2005}]{Gorski_2005}
Gorski K.~M.,  Hivon E.,  Banday A.~J.,  Wandelt B.~D.,  Hansen F.~K.,  Reinecke M.,   Bartelmann M.,  2005, \mn@doi [\apj] {10.1086/427976}, 622, 759

\bibitem[\protect\citeauthoryear{{Gragg}}{{Gragg}}{1965}]{1965SJNA....2..384G}
{Gragg} W.~B.,  1965, \mn@doi [SIAM Journal on Numerical Analysis] {10.1137/0702030}, \href {https://ui.adsabs.harvard.edu/abs/1965SJNA....2..384G} {2, 384}

\bibitem[\protect\citeauthoryear{{Greene} \& {Ho}}{{Greene} \& {Ho}}{2006}]{2006ApJ...641L..21G}
{Greene} J.~E.,  {Ho} L.~C.,  2006, \mn@doi [\apjl] {10.1086/500507}, \href {https://ui.adsabs.harvard.edu/abs/2006ApJ...641L..21G} {641, L21}

\bibitem[\protect\citeauthoryear{{Greene} \& {Ho}}{{Greene} \& {Ho}}{2007a}]{2007ApJ...656...84G}
{Greene} J.~E.,  {Ho} L.~C.,  2007a, \mn@doi [\apj] {10.1086/509064}, \href {https://ui.adsabs.harvard.edu/abs/2007ApJ...656...84G} {656, 84}

\bibitem[\protect\citeauthoryear{{Greene} \& {Ho}}{{Greene} \& {Ho}}{2007b}]{2007ApJ...670...92G}
{Greene} J.~E.,  {Ho} L.~C.,  2007b, \mn@doi [\apj] {10.1086/522082}, \href {https://ui.adsabs.harvard.edu/abs/2007ApJ...670...92G} {670, 92}

\bibitem[\protect\citeauthoryear{{Greene}, {Strader}  \& {Ho}}{{Greene} et~al.}{2020}]{Greene2020ARA&A..58..257G}
{Greene} J.~E.,  {Strader} J.,   {Ho} L.~C.,  2020, \mn@doi [\araa] {10.1146/annurev-astro-032620-021835}, \href {https://ui.adsabs.harvard.edu/abs/2020ARA&A..58..257G} {58, 257}

\bibitem[\protect\citeauthoryear{{Habouzit}, {Volonteri}  \& {Dubois}}{{Habouzit} et~al.}{2017}]{2017MNRAS.468.3935H}
{Habouzit} M.,  {Volonteri} M.,   {Dubois} Y.,  2017, \mn@doi [\mnras] {10.1093/mnras/stx666}, \href {https://ui.adsabs.harvard.edu/abs/2017MNRAS.468.3935H} {468, 3935}

\bibitem[\protect\citeauthoryear{{Hernquist}}{{Hernquist}}{1990}]{1990ApJ...356..359H}
{Hernquist} L.,  1990, \mn@doi [\apj] {10.1086/168845}, \href {https://ui.adsabs.harvard.edu/abs/1990ApJ...356..359H} {356, 359}

\bibitem[\protect\citeauthoryear{{Hislop}, {Naab}, {Steinwandel}, {Lah{\'e}n}, {Irodotou}, {Johansson}  \& {Walch}}{{Hislop} et~al.}{2022}]{2022MNRAS.509.5938H}
{Hislop} J.~M.,  {Naab} T.,  {Steinwandel} U.~P.,  {Lah{\'e}n} N.,  {Irodotou} D.,  {Johansson} P.~H.,   {Walch} S.,  2022, \mn@doi [\mnras] {10.1093/mnras/stab3347}, \href {https://ui.adsabs.harvard.edu/abs/2022MNRAS.509.5938H} {509, 5938}

\bibitem[\protect\citeauthoryear{{Hopkins}, {Quataert}  \& {Murray}}{{Hopkins} et~al.}{2012}]{2012MNRAS.421.3488H}
{Hopkins} P.~F.,  {Quataert} E.,   {Murray} N.,  2012, \mn@doi [\mnras] {10.1111/j.1365-2966.2012.20578.x}, \href {https://ui.adsabs.harvard.edu/abs/2012MNRAS.421.3488H} {421, 3488}

\bibitem[\protect\citeauthoryear{{Hopkins} et~al.,}{{Hopkins} et~al.}{2024}]{2024OJAp....7E..18H}
{Hopkins} P.~F.,  et~al., 2024, \mn@doi [The Open Journal of Astrophysics] {10.21105/astro.2309.13115}, \href {https://ui.adsabs.harvard.edu/abs/2024OJAp....7E..18H} {7, 18}

\bibitem[\protect\citeauthoryear{Hoyer, Neumayer, Georgiev, Seth  \& Greene}{Hoyer et~al.}{2021}]{Hoyer_2021}
Hoyer N.,  Neumayer N.,  Georgiev I.~Y.,  Seth A.~C.,   Greene J.~E.,  2021, \mn@doi [\mnras] {10.1093/mnras/stab2277}, 507, 3246–3266

\bibitem[\protect\citeauthoryear{Hoyer, Neumayer, Seth, Georgiev  \& Greene}{Hoyer et~al.}{2023}]{Hoyer_2023}
Hoyer N.,  Neumayer N.,  Seth A.~C.,  Georgiev I.~Y.,   Greene J.~E.,  2023, \mn@doi [\mnras] {10.1093/mnras/stad220}, 520, 4664–4682

\bibitem[\protect\citeauthoryear{Hoyer, Arcodia, Bonoli, Merloni, Neumayer, Zhang  \& Comparat}{Hoyer et~al.}{2024a}]{hoyer2024massiveblackholesnuclear}
Hoyer N.,  Arcodia R.,  Bonoli S.,  Merloni A.,  Neumayer N.,  Zhang Y.,   Comparat J.,  2024a (\mn@eprint {arXiv} {2401.17288})

\bibitem[\protect\citeauthoryear{{Hoyer}, {Arcodia}, {Bonoli}, {Merloni}, {Neumayer}, {Zhang}  \& {Comparat}}{{Hoyer} et~al.}{2024b}]{2024A&A...682A..36H}
{Hoyer} N.,  {Arcodia} R.,  {Bonoli} S.,  {Merloni} A.,  {Neumayer} N.,  {Zhang} Y.,   {Comparat} J.,  2024b, \mn@doi [\aap] {10.1051/0004-6361/202347665}, \href {https://ui.adsabs.harvard.edu/abs/2024A&A...682A..36H} {682, A36}

\bibitem[\protect\citeauthoryear{Hu}{Hu}{2016}]{ediss19451}
Hu C.-Y.,  2016, Star formation and molecular hydrogen in dwarf galaxies, \url {http://nbn-resolving.de/urn:nbn:de:bvb:19-194516}

\bibitem[\protect\citeauthoryear{{Hu}}{{Hu}}{2019}]{2019MNRAS.483.3363H}
{Hu} C.-Y.,  2019, \mn@doi [\mnras] {10.1093/mnras/sty3252}, \href {https://ui.adsabs.harvard.edu/abs/2019MNRAS.483.3363H} {483, 3363}

\bibitem[\protect\citeauthoryear{Hu, Naab, Walch, Moster  \& Oser}{Hu et~al.}{2014}]{Hu_2014}
Hu C.-Y.,  Naab T.,  Walch S.,  Moster B.~P.,   Oser L.,  2014, \mn@doi [\mnras] {10.1093/mnras/stu1187}, 443, 1173

\bibitem[\protect\citeauthoryear{Hu, Naab, Walch, Glover  \& Clark}{Hu et~al.}{2016}]{Hu_2016}
Hu C.-Y.,  Naab T.,  Walch S.,  Glover S. C.~O.,   Clark P.~C.,  2016, \mn@doi [\mnras] {10.1093/mnras/stw544}, 458, 3528

\bibitem[\protect\citeauthoryear{Hu, Naab, Glover, Walch  \& Clark}{Hu et~al.}{2017}]{Hu_2017}
Hu C.-Y.,  Naab T.,  Glover S. C.~O.,  Walch S.,   Clark P.~C.,  2017, \mn@doi [\mnras] {10.1093/mnras/stx1773}, 471, 2151

\bibitem[\protect\citeauthoryear{Häberle et~al.,}{Häberle et~al.}{2024}]{H_berle_2024}
Häberle M.,  et~al., 2024, \mn@doi [Nature] {10.1038/s41586-024-07511-z}, 631, 285–288

\bibitem[\protect\citeauthoryear{Ibata et~al.,}{Ibata et~al.}{2009}]{Ibata_2009}
Ibata R.,  et~al., 2009, \mn@doi [\apj] {10.1088/0004-637x/699/2/l169}, 699, L169–L173

\bibitem[\protect\citeauthoryear{Inayoshi, Visbal  \& Haiman}{Inayoshi et~al.}{2020}]{Inayoshi_2020}
Inayoshi K.,  Visbal E.,   Haiman Z.,  2020, \mn@doi [\araa] {10.1146/annurev-astro-120419-014455}, 58, 27–97

\bibitem[\protect\citeauthoryear{Islam, Taylor  \& Silk}{Islam et~al.}{2003}]{Islam_2003}
Islam R.~R.,  Taylor J.~E.,   Silk J.,  2003, \mn@doi [\mnras] {10.1046/j.1365-8711.2003.06329.x}, 340, 647–656

\bibitem[\protect\citeauthoryear{{Karakas}}{{Karakas}}{2010}]{2010MNRAS.403.1413K}
{Karakas} A.~I.,  2010, \mn@doi [\mnras] {10.1111/j.1365-2966.2009.16198.x}, \href {https://ui.adsabs.harvard.edu/abs/2010MNRAS.403.1413K} {403, 1413}

\bibitem[\protect\citeauthoryear{{Kaviraj}, {Martin}  \& {Silk}}{{Kaviraj} et~al.}{2019}]{2019MNRAS.489L..12K}
{Kaviraj} S.,  {Martin} G.,   {Silk} J.,  2019, \mn@doi [\mnras] {10.1093/mnrasl/slz102}, \href {https://ui.adsabs.harvard.edu/abs/2019MNRAS.489L..12K} {489, L12}

\bibitem[\protect\citeauthoryear{{Kormendy} \& {Ho}}{{Kormendy} \& {Ho}}{2013}]{2013ARA&A..51..511K}
{Kormendy} J.,  {Ho} L.~C.,  2013, \mn@doi [\araa] {10.1146/annurev-astro-082708-101811}, \href {https://ui.adsabs.harvard.edu/abs/2013ARA&A..51..511K} {51, 511}

\bibitem[\protect\citeauthoryear{{Koudmani}, {Sijacki}, {Bourne}  \& {Smith}}{{Koudmani} et~al.}{2019}]{2019MNRAS.484.2047K}
{Koudmani} S.,  {Sijacki} D.,  {Bourne} M.~A.,   {Smith} M.~C.,  2019, \mn@doi [\mnras] {10.1093/mnras/stz097}, \href {https://ui.adsabs.harvard.edu/abs/2019MNRAS.484.2047K} {484, 2047}

\bibitem[\protect\citeauthoryear{Koudmani, Henden  \& Sijacki}{Koudmani et~al.}{2021}]{Koudmani_2021}
Koudmani S.,  Henden N.~A.,   Sijacki D.,  2021, \mn@doi [\mnras] {10.1093/mnras/stab677}, 503, 3568–3591

\bibitem[\protect\citeauthoryear{Koudmani, Sijacki  \& Smith}{Koudmani et~al.}{2022}]{Koudmani_2022}
Koudmani S.,  Sijacki D.,   Smith M.~C.,  2022, \mn@doi [\mnras] {10.1093/mnras/stac2252}, 516, 2112

\bibitem[\protect\citeauthoryear{{Kritos}, {Berti}  \& {Silk}}{{Kritos} et~al.}{2023}]{2023PhRvD.108h3012K}
{Kritos} K.,  {Berti} E.,   {Silk} J.,  2023, \mn@doi [\prd] {10.1103/PhysRevD.108.083012}, \href {https://ui.adsabs.harvard.edu/abs/2023PhRvD.108h3012K} {108, 083012}

\bibitem[\protect\citeauthoryear{{Kroupa}}{{Kroupa}}{2001}]{2001MNRAS.322..231K}
{Kroupa} P.,  2001, \mn@doi [\mnras] {10.1046/j.1365-8711.2001.04022.x}, \href {https://ui.adsabs.harvard.edu/abs/2001MNRAS.322..231K} {322, 231}

\bibitem[\protect\citeauthoryear{Kızıltan, Baumgardt  \& Loeb}{Kızıltan et~al.}{2017}]{K_z_ltan_2017}
Kızıltan B.,  Baumgardt H.,   Loeb A.,  2017, \mn@doi [Nature] {10.1038/nature21361}, 542, 203–205

\bibitem[\protect\citeauthoryear{{Lah{\'e}n}, {Naab}, {Johansson}, {Elmegreen}, {Hu}  \& {Walch}}{{Lah{\'e}n} et~al.}{2020}]{2020ApJ...904...71L}
{Lah{\'e}n} N.,  {Naab} T.,  {Johansson} P.~H.,  {Elmegreen} B.,  {Hu} C.-Y.,   {Walch} S.,  2020, \mn@doi [\apj] {10.3847/1538-4357/abc001}, \href {https://ui.adsabs.harvard.edu/abs/2020ApJ...904...71L} {904, 71}

\bibitem[\protect\citeauthoryear{Lah{\'{e}}n, Naab  \& Kauffmann}{Lah{\'{e}}n et~al.}{2022}]{Lah_n_2022}
Lah{\'{e}}n N.,  Naab T.,   Kauffmann G.,  2022, \mn@doi [\mnras] {10.1093/mnras/stac1594}, 514, 4560

\bibitem[\protect\citeauthoryear{Lah{\'{e}}n et~al.,}{Lah{\'{e}}n et~al.}{2023}]{Lah_n_2023}
Lah{\'{e}}n N.,  et~al., 2023, \mn@doi [\mnras] {10.1093/mnras/stad1147}, 522, 3092

\bibitem[\protect\citeauthoryear{{Lah{\'e}n}, {Naab}  \& {Sz{\'e}csi}}{{Lah{\'e}n} et~al.}{2024}]{2024MNRAS.530..645L}
{Lah{\'e}n} N.,  {Naab} T.,   {Sz{\'e}csi} D.,  2024, \mn@doi [\mnras] {10.1093/mnras/stae904}, \href {https://ui.adsabs.harvard.edu/abs/2024MNRAS.530..645L} {530, 645}

\bibitem[\protect\citeauthoryear{Latimer, Reines, Bogdan  \& Kraft}{Latimer et~al.}{2021}]{Latimer_2021}
Latimer L.~J.,  Reines A.~E.,  Bogdan A.,   Kraft R.,  2021, \mn@doi [\apj] {10.3847/2041-8213/ac3af6}, 922, L40

\bibitem[\protect\citeauthoryear{{Lejeune}, {Cuisinier}  \& {Buser}}{{Lejeune} et~al.}{1997}]{1997A&AS..125..229L}
{Lejeune} T.,  {Cuisinier} F.,   {Buser} R.,  1997, \mn@doi [\aaps] {10.1051/aas:1997373}, \href {http://adsabs.harvard.edu/abs/1997A%26AS..125..229L} {125, 229}

\bibitem[\protect\citeauthoryear{{Lejeune}, {Cuisinier}  \& {Buser}}{{Lejeune} et~al.}{1998}]{1998A&AS..130...65L}
{Lejeune} T.,  {Cuisinier} F.,   {Buser} R.,  1998, \mn@doi [\aaps] {10.1051/aas:1998405}, \href {http://adsabs.harvard.edu/abs/1998A%26AS..130...65L} {130, 65}

\bibitem[\protect\citeauthoryear{{Liao}, {Irodotou}, {Johansson}, {Naab}, {Rizzuto}, {Hislop}, {Rawlings}  \& {Wright}}{{Liao} et~al.}{2024a}]{2024MNRAS.528.5080L}
{Liao} S.,  {Irodotou} D.,  {Johansson} P.~H.,  {Naab} T.,  {Rizzuto} F.~P.,  {Hislop} J.~M.,  {Rawlings} A.,   {Wright} R.~J.,  2024a, \mn@doi [\mnras] {10.1093/mnras/stae360}, \href {https://ui.adsabs.harvard.edu/abs/2024MNRAS.528.5080L} {528, 5080}

\bibitem[\protect\citeauthoryear{{Liao}, {Irodotou}, {Johansson}, {Naab}, {Rizzuto}, {Hislop}, {Wright}  \& {Rawlings}}{{Liao} et~al.}{2024b}]{2024MNRAS.530.4058L}
{Liao} S.,  {Irodotou} D.,  {Johansson} P.~H.,  {Naab} T.,  {Rizzuto} F.~P.,  {Hislop} J.~M.,  {Wright} R.~J.,   {Rawlings} A.,  2024b, \mn@doi [\mnras] {10.1093/mnras/stae1123}, \href {https://ui.adsabs.harvard.edu/abs/2024MNRAS.530.4058L} {530, 4058}

\bibitem[\protect\citeauthoryear{{L{\"u}tzgendorf} et~al.,}{{L{\"u}tzgendorf} et~al.}{2013a}]{2013A&A...552A..49L}
{L{\"u}tzgendorf} N.,  et~al., 2013a, \mn@doi [\aap] {10.1051/0004-6361/201220307}, \href {https://ui.adsabs.harvard.edu/abs/2013A&A...552A..49L} {552, A49}

\bibitem[\protect\citeauthoryear{{L{\"u}tzgendorf} et~al.,}{{L{\"u}tzgendorf} et~al.}{2013b}]{2013A&A...555A..26L}
{L{\"u}tzgendorf} N.,  et~al., 2013b, \mn@doi [\aap] {10.1051/0004-6361/201321183}, \href {https://ui.adsabs.harvard.edu/abs/2013A&A...555A..26L} {555, A26}

\bibitem[\protect\citeauthoryear{Lützgendorf, Gebhardt, Baumgardt, Noyola, Neumayer, Kissler-Patig  \& de Zeeuw}{Lützgendorf et~al.}{2015}]{L_tzgendorf_2015}
Lützgendorf N.,  Gebhardt K.,  Baumgardt H.,  Noyola E.,  Neumayer N.,  Kissler-Patig M.,   de Zeeuw T.,  2015, \mn@doi [\aap] {10.1051/0004-6361/201425524}, 581, A1

\bibitem[\protect\citeauthoryear{Ma, Hopkins, Ma, Anglés-Alcázar, Faucher-Giguère  \& Kelley}{Ma et~al.}{2021}]{Ma_2021}
Ma L.,  Hopkins P.~F.,  Ma X.,  Anglés-Alcázar D.,  Faucher-Giguère C.-A.,   Kelley L.~Z.,  2021, \mn@doi [\mnras] {10.1093/mnras/stab2713}, 508, 1973–1985

\bibitem[\protect\citeauthoryear{{Mannerkoski}, {Johansson}, {Rantala}, {Naab}  \& {Liao}}{{Mannerkoski} et~al.}{2021}]{2021ApJ...912L..20M}
{Mannerkoski} M.,  {Johansson} P.~H.,  {Rantala} A.,  {Naab} T.,   {Liao} S.,  2021, \mn@doi [\apjl] {10.3847/2041-8213/abf9a5}, \href {https://ui.adsabs.harvard.edu/abs/2021ApJ...912L..20M} {912, L20}

\bibitem[\protect\citeauthoryear{Mannerkoski, Rawlings, Johansson, Naab, Rantala, Springel, Irodotou  \& Liao}{Mannerkoski et~al.}{2023}]{Mannerkoski_2023}
Mannerkoski M.,  Rawlings A.,  Johansson P.~H.,  Naab T.,  Rantala A.,  Springel V.,  Irodotou D.,   Liao S.,  2023, \mn@doi [\mnras] {10.1093/mnras/stad2139}, 524, 4062

\bibitem[\protect\citeauthoryear{{Mapelli}}{{Mapelli}}{2007}]{2007MNRAS.376.1317M}
{Mapelli} M.,  2007, \mn@doi [\mnras] {10.1111/j.1365-2966.2007.11528.x}, \href {https://ui.adsabs.harvard.edu/abs/2007MNRAS.376.1317M} {376, 1317}

\bibitem[\protect\citeauthoryear{{Mehta}, {Regan}  \& {Prole}}{{Mehta} et~al.}{2024}]{2024arXiv240908326M}
{Mehta} D.,  {Regan} J.~A.,   {Prole} L.,  2024, \mn@doi [arXiv e-prints] {10.48550/arXiv.2409.08326}, \href {https://ui.adsabs.harvard.edu/abs/2024arXiv240908326M} {p. arXiv:2409.08326}

\bibitem[\protect\citeauthoryear{{Mezcua} \& {Dom{\'\i}nguez S{\'a}nchez}}{{Mezcua} \& {Dom{\'\i}nguez S{\'a}nchez}}{2020}]{2020ApJ...898L..30M}
{Mezcua} M.,  {Dom{\'\i}nguez S{\'a}nchez} H.,  2020, \mn@doi [\apjl] {10.3847/2041-8213/aba199}, \href {https://ui.adsabs.harvard.edu/abs/2020ApJ...898L..30M} {898, L30}

\bibitem[\protect\citeauthoryear{Mezcua \& S{\'{a}}nchez}{Mezcua \& S{\'{a}}nchez}{2020}]{Mezcua_2020}
Mezcua M.,  S{\'{a}}nchez H.~D.,  2020, \mn@doi [\apj] {10.3847/2041-8213/aba199}, 898, L30

\bibitem[\protect\citeauthoryear{Mezcua, Civano, Marchesi, Suh, Fabbiano  \& Volonteri}{Mezcua et~al.}{2018}]{Mezcua_2018}
Mezcua M.,  Civano F.,  Marchesi S.,  Suh H.,  Fabbiano G.,   Volonteri M.,  2018, \mn@doi [\mnras] {10.1093/mnras/sty1163}, 478, 2576

\bibitem[\protect\citeauthoryear{Mezcua, Suh  \& Civano}{Mezcua et~al.}{2019}]{Mezcua_2019}
Mezcua M.,  Suh H.,   Civano F.,  2019, \mn@doi [\mnras] {10.1093/mnras/stz1760}, 488, 685

\bibitem[\protect\citeauthoryear{Mezcua, Siudek, Suh, Valiante, Spinoso  \& Bonoli}{Mezcua et~al.}{2023}]{Mezcua_2023}
Mezcua M.,  Siudek M.,  Suh H.,  Valiante R.,  Spinoso D.,   Bonoli S.,  2023, \mn@doi [\apj] {10.3847/2041-8213/acae25}, 943, L5

\bibitem[\protect\citeauthoryear{{Mikkola} \& {Merritt}}{{Mikkola} \& {Merritt}}{2006}]{2006MNRAS.372..219M}
{Mikkola} S.,  {Merritt} D.,  2006, \mn@doi [\mnras] {10.1111/j.1365-2966.2006.10854.x}, \href {https://ui.adsabs.harvard.edu/abs/2006MNRAS.372..219M} {372, 219}

\bibitem[\protect\citeauthoryear{{Mikkola} \& {Merritt}}{{Mikkola} \& {Merritt}}{2008}]{2008AJ....135.2398M}
{Mikkola} S.,  {Merritt} D.,  2008, \mn@doi [\aj] {10.1088/0004-6256/135/6/2398}, \href {https://ui.adsabs.harvard.edu/abs/2008AJ....135.2398M} {135, 2398}

\bibitem[\protect\citeauthoryear{{Mikkola} \& {Tanikawa}}{{Mikkola} \& {Tanikawa}}{1999}]{1999MNRAS.310..745M}
{Mikkola} S.,  {Tanikawa} K.,  1999, \mn@doi [\mnras] {10.1046/j.1365-8711.1999.02982.x}, \href {https://ui.adsabs.harvard.edu/abs/1999MNRAS.310..745M} {310, 745}

\bibitem[\protect\citeauthoryear{{Milosavljevi{\'c}}}{{Milosavljevi{\'c}}}{2004}]{2004ApJ...605L..13M}
{Milosavljevi{\'c}} M.,  2004, \mn@doi [\apjl] {10.1086/420696}, \href {https://ui.adsabs.harvard.edu/abs/2004ApJ...605L..13M} {605, L13}

\bibitem[\protect\citeauthoryear{{Nelson} \& {Langer}}{{Nelson} \& {Langer}}{1997}]{1997ApJ...482..796N}
{Nelson} R.~P.,  {Langer} W.~D.,  1997, \mn@doi [\apj] {10.1086/304167}, \href {https://ui.adsabs.harvard.edu/abs/1997ApJ...482..796N} {482, 796}

\bibitem[\protect\citeauthoryear{Neumayer, Seth  \& Böker}{Neumayer et~al.}{2020}]{Neumayer_2020}
Neumayer N.,  Seth A.,   Böker T.,  2020, \mn@doi [The Astronomy and Astrophysics Review] {10.1007/s00159-020-00125-0}, 28

\bibitem[\protect\citeauthoryear{{Ni} et~al.,}{{Ni} et~al.}{2022}]{2022MNRAS.513..670N}
{Ni} Y.,  et~al., 2022, \mn@doi [\mnras] {10.1093/mnras/stac351}, \href {https://ui.adsabs.harvard.edu/abs/2022MNRAS.513..670N} {513, 670}

\bibitem[\protect\citeauthoryear{Pacucci, Mezcua  \& Regan}{Pacucci et~al.}{2021}]{Pacucci_2021}
Pacucci F.,  Mezcua M.,   Regan J.~A.,  2021, \mn@doi [\apj] {10.3847/1538-4357/ac1595}, 920, 134

\bibitem[\protect\citeauthoryear{{Partmann}, {Naab}, {Rantala}, {Genina}, {Mannerkoski}  \& {Johansson}}{{Partmann} et~al.}{2024}]{2024MNRAS.532.4681P}
{Partmann} C.,  {Naab} T.,  {Rantala} A.,  {Genina} A.,  {Mannerkoski} M.,   {Johansson} P.~H.,  2024, \mn@doi [\mnras] {10.1093/mnras/stae1712}, \href {https://ui.adsabs.harvard.edu/abs/2024MNRAS.532.4681P} {532, 4681}

\bibitem[\protect\citeauthoryear{Pechetti, Seth, Neumayer, Georgiev, Kacharov  \& den Brok}{Pechetti et~al.}{2020}]{Pechetti_2020}
Pechetti R.,  Seth A.,  Neumayer N.,  Georgiev I.,  Kacharov N.,   den Brok M.,  2020, \mn@doi [\apj] {10.3847/1538-4357/abaaa7}, 900, 32

\bibitem[\protect\citeauthoryear{Pfister, Volonteri, Dubois, Dotti  \& Colpi}{Pfister et~al.}{2019}]{Pfister_2019}
Pfister H.,  Volonteri M.,  Dubois Y.,  Dotti M.,   Colpi M.,  2019, \mn@doi [\mnras] {10.1093/mnras/stz822}, 486, 101–111

\bibitem[\protect\citeauthoryear{{Portegies Zwart}, {Baumgardt}, {Hut}, {Makino}  \& {McMillan}}{{Portegies Zwart} et~al.}{2004}]{2004Natur.428..724P}
{Portegies Zwart} S.~F.,  {Baumgardt} H.,  {Hut} P.,  {Makino} J.,   {McMillan} S. L.~W.,  2004, \mn@doi [\nat] {10.1038/nature02448}, \href {https://ui.adsabs.harvard.edu/abs/2004Natur.428..724P} {428, 724}

\bibitem[\protect\citeauthoryear{{Preto} \& {Tremaine}}{{Preto} \& {Tremaine}}{1999}]{1999AJ....118.2532P}
{Preto} M.,  {Tremaine} S.,  1999, \mn@doi [\aj] {10.1086/301102}, \href {https://ui.adsabs.harvard.edu/abs/1999AJ....118.2532P} {118, 2532}

\bibitem[\protect\citeauthoryear{{Rantala}, {Pihajoki}, {Johansson}, {Naab}, {Lah{\'e}n}  \& {Sawala}}{{Rantala} et~al.}{2017}]{Rantala_2017}
{Rantala} A.,  {Pihajoki} P.,  {Johansson} P.~H.,  {Naab} T.,  {Lah{\'e}n} N.,   {Sawala} T.,  2017, \mn@doi [\apj] {10.3847/1538-4357/aa6d65}, \href {https://ui.adsabs.harvard.edu/abs/2017ApJ...840...53R} {840, 53}

\bibitem[\protect\citeauthoryear{{Rantala}, {Johansson}, {Naab}, {Thomas}  \& {Frigo}}{{Rantala} et~al.}{2018}]{2018ApJ...864..113R}
{Rantala} A.,  {Johansson} P.~H.,  {Naab} T.,  {Thomas} J.,   {Frigo} M.,  2018, \mn@doi [\apj] {10.3847/1538-4357/aada47}, \href {https://ui.adsabs.harvard.edu/abs/2018ApJ...864..113R} {864, 113}

\bibitem[\protect\citeauthoryear{{Rantala}, {Pihajoki}, {Mannerkoski}, {Johansson}  \& {Naab}}{{Rantala} et~al.}{2020}]{2020MNRAS.492.4131R}
{Rantala} A.,  {Pihajoki} P.,  {Mannerkoski} M.,  {Johansson} P.~H.,   {Naab} T.,  2020, \mn@doi [\mnras] {10.1093/mnras/staa084}, \href {https://ui.adsabs.harvard.edu/abs/2020MNRAS.492.4131R} {492, 4131}

\bibitem[\protect\citeauthoryear{Rantala, Naab  \& Lahén}{Rantala et~al.}{2024}]{rantala2024frostclusters}
Rantala A.,  Naab T.,   Lahén N.,  2024 (\mn@eprint {arXiv} {2403.10602})

\bibitem[\protect\citeauthoryear{{Rathjen} et~al.,}{{Rathjen} et~al.}{2021}]{2021MNRAS.504.1039R}
{Rathjen} T.-E.,  et~al., 2021, \mn@doi [\mnras] {10.1093/mnras/stab900}, \href {https://ui.adsabs.harvard.edu/abs/2021MNRAS.504.1039R} {504, 1039}

\bibitem[\protect\citeauthoryear{{Rees}}{{Rees}}{1988}]{1988Natur.333..523R}
{Rees} M.~J.,  1988, \mn@doi [\nat] {10.1038/333523a0}, \href {https://ui.adsabs.harvard.edu/abs/1988Natur.333..523R} {333, 523}

\bibitem[\protect\citeauthoryear{Reines}{Reines}{2022}]{Reines_2022}
Reines A.~E.,  2022, \mn@doi [Nature Astronomy] {10.1038/s41550-021-01556-0}, 6, 26–34

\bibitem[\protect\citeauthoryear{Reines \& Deller}{Reines \& Deller}{2012}]{Reines_2012}
Reines A.~E.,  Deller A.~T.,  2012, \mn@doi [\apj] {10.1088/2041-8205/750/1/l24}, 750, L24

\bibitem[\protect\citeauthoryear{Reines, Condon, Darling  \& Greene}{Reines et~al.}{2020}]{Reines_2020}
Reines A.~E.,  Condon J.~J.,  Darling J.,   Greene J.~E.,  2020, \mn@doi [\apj] {10.3847/1538-4357/ab4999}, 888, 36

\bibitem[\protect\citeauthoryear{Rizzuto, Naab, Spurzem, Arca-Sedda, Giersz, Ostriker  \& Banerjee}{Rizzuto et~al.}{2022}]{Rizzuto_2022}
Rizzuto F.~P.,  Naab T.,  Spurzem R.,  Arca-Sedda M.,  Giersz M.,  Ostriker J.~P.,   Banerjee S.,  2022, \mn@doi [\mnras] {10.1093/mnras/stac231}, 512, 884–898

\bibitem[\protect\citeauthoryear{Sacchi, Bogdan, Chadayammuri  \& Ricarte}{Sacchi et~al.}{2024}]{sacchi2024xray}
Sacchi A.,  Bogdan A.,  Chadayammuri U.,   Ricarte A.,  2024 (\mn@eprint {arXiv} {2406.01707})

\bibitem[\protect\citeauthoryear{Schutte \& Reines}{Schutte \& Reines}{2022}]{Schutte_2022}
Schutte Z.,  Reines A.~E.,  2022, \mn@doi [Nature] {10.1038/s41586-021-04215-6}, 601, 329–333

\bibitem[\protect\citeauthoryear{Sharma, Brooks, Somerville, Tremmel, Bellovary, Wright  \& Quinn}{Sharma et~al.}{2020}]{Sharma_2020}
Sharma R.~S.,  Brooks A.~M.,  Somerville R.~S.,  Tremmel M.,  Bellovary J.,  Wright A.~C.,   Quinn T.~R.,  2020, \mn@doi [\apj] {10.3847/1538-4357/ab960e}, 897, 103

\bibitem[\protect\citeauthoryear{Shi, Kremer, Grudić, Gerling-Dunsmore  \& Hopkins}{Shi et~al.}{2022}]{Shi_2022}
Shi Y.,  Kremer K.,  Grudić M.~Y.,  Gerling-Dunsmore H.~J.,   Hopkins P.~F.,  2022, \mn@doi [\mnras] {10.1093/mnras/stac3245}, 518, 3606–3621

\bibitem[\protect\citeauthoryear{{Shi}, {Kremer}  \& {Hopkins}}{{Shi} et~al.}{2024a}]{2024arXiv240512164S}
{Shi} Y.,  {Kremer} K.,   {Hopkins} P.~F.,  2024a, \mn@doi [arXiv e-prints] {10.48550/arXiv.2405.12164}, \href {https://ui.adsabs.harvard.edu/abs/2024arXiv240512164S} {p. arXiv:2405.12164}

\bibitem[\protect\citeauthoryear{{Shi}, {Kremer}  \& {Hopkins}}{{Shi} et~al.}{2024b}]{2024ApJ...969L..31S}
{Shi} Y.,  {Kremer} K.,   {Hopkins} P.~F.,  2024b, \mn@doi [\apjl] {10.3847/2041-8213/ad5a95}, \href {https://ui.adsabs.harvard.edu/abs/2024ApJ...969L..31S} {969, L31}

\bibitem[\protect\citeauthoryear{Silk}{Silk}{2017}]{Silk_2017}
Silk J.,  2017, \mn@doi [\apj] {10.3847/2041-8213/aa67da}, 839, L13

\bibitem[\protect\citeauthoryear{Sivasankaran et~al.,}{Sivasankaran et~al.}{2022}]{Sivasankaran_2022}
Sivasankaran A.,  et~al., 2022, \mn@doi [\mnras] {10.1093/mnras/stac2759}, 517, 4752–4767

\bibitem[\protect\citeauthoryear{{Smith}, {Bryan}, {Somerville}, {Hu}, {Teyssier}, {Burkhart}  \& {Hernquist}}{{Smith} et~al.}{2021}]{2021MNRAS.506.3882S}
{Smith} M.~C.,  {Bryan} G.~L.,  {Somerville} R.~S.,  {Hu} C.-Y.,  {Teyssier} R.,  {Burkhart} B.,   {Hernquist} L.,  2021, \mn@doi [\mnras] {10.1093/mnras/stab1896}, \href {https://ui.adsabs.harvard.edu/abs/2021MNRAS.506.3882S} {506, 3882}

\bibitem[\protect\citeauthoryear{Springel}{Springel}{2005}]{Springel_2005}
Springel V.,  2005, \mn@doi [\mnras] {10.1111/j.1365-2966.2005.09655.x}, 364, 1105

\bibitem[\protect\citeauthoryear{{Springel}, {Di Matteo}  \& {Hernquist}}{{Springel} et~al.}{2005}]{2005MNRAS.361..776S}
{Springel} V.,  {Di Matteo} T.,   {Hernquist} L.,  2005, \mn@doi [\mnras] {10.1111/j.1365-2966.2005.09238.x}, \href {https://ui.adsabs.harvard.edu/abs/2005MNRAS.361..776S} {361, 776}

\bibitem[\protect\citeauthoryear{Steinwandel, Moster, Naab, Hu  \& Walch}{Steinwandel et~al.}{2020}]{Steinwandel_2020}
Steinwandel U.~P.,  Moster B.~P.,  Naab T.,  Hu C.-Y.,   Walch S.,  2020, \mn@doi [\mnras] {10.1093/mnras/staa821}, 495, 1035

\bibitem[\protect\citeauthoryear{Trebitsch, Volonteri, Dubois  \& Madau}{Trebitsch et~al.}{2018}]{Trebitsch_2018}
Trebitsch M.,  Volonteri M.,  Dubois Y.,   Madau P.,  2018, \mn@doi [\mnras] {10.1093/mnras/sty1406}, 478, 5607–5625

\bibitem[\protect\citeauthoryear{{Tremaine}, {Ostriker}  \& {Spitzer}}{{Tremaine} et~al.}{1975}]{1975ApJ...196..407T}
{Tremaine} S.~D.,  {Ostriker} J.~P.,   {Spitzer} L. J.,  1975, \mn@doi [\apj] {10.1086/153422}, \href {https://ui.adsabs.harvard.edu/abs/1975ApJ...196..407T} {196, 407}

\bibitem[\protect\citeauthoryear{Tremmel, Karcher, Governato, Volonteri, Quinn, Pontzen, Anderson  \& Bellovary}{Tremmel et~al.}{2017}]{Tremmel_2017}
Tremmel M.,  Karcher M.,  Governato F.,  Volonteri M.,  Quinn T.~R.,  Pontzen A.,  Anderson L.,   Bellovary J.,  2017, \mn@doi [\mnras] {10.1093/mnras/stx1160}, 470, 1121–1139

\bibitem[\protect\citeauthoryear{{{\"U}bler} et~al.,}{{{\"U}bler} et~al.}{2023}]{2023A&A...677A.145U}
{{\"U}bler} H.,  et~al., 2023, \mn@doi [\aap] {10.1051/0004-6361/202346137}, \href {https://ui.adsabs.harvard.edu/abs/2023A&A...677A.145U} {677, A145}

\bibitem[\protect\citeauthoryear{Volonteri \& Perna}{Volonteri \& Perna}{2005}]{Volonteri_2005}
Volonteri M.,  Perna R.,  2005, \mn@doi [\mnras] {10.1111/j.1365-2966.2005.08832.x}, 358, 913–922

\bibitem[\protect\citeauthoryear{{Volonteri}, {Madau}  \& {Haardt}}{{Volonteri} et~al.}{2003}]{2003ApJ...593..661V}
{Volonteri} M.,  {Madau} P.,   {Haardt} F.,  2003, \mn@doi [\apj] {10.1086/376722}, \href {https://ui.adsabs.harvard.edu/abs/2003ApJ...593..661V} {593, 661}

\bibitem[\protect\citeauthoryear{Weisz, Skillman, Cannon, Dolphin, Kennicutt, Lee  \& Walter}{Weisz et~al.}{2009}]{Weisz_2009}
Weisz D.~R.,  Skillman E.~D.,  Cannon J.~M.,  Dolphin A.~E.,  Kennicutt R.~C.,  Lee J.,   Walter F.,  2009, \mn@doi [\apj] {10.1088/0004-637x/704/2/1538}, 704, 1538–1569

\bibitem[\protect\citeauthoryear{{Westera}, {Lejeune}, {Buser}, {Cuisinier}  \& {Bruzual}}{{Westera} et~al.}{2002}]{2002A&A...381..524W}
{Westera} P.,  {Lejeune} T.,  {Buser} R.,  {Cuisinier} F.,   {Bruzual} G.,  2002, \mn@doi [\aap] {10.1051/0004-6361:20011493}, \href {https://ui.adsabs.harvard.edu/abs/2002A&A...381..524W} {381, 524}

\bibitem[\protect\citeauthoryear{{Wiersma}, {Schaye}  \& {Smith}}{{Wiersma} et~al.}{2009}]{2009MNRAS.393...99W}
{Wiersma} R. P.~C.,  {Schaye} J.,   {Smith} B.~D.,  2009, \mn@doi [\mnras] {10.1111/j.1365-2966.2008.14191.x}, \href {https://ui.adsabs.harvard.edu/abs/2009MNRAS.393...99W} {393, 99}

\bibitem[\protect\citeauthoryear{Zaw, Rosenthal, Katkov, Gelfand, Chen, Greenhill, Brisken  \& Noori}{Zaw et~al.}{2020}]{Zaw_2020}
Zaw I.,  Rosenthal M.~J.,  Katkov I.~Y.,  Gelfand J.~D.,  Chen Y.-P.,  Greenhill L.~J.,  Brisken W.,   Noori H.~A.,  2020, \mn@doi [\apj] {10.3847/1538-4357/ab9944}, 897, 111

\bibitem[\protect\citeauthoryear{Zheng, Shi, Bian, Yu, Wang, Chen, Li  \& Gu}{Zheng et~al.}{2023}]{Zheng_2023}
Zheng Z.,  Shi Y.,  Bian F.,  Yu X.,  Wang J.,  Chen J.,  Li X.,   Gu Q.,  2023, \mn@doi [\mnras] {10.1093/mnras/stad1642}, 523, 3274–3285

\makeatother
\end{thebibliography}

\bsp	
\label{lastpage}
\end{document}